\documentclass[11pt]{article}
\pdfoutput=1
\usepackage[text={17cm,23.4cm},centering]{geometry}
\usepackage{graphicx}
\usepackage{amsmath,amssymb,amsthm,upgreek}
\usepackage{authblk}
\usepackage[format=hang,indention=-3em,textfont={small,it},labelfont={small,bf},width=0.92\textwidth]{caption}

\newtheoremstyle{mythm}{2ex}{2ex}{\itshape}{}{\color{darkgray}\normalfont\normalsize\bfseries\sffamily\slshape}{}{.5em}{\thmname{#1}\thmnumber{ #2}\thmnote{ (#3)}}
\newtheorem{prop}{Proposition}
\newtheorem{remark}{Remark}
\newtheorem{lemma}{Lemma}
\newcommand{\periodafter}[1]{#1.}
\usepackage{titlesec}
\usepackage{titletoc}
\titleformat{\section}{\color{RoyalBlue3}\normalfont\large\sffamily\bfseries}{\thesection}{1em}{}
\titleformat{\subsection}[runin]{\color{RoyalBlue4}\normalfont\normalsize\sffamily\bfseries}{\thesubsection}{1em}{\periodafter}
\titleformat{\subsubsection}[runin]{\color{RoyalBlue4}\normalfont\normalsize\sffamily\bfseries\itshape}{\thesubsubsection}{1em}{\periodafter}
\titleformat{\paragraph}[runin]{\color{RoyalBlue4}\normalfont\normalsize\slshape\bfseries\sffamily}{}{}{\periodafter}
\titlespacing{\section}{0pt}{3.5ex plus 1ex minus .2ex}{2.3ex plus .2ex}
\titlespacing{\subsection}{0em}{3.5ex plus 1ex minus .2ex}{1.5ex plus .2ex}
\titlespacing{\subsubsection}{1.3em}{3.25ex plus 1ex minus .2ex}{1ex plus .2ex}
\titlespacing{\paragraph}{0pt}{3ex plus 1ex minus .2ex}{1em}
\titlespacing*{\subparagraph}{1.3em}{2ex plus 1ex minus .2ex}{1em}
\usepackage[usenames,dvipsnames,x11names]{xcolor}

\graphicspath{{figures/}}

\makeatletter
\newcommand{\proofstep}[1]{%
\par
\addvspace{\smallskipamount}
\textit{#1\@addpunct{.}}\enspace\ignorespaces}
\makeatother

\usepackage{autonum}

\usepackage[notref,notcite,color,final]{showkeys}      
\definecolor{refkey}{named}{blue}
\definecolor{labelkey}{named}{blue}
\let\oldbibliography\thebibliography
\renewcommand{\thebibliography}[1]{%
  \oldbibliography{#1}%
  \setlength{\itemsep}{1ex}%
}
\newcommand{\hh}{\hspace*{0.2pt}}
\newcommand{\exs}{\hspace*{0.6pt}}
\newcommand{\alphad}{\dot{\alpha}}

\newcommand{\bfa} {\boldsymbol{a}}
\newcommand{\bfA} {\boldsymbol{A}}
\newcommand{\bfAH} {\widehat{\boldsymbol{A}}{}}
\newcommand{\bfalb}{\bar{\bfal}{}}
\newcommand{\bfalbH}{\overline{\bfalH}}
\newcommand{\bfal}{\boldsymbol{\alpha}}
\newcommand{\bfald}{\dot{\bfal}}
\newcommand{\bff}{\boldsymbol{f}}
\newcommand{\bfalH}{\widehat{\bfal}}
\newcommand{\bfalp}{\bfal_{\mathsf{p}}}

\newcommand{\bfAp}{\bfA_{\mathsf{p}}}
\newcommand{\bfbe}{\boldsymbol{\beta}}
\newcommand{\bfbep}{\boldsymbol{\beta}_{\mathsf{p}}}
\newcommand{\bfe} {\boldsymbol{e}}
\newcommand{\bfepsb} {\bar{\bfeps}{}}
\newcommand{\bfeps}{\boldsymbol{\varepsilon}}
\newcommand{\bfepsd}{\dot{\bfeps}}
\newcommand{\bfeta}{\boldsymbol{\eta}}
\newcommand{\bfetap}{\boldsymbol{\eta}_{\mathsf{p}}}
\newcommand{\bfetaR}{\bfeta\Rsub}

\newcommand{\bfg} {\boldsymbol{g}}
\newcommand{\bfd} {\boldsymbol{d}}
\newcommand{\bfI} {\boldsymbol{I}}
\newcommand{\bfn} {\boldsymbol{n}}
\newcommand{\bfr} {\boldsymbol{r}}
\newcommand{\bfs} {\boldsymbol{s}}

\newcommand{\bfsig}{\boldsymbol{\sigma}}

\newcommand{\bfsige}{\bfsig^{\mathrm{e}}}
\newcommand{\bfsigep}{\bfsige_{\mathsf{p}}}
\newcommand{\bfsigHe}{\widehat{\bfsig}{}^{\mathrm{e}}}
\newcommand{\bfsigHv}{\widehat{\bfsig}{}^{\mathrm{v}}}
\newcommand{\bfsigv}{\bfsig^{\mathrm{v}}}
\newcommand{\bfsigvp}{\bfsigv_{\mathsf{p}}}
\newcommand{\bfsR}{\bfs\Rsub}
\newcommand{\bfsigR}{\bfsig\!\Rsub}

\newcommand{\bfubH}{\overline{\bfuH}}
\newcommand{\bfu} {\boldsymbol{u}}
\newcommand{\bfuH} {\widehat{\boldsymbol{u}}{}}
\newcommand{\bfum} {\bfu\obs}
\newcommand{\bfup}{\bfu_{\mathsf{p}}}
\newcommand{\bfvbb} {\overline{\boldsymbol{v}}{}}
\newcommand{\bfv} {\boldsymbol{v}}
\newcommand{\bfwb} {\bar{\boldsymbol{w}}{}}
\newcommand{\bfwbH}{\overline{\bfwH}}
\newcommand{\bfw} {\boldsymbol{w}}
\newcommand{\bfwH} {\widehat{\boldsymbol{w}}{}}
\newcommand{\bfwp}{\bfw_{\mathsf{p}}}
\newcommand{\bfwR}{\bfw\Rsub}
\newcommand{\bfwRd}{\dot{\bfw}\Rsub}

\newcommand{\bfxbH}{\overline{\bfxH}}
\newcommand{\bfx} {\boldsymbol{x}}
\newcommand{\bfxH} {\widehat{\boldsymbol{x}}{}}
\newcommand{\bfze}{\mathbf{0}}

\newcommand{\bsfp} {\boldsymbol{\mathsf{p}}}
\newcommand{\bsfpH} {\hat{\bsfp}}
\newcommand{\bsfu} {\boldsymbol{\mathsf{u}}}
\newcommand{\bsfv} {\boldsymbol{\mathsf{v}}}
\newcommand{\bsfX} {\boldsymbol{\mathsf{X}}}

\newcommand{\bsfXp}{\bsfX_{\mathsf{p}}}
\newcommand{\bsfY} {\boldsymbol{\mathsf{Y}}}
\newcommand{\Bsub}{_{\text{\tiny B}}}
\newcommand{\CA}{\CS^{\star}_{\!A}}
\newcommand{\Ca}{\CS_{\alpha}}

\newcommand{\CaS}{\Ca\Ssup}
\newcommand{\Ccal}{\mathcal{C}}

\newcommand{\Ce}{\CS_{\eps}}

\newcommand{\CeS}{\Ce\Ssup}
\newcommand{\CH}{\widehat{\CS}}
\newcommand{\CHT}{\widehat{\CS}{}\Tsup}
\newcommand{\Cm}{\CS_{\mathrm{m}}}
\newcommand{\Cms}{\CS^{\star}_{\mathrm{m}}}
\newcommand{\CmsT}{\CS^{\star}_{\mathrm{m}}\!\nes\Tsup}
\newcommand{\CmT}{\CS_{\mathrm{m}}}
\newcommand{\Cs}{\CS^{\star}_{\sigma}}

\newcommand{\CS}{\text{\boldmath $\mathcal{C}$}}
\newcommand{\DA}{\DS^{\star}_{\!A}}
\newcommand{\Da}{\DS_{\nes\alpha}}

\newcommand{\DaS}{\Da\Ssup}

\newcommand{\De}{\DS_{\eps}}

\newcommand{\DeS}{\De\Ssup}
\newcommand{\del}[1][]{\partial_{#1}}
\newcommand{\demi} {\tfrac{1}{2}}

\newcommand{\dip} {\! :\!}
\newcommand{\Dm}{\DS_{\mathrm{m}}}
\newcommand{\DmT}{\DS_{\mathrm{m}}}
\newcommand{\Dms}{\DS^{\star}_{\mathrm{m}}}
\newcommand{\DmsT}{\DS^{\star}_{\mathrm{m}}\!\nes\Tsup}
\newcommand{\dotp}{\raisebox{1pt}{\hspace*{1pt}\scalebox{0.45}{$\bullet$}}\hspace*{1pt}}

\newcommand{\DS}{\mbox{\boldmath $\mathcal{D}$}}
\newcommand{\Ds}{\DS^{\star}_{\sigma}}

\newcommand{\dS}{\,\text{d}\Gamma}
\newcommand{\dtau}{\,\text{d}\tau}
\newcommand{\dth}{\;\text{d}\theta}
\newcommand{\dt}{\,\text{d}t}
\newcommand{\dV}{\;\text{d}\Omega}
\newcommand{\Ecale}{\Ecal{}^{\mathrm{e}}}
\newcommand{\Ecalec}{\check{\Ecal}{}^{\mathrm{e}}}
\newcommand{\Ecalvc}{\check{\Ecal}{}^{\mathrm{v}}}
\newcommand{\Ecalc}{\check{\Ecal}}
\newcommand{\Ecal}{\mathcal{E}}

\newcommand{\EcalR}{\widetilde{\Ecal}}
\newcommand{\Ecalv}{\Ecal{}^{\mathrm{v}}}
\newcommand{\ECR}{_{\text{\tiny ECR}}}
\newcommand{\epsd}{\dot{\eps}}
\newcommand{\eps}{\varepsilon}
\newcommand{\Eq}[1]{\stackrel{\text{#1}}{=}}
\newcommand{\Fcal}{\mathcal{F}}
\newcommand{\bdot}{\boldsymbol{\cdot}}
\newcommand{\FS}{\ensuremath{\mbox{\boldmath $\mathcal{F}$}}}
\newcommand{\FSR}{\FS\Rsub}
\newcommand{\Gc}{\G_{c}}
\newcommand{\GD}{\Gamma_{\!\text{D}}}
\newcommand{\G}{\Gamma}
\newcommand{\GN}{\Gamma_{\!\text{N}}}

\newcommand{\iG} {\int_{\G}}
\newcommand{\inv}[1]{\dfrac{1}{#1}}
\newcommand{\iO}{\int_{\OO}}
\newcommand{\IS}{\mbox{\boldmath $\mathcal{I}$}}
\newcommand{\Isub}{_{\text{\tiny I}}}
\newcommand{\JS}{\mbox{\boldmath $\mathcal{J}$}}
\newcommand{\KS}{\mbox{\boldmath $\mathcal{K}$}}
\newcommand{\Lambdak}{\Lambda_{\kappa}}
\newcommand{\LambdakL}{\Lambda_{\kappa,L}}
\newcommand{\LambdaRk}{\widetilde{\Lambda}{}_{\kappa}}
\newcommand{\lbra}{\big\langle}
\newcommand{\Lcal}{\mathcal{L}}
\newcommand{\LcalB}{\breve{\Lcal}}
\newcommand{\lcb}{\big\{}
\newcommand{\Lcb}{\Big\{}

\newcommand{\lpar}{\big(}
\newcommand{\lsqb}{\big[}
\newcommand{\LSsub}{_{\text{\tiny LS}}}
\newcommand{\Mcal}{\mathcal{M}}

\newcommand{\McalRT}{\widetilde{\Mcal}_{\nes\textit{\tiny T}}}

\newcommand{\obs}{_{\text{obs}}}
\newcommand{\uiobs}{u_{i,\text{obs}}}
\newcommand{\oo}{\omega}
\newcommand{\OO}{\Omega}
\newcommand{\PS}{\mbox{\boldmath $\mathcal{P}$}}
\newcommand{\QS}{\mbox{\boldmath $\mathcal{Q}$}}
\newcommand{\rabs}{\big|}
\newcommand{\Rbb} {\mathbb{R}}
\newcommand{\rbra}{\big\rangle}
\newcommand{\rcb}{\big\}}
\newcommand{\Rcb}{\Big\}}
\newcommand{\rmi}{\mathrm{i}}

\newcommand{\rpar}{\big)}
\newcommand{\rsqb}{\big]}
\newcommand{\Rsub}{_{\text{\tiny R}}}
\newcommand{\sfp} {\mathsf{p}}

\newcommand{\shcup}{\hspace*{-0.1em}\cup\hspace*{-0.1em}}
\newcommand{\shdeq}{\hspace*{-0.1em}:=\hspace*{-0.1em}}
\newcommand{\sheq}{\hspace*{-0.1em}=\hspace*{-0.1em}}
\newcommand{\shgeq}{\hspace*{-0.1em}\geqslant\hspace*{-0.1em}}
\newcommand{\shg}{\hspace*{-0.1em}>\hspace*{-0.1em}}
\newcommand{\shin}{\hspace*{-0.1em}\in\hspace*{-0.1em}}
\newcommand{\shleq}{\hspace*{-0.1em}\leqslant\hspace*{-0.1em}}
\newcommand{\shm}{\hspace*{-0.1em}-\hspace*{-0.1em}}
\newcommand{\shp}{\hspace*{-0.1em}+\hspace*{-0.1em}}
\newcommand{\shsetm}{\hspace*{-0.1em}\setminus\hspace*{-0.1em}}
\newcommand{\shsubs}{\hspace*{-0.1em}\subset\hspace*{-0.1em}}
\newcommand{\shtimes}{\hspace*{-0.1em}\times\hspace*{-0.1em}}
\newcommand{\sige}{\sigma^{\mathrm{e}}}
\newcommand{\sigv}{\sigma^{\mathrm{v}}}
\newcommand{\sip} {\! \cdot\!}
\renewcommand{\SS}{\mbox{\boldmath $\mathcal{S}$}}
\newcommand{\Ssub}{_{\text{\tiny S}}}
\newcommand{\Ssup}{^{\text{\tiny S}}}
\newcommand{\blsup}{^{\text{\tiny (0)}}}
\newcommand{\trsup}{^{\text{\tiny (1)}}}
\newcommand{\suite}[1][0ex]{\notag \\[#1] & \mbox{}\hspace{15pt}}
\newcommand{\tdemi} {\tfrac{1}{2}\exs}
\newcommand{\tens}{\hspace*{-1pt}\otimes\hspace*{-1pt}}
\newcommand{\tiers} {\tfrac{1}{3}}
\newcommand{\Tr}{\hh\text{tr}\hspace*{0.7pt}}
\newcommand{\Tsup}{^{\text{\tiny T}}}
\newcommand{\Ucal}{\mathcal{U}}
\newcommand{\Vcal}{\mathcal{V}}
\newcommand{\Wcal}{\mathcal{W}}
\renewcommand{\del}[1]{\partial_{#1}}
\renewcommand{\Re}{\text{Re}}
\newcommand{\McalT}{\Mcal_{\nes\textit{\tiny T}}}

\newcommand{\Tinv}{T^{-1}}
\newcommand{\nes}{\hspace*{-0.6pt}}

\title{{\color{RoyalBlue3}\normalfont\Large\sffamily\bfseries Error-in-constitutive-relation (ECR) framework for the characterization of linear viscoelastic solids}}

\author[1]{Marc Bonnet}
\author[2]{Prasanna Salasiya}
\author[2]{Bojan B. Guzina\thanks{Corresponding author (guzin001@umn.edu)}} 
\affil[1]{{\small POEMS (CNRS-INRIA-ENSTA), Dept. of Applied Mathematics, ENSTA Paris, France}}
\affil[2]{\small{Dept. of Civil, Environmental, and Geo-Engineering, University of Minnesota, Twin Cities}}

\date{\today}

\begin{document}

\maketitle

\begin{abstract}
\noindent We develop an error-in-constitutive-relation (ECR) approach toward the full-field characterization of linear viscoelastic solids described within  the framework of standard generalized materials. To this end, we formulate the viscoelastic behavior in terms of the (Helmholtz) free energy potential and a dissipation potential. Assuming the availability of full-field interior kinematic data, the constitutive mismatch between the kinematic quantities (strains and internal thermodynamic variables) and their ``stress'' counterparts (Cauchy stress tensor and that of thermodynamic tensions), commonly referred to as the ECR functional, is established with the aid of Legendre-Fenchel gap functionals linking the thermodynamic potentials to their energetic conjugates. We then proceed by introducing the modified ECR (MECR) functional as a linear combination between its ECR parent and the kinematic data misfit, computed for a trial set of constitutive parameters. The affiliated stationarity conditions then yield two coupled evolution problems, namely (i) the forward evolution problem for the (trial) displacement field driven by the constitutive mismatch, and (ii) the backward evolution problem for the adjoint field driven by the data mismatch. This allows us to establish compact expressions for the MECR functional and its gradient with respect to the viscoelastic constitutive parameters. For generality, the formulation is established assuming both time-domain (i.e. transient) and frequency-domain data. We illustrate the developments in a two-dimensional setting by pursuing the multi-frequency MECR reconstruction of (i) piecewise-homogeneous standard linear solid, and (b) smoothly-varying Jeffreys viscoelastic material.
\end{abstract} 
\newpage

\section{Introduction}

\noindent Material characterization of lossy solids through the lens of linear viscoelasticity carries a broad range of applications across sciences and engineering disciplines. In geophysics and geomechanics \cite{guzina2005spectral,xia2014estimation,zong2015complex,chen2018estimating}, for instance, effective parameters of energy dissipation -- often synthesized via the reciprocal P- and S-wave quality factors -- are known to relate to hydrocarbon reservoir parameters, the state of crust and uppermost mantle, tectonic evolution, and hydrogeology. In medicine, on the other hand, viscoelastic properties are known to correlate with tissue pathology, see e.g. \cite{mariappan2010magnetic,parker2005unified}. One potential application that motivates this study is that of mineral CO$_2$ storage~\cite{snaebjornsdottir2020carbon}, which puts a spotlight on the (presently unknown) mechanical fingerprint of reactive flow in mafic and ultramafic rocks. When dealing with lossy solids, the conventional approach of linear elasticity to non-invasive material interrogation is clearly inadequate for it ignores the dissipation-induced wave attenuation and dispersion~\cite{charara2005full}. 

In certain applications such as medical diagnosis, the use of ultrasound and magnetic resonance imaging facilitates the remote sensing of \emph{interior} deformation (or motion) data \mbox{\cite{sigrist2017ultrasound,mariappan2010magnetic}}, which can then be used as sensory input for material characterization. In experimental mechanics, a similar claim can be made in terms of laser Doppler vibrometer observations~\cite{tokmashev2013experimental,pourahmadian2018elastic} of the plane-stress ultrasonic wavefields propagating through slab-like solid specimens. The class of inverse solutions that operate on this premise are commonly known as \emph{elastography} techniques, see~\cite{greenleaf2003selected,parker2005unified} for reviews. In the context of viscoelastic characterization, there are a variety of avenues to elastography, including direct algebraic inversion \cite{oliphant2001complex,sinkus2005imaging,pourahmadian2018elastic}, adjoint state techniques \cite{yuan2010reconstruction,zhang2012solution,tan2016gradient}, and error-in-constitutive-relation (ECR) approach~\cite{diaz2015modified}. While the algebraic inversion methods postulate spatial differentiability of the interior data which makes them susceptible to measurement errors, all existing approaches to viscoelastic elastography rely on the premises of either (i) time-harmonic motion or (ii) specific constitutive model, or both. Thanks to the correspondence principle in linear  viscoelasticity~\cite{findley2013creep}, the premise of time-harmonic motion is particularly helpful for it permits generalization of the elastic solutions by allowing the moduli to be complex-valued. 

To help establish a universal framework for the viscoelastic characterization of lossy solids from the interior deformation data we pursue the modified ECR framework, in both time- and frequency-domain, by formulating the viscoelastic constitutive behavior as that of a standard generalized material (SGM)~\cite{Halp75,ger:nqs:suq:83}. The SGM framework describes a broad family of lossy constitutive models that are formulated via two convex thermodynamic potentials, namely the Helmholtz free energy and dissipation potential. In short, the ECR approach \cite{ladeveze:83} to elastography relies on minimization of a functional that quantifies the misfit in a constitutive relation that connects kinematically-admissible strains and physically-admissible stresses satisfying the balance of linear momentum. More recently, researchers have pursued a modified ECR (MECR) functional \cite{lad:ned:rey:94,allix:05,feissel:allix:06,B-2016-07,B-2012-2,diaz2015modified} -- adopted in this study - that endows its ECR predecessor with a penalty term featuring the (interior) kinematic data misfit. 

We begin the analysis in Section~\ref{SGM} by describing a linear viscoelastic solid within the SGM framework featuring the relevant kinematic quantities and their energetic conjugates. To help establish the MECR method in a thermodynamic setting, in Section~\ref{sec:ECR} we formulate the conjugate (free energy and dissipation) potentials via the Legendre-Fenchel transform, which then motivates introduction of the respective (constitutive) gap densities. Section~\ref{sec:ECR:min} formulates the constrained optimization problem for the MECR-based cost functional, whose Lagrangian combines the ECR functional, a quadratic penalty term quantifying the kinematic data misfit, and a weak statement of the balance of linear momentum.  The affiliated stationarity conditions then yield two coupled evolution problems, namely (i) the forward evolution problem for the (trial) displacement field driven by the constitutive mismatch, and (ii) the backward evolution problem for the adjoint field driven by the data mismatch. Section~\ref{tharmonic} and Section~\ref{vispecial} specialize the MECR method to time-harmonic interior data and a pair of classical viscoelastic models, respectively. The analysis is illustrated in Section~\ref{numres} via two-dimensional reconstruction of (i) piecewise-uniform and (ii) smoothly-graded viscoelastic solid. 

\section{Linear viscoelastic solid as a standard generalized material} \label{SGM}

\noindent \emph{Standard generalized material} (SGM) describes a broad family of constitutive models for solids that are formulated via two convex thermodynamic potentials, namely the free energy $\psi$ and dissipation potential $\varphi$. The seminal paper on SGMs appears to be~\cite{Halp75}; an early review of the related continuum thermodynamics concepts is available in~\cite{ger:nqs:suq:83}, and these ideas also appear in monographs such as~\cite[Sec.~5.1]{maugin:92}, \cite[Chap.~2]{son:book} and~\cite[Chaps.~1,\,2,\,7]{simo}. Hereon we assume small-deformation conditions, implying in particular that the material time derivative is well approximated by the partial time derivative. Letting $\bfu$ denote the time history of the germane displacement field, constitutive models belonging to the SGM class are formulated in terms of two convex potentials, namely the Helmholtz free energy $\psi=\psi(\bfeps,\bfal)$ and a dissipation potential $\varphi=\varphi(\bfepsd,\bfald)$, where $\bfeps=\bfeps[\bfu]$ is the linearized strain tensor, $\bfal$ is a tensor of internal thermodynamic variables, and the over-dot symbol stands for the partial time derivative. For the present case of linear viscoelasticity, both potentials $\psi$ and $\varphi$ are differentiable (and in fact quadratic) functions of their arguments. On decomposing the Cauchy stress tensor $\bfsig$ caused by~$\bfu$ into a reversible component $\bfsige$ and irreversible component $\bfsigv$ as 
\begin{equation}
\bfsig[\bfu] = \bfsige[\bfu] +\bfsigv[\bfu],   \label{stress0}
\end{equation}
the SGM constitutive relationships then yield $\bfsige$, $\bfsigv$ and the tensor of thermodynamic tensions $\bfA$ (energetically conjugate to~$\bfal$) as
\begin{equation}
\bfsige[\bfu] = \del{\eps}\psi, \qquad\quad \bfsigv[\bfu]=\del{\epsd}\varphi, \qquad\quad
\bfA[\bfu] = -\del{\alpha}\psi = \del{\alphad}\varphi. \label{stress}
\end{equation}
Assuming isothermal conditions (which is our premise going forward), the power per unit volume $\mathfrak{D}=\mathfrak{T\dot{S}}$ ($\mathfrak{T}=\,$temperature, $\mathfrak{S}=\,$specific entropy) dissipated by a material obeying~\eqref{stress} reads
\begin{equation}
\mathfrak{D} = \bfsigv\dip\bfepsd + \bfA\dip\bfald, \label{dissip}
\end{equation}
noting that the sign convention in the expressions for~$\bfA[\bfu]$ is chosen so that $\mathfrak{D}$ has the form~\eqref{dissip}. 

\subsection{Linear viscoelastic solid} \label{sec:generic}
Within the SGM formalism, we consider the class of viscoelastic solids described by the free-energy potential and a dissipation potential given respectively by 
\begin{equation}
\begin{aligned}
 \psi(\bfeps,\bfal)
 &= \demi\bigl( \bfeps\dip\Ce\dip\bfeps + 2\bfeps\dip\CmT\dip\bfal + \bfal\dip\Ca\dip\bfal \bigr), \\*[2mm]
 \varphi(\bfepsd,\bfald)
 &= \demi\bigl( \bfepsd\dip\De\dip\bfepsd + 2\bfepsd\dip\DmT\dip\bfald + \bfald\dip\Da\dip\bfald \bigr),
\end{aligned} \label{phi:psi:def}
\end{equation}
where (i) the second-order symmetric tensor $\bfal$ (i.e.~``viscoelastic strain'' tensor) collects internal variables of the model; (ii)  $\Ce,\Ca,\De$ and~$\Da$ are fourth-order tensors, characterized by major and minor symmetries, that define non-negative quadratic forms over the second-order symmetric tensors; and (iii)  $\Cm$ and~$\Dm$ are the fourth-order tensors carrying minor symmetries and must be such that $\psi(\bfeps,\bfal)$ and $\varphi(\bfepsd,\bfald)$ are positive -- and so convex -- functions. For clarity of exposition, we hereon assume that $\Cm$ and~$\Dm$ also carry the major symmetry. This (unnecessary but convenient) simplifying hypothesis covers most cases of practical interest. Finally, tensors $\Ce$ and $\Da$ are assumed to be invertible and hence positive definite. 

\subsection{Constitutive relations}

The constitutive equations~(\ref{stress}) imply $\del{\alpha}\psi+\del{\alphad}\varphi=\bfze$, i.e.
\begin{equation}
  \Da\dip\bfald + \Ca\dip\bfal + \Dm\dip\bfepsd + \Cm\dip\bfeps = \bfze. \label{evol}
\end{equation}
For a given strain history $\bfeps(t)$, the above identity is a tensor-valued first-order ordinary differential equation (ODE) for the internal variable history $\bfal(t)$. Solving this tensor ODE by means of Duhamel's formula yields, under the assumption of quiescent past, the viscous strain $\bfal$ as a linear functional on the strain history up to time $t$: we have
\begin{equation}
  \bfal(t) = -\FS[\Cm\dip\bfeps\shp\Dm\dip\bfepsd](t), \label{alpha:expr}
\end{equation}
where the tensor-valued linear convolution operator $\bfs\mapsto\FS[\bfs]$ is given by
\begin{equation}
\FS[\bfs](t) = \int_{0}^{t} \exp[-\QS(t\shm\tau)]\dip\Da^{-1}\dip\bfs(\tau) \dtau, \qquad
\QS\shdeq\Da^{-1}\dip\Ca  \label{KS:def}
\end{equation}
(the exponential of a tensor being defined like that of a matrix) and verifies
\begin{equation}
  \Da\dip\dot{\FS}[\bfs] + \Ca\dip\FS[\bfs] - \bfs = \bf0 \label{sol:prop}. 
\end{equation}
On integrating by parts the term featuring $\bfepsd$, the expression~\eqref{alpha:expr} of $\bfal$ is recast as
\begin{equation}
\bfal[\bfu](t) = -\Da^{-1}\dip\Dm\dip\bfeps(t) - \FS[\CH\dip\bfeps](t), \qquad\text{with}\quad
\CH:=\Cm-\Ca\dip\Da^{-1}\dip\Dm \label{alpha(t)}. \label{alpha(t)}
\end{equation}

By~\eqref{stress} and~\eqref{phi:psi:def}, the viscoelastic stress tensor $\bfsig[\bfu](t)$ due to given strain history $\bfeps[\bfu](\tau)$, $\tau\shleq t$ reads 
\begin{equation}
  \bfsig[\bfu](t)
 = \del{\eps}\psi + \del{\epsd}\varphi
 = \De\dip\bfepsd + \Ce\dip\bfeps + \DmT\dip\bfald[\bfu] + \CmT\dip\bfal[\bfu]. \label{sigma}
\end{equation}
Then, eliminating $\bfald[\bfu]$ in the above formula by way of~\eqref{evol}, using~\eqref{alpha(t)} and
rearranging terms, one finds
\begin{equation}
  \bfsig[\bfu](t)
 = \CS\Isub \dip\bfeps(t) + \DS\Isub\dip\bfepsd(t) - \CHT\!\dip\FS[\CH\dip\bfeps](t), \label{sig:expr}
\end{equation}
where $\CH$ is given by~\eqref{alpha(t)} and
\begin{equation}
  \CS\Isub = \Ce - \CmT\dip\Ca^{-1}\dip\Cm + \CHT\dip\Ca^{-1}\dip\CH, \qquad\qquad \DS\Isub = \De-\DmT\dip\Da^{-1}\dip\Dm
\label{C:inst}
\end{equation}
signify respectively the \emph{instantaneous} elasticity and viscosity tensors that are characterized by major (and minor) symmetries.

\begin{remark} \label{symm1}
In general, the assumptions made in Section~\ref{sec:generic} do not guarantee the major symmetry of tensors $\QS$ and $\CH$; however they do when the material tensors in~\eqref{phi:psi:def} are isotropic.
\end{remark}

\subsection{Reciprocity property}

The viscoelastic constitutive model~\eqref{sig:expr} carries the following reciprocity property (an extension of the symmetry relationship in linear elasticity) and plays a key role in this study. See Appendix~\ref{reciprocity:proof}, electronic supplementary material (ESM) for proof.  

\begin{lemma}[Stress-strain reciprocity]\label{reciprocity}
Let $\bfu,\bfw$ be two time-dependent displacement fields defined over the time interval $[0,T]$. Let $\bfwR(\bfx,\dotp):=\bfw(\bfx,T\shm\dotp)$ and $\bfsigR(\bfx,\dotp):=\bfsig(\bfx,T\shm\dotp)$ be the time-reversed versions of $\bfw$ and~$\bfsig$, respectively. Then, the strain histories $\bfeps[\bfu],\bfeps[\bfw]$ and viscoelastic stress histories $\bfsig[\bfu],\bfsigR[\bfwR]$ obey the reciprocity identity
\begin{equation}
\int_0^T \big( \bfsig[\bfu]\dip\bfeps[\bfw] - \bfsigR[\bfwR]\dip\bfeps[\bfu] \big) \dt
  \,=\, \lpar \bfeps[\bfw]\dip\DS\Isub\dip\bfeps[\bfu] \rpar \rabs^T_0.
\end{equation}
\end{lemma}

\section{Error in constitutive relation} \label{sec:ECR}

\noindent To resolve the SGM material parameters featured in~\eqref{phi:psi:def} from the full-field sensory data~$\bfum$, the general idea is (as usual) to minimize a misfit between the physical observations and their (viscoelastic) simulations~$\bfu$. In this work we endow the objective function, using Legendre-Fenchel constitutive gap functionals, with a misfit between (i) the \emph{viscoelastic} stress field $\bfsig[\bfu]$ that is compatible with the assumed constitutive model, and (ii) the ``\emph{optimal}'' stress field $\bfsig$ that solves the stationarity problem which accounts for both the constitutive gap, the data misfit, and the constraint that $\bfsig$ must satisfy the balance of linear momentum. By analogy to~\eqref{stress0}, we shall make use of the decomposition $\bfsig=\bfsige+\bfsigv$ in that, loosely speaking, $\bfsig\to\bfsig[\bfu]$ implies $\bfsige\to\bfsige[\bfu]$ and~$\bfsigv\to\bfsigv[\bfu]$.

\subsection{Legendre-Fenchel constitutive gap functionals}

Error in constitutive relation (ECR) functionals for history-dependent constitutive models belonging to the SGM class can be expressed in terms of the Legendre-Fenchel gaps associated with potentials $\psi$ and $\varphi$. These gaps involve the conjugate potentials $\psi^{\star}$ and $\varphi^{\star}$, defined as the Legendre-Fenchel transforms of $\psi$ and $\varphi$ (see e.g. \cite[App.~2]{maugin:92}), namely
\begin{subequations}
\begin{align}
  \psi^{\star}(\bfsige,\bfA)
 &= \max_{\eps,\alpha} \bigl[ \bfsige\dip\bfeps-\bfA\dip\bfal - \psi(\bfeps,\bfal) \bigr],
\label{psi:star} \\
  \varphi^{\star}(\bfsigv,\bfA)
 &= \max_{\epsd,\alphad} \bigl[ \bfsigv\dip\bfepsd+\bfA\dip\bfald
   - \varphi(\bfepsd,\bfald) \bigr]. \label{phi:star}
\end{align}
\end{subequations}
From~\eqref{stress}, it is seen that the right-hand side of~\eqref{psi:star} features the negative of enthalpy,  which motivates the use of the Legendre-Fenchel gaps. The assumed convexity of $\psi$ and $\varphi$ ensures that the maximization problems entering definitions~(\ref{psi:star},b) are solvable. The Legendre-Fenchel gaps $\epsilon_\psi$ and $\epsilon_\varphi$ affiliated respectively with potentials $\psi$ and $\varphi$ are then defined as
\begin{subequations}
\begin{align}
  \epsilon_\psi(\bfeps,\bfal,\bfsige,\bfA)
 &:=\, \psi^{\star}(\bfsige,\bfA)  + \psi(\bfeps,\bfal) - \bfsige\dip\bfeps +\bfA\dip\bfal, \label{LF:gap:def:psi} \\
  \epsilon_\varphi(\bfepsd,\bfald,\bfsigv,\bfA)
 &:=\, \varphi^{\star}(\bfsigv,\bfA) + \varphi(\bfepsd,\bfald)  - \bfsigv\dip\bfepsd-\bfA\dip\bfald, 
\label{LF:gap:def:phi}
\end{align}
\end{subequations}
noting in particular the physical relevance of the last three terms in~\eqref{LF:gap:def:psi} as the Gibbs free energy. The potentials $\psi$ and $\varphi$ being convex, the Legendre-Fenchel gaps enjoy the  key properties
\begin{subequations}
\begin{align}
  &\epsilon_\psi \geqslant 0, \qquad \epsilon_\psi = 0 
  ~ \Longleftrightarrow ~\{\bfsige=\del{\eps}\psi,~  \bfA=-\del{\alpha}\psi \}, \label{LF:gap:prop:psi} \\
  &\epsilon_\varphi \geqslant 0, \qquad \epsilon_\varphi = 0 
  ~ \Longleftrightarrow ~ \{ \bfsigv=\del{\epsd}\varphi,~ \bfA=\del{\alphad}\varphi \}, \label{LF:gap:prop:phi}
\end{align} \label{gaps-all}
\end{subequations}
i.e. they are non-negative and vanish if and only if the thermodynamic variables satisfy the constitutive relations~\eqref{stress}. To demonstrate~\eqref{LF:gap:prop:psi}, we note that $\epsilon_\psi\geqslant \psi^{\star}(\bfsige,\bfA)+\min_{\eps,\alpha} \bigl[  \psi(\bfeps,\bfal) - \bfsige\dip\bfeps +\bfA\dip\bfal \bigr]=0$ by the definition~\eqref{psi:star} of $\psi^{\star}$. Moreover, $\epsilon_\psi\sheq0$ when the optimality conditions for problem~\eqref{psi:star} (namely $\bfsige=\del{\eps}\psi$ and $\bfA=\del{\alpha}\psi$) are satisfied, i.e. when the featured variables are connected by the constitutive relations. Analogous argument applies to~\eqref{LF:gap:prop:phi}.

Thanks to~\eqref{gaps-all}, the Legendre-Fenchel gaps $\epsilon_\psi$ and $\epsilon_\varphi$ allow us to quantify the local (in space and time) constitutive mismatch between the ``strain'' tensor variables $(\bfeps,\bfal)$ and their ``stress'' counterparts ($\bfsig\sheq\bfsige+\bfsigv,\bfA$) for a given viscoelastic material~\eqref{phi:psi:def}. The resulting ECR density can be defined by either
\begin{equation}
\begin{aligned}  
& \text{(a) \ } e\ECR(\bfx,t) = \epsilon_\psi(\bfx,t) + T\epsilon_\varphi(\bfx,t), \qquad \\
& \text{(b) \ } e\ECR(\bfx,t) = \epsilon_\psi(\bfx,t) + \int_{0}^{t} \epsilon_\varphi(\bfx,\tau) \dtau, \quad 
  (\bfx,t)\in\OO\shtimes[0,T] \label{ECR:density}
\end{aligned} 
\end{equation}
where $\epsilon_\psi$ and $\epsilon_\varphi$ are functions of space and time via thermodynamic variables $\bfeps,\bfsig$ etc., with the scaling factor $T$ in case (a) introduced for dimensional consistency. In the sequel, we focus on the form (a) of $e\ECR$, which leads to the stationarity conditions that are somewhat easier to derive. For completeness, we note that the ECR functionals based on the Legendre-Fenchel gap associated with thermodynamic potentials have been formulated elsewhere for nonlinear (e.g. elastoplastic) constitutive models, see e.g.~\cite{lad:moe:97} for FEM error estimation and~\cite{marchand:18} for material identification problems.\enlargethispage*{7ex}

\subsection{Conjugated potentials for linear viscoelastic materials}

Since the potentials $\psi$ and~$\varphi$ are by premise quadratic, their conjugates are also quadratic functions (of the dual variables). For the generic constitutive model given by~\eqref{phi:psi:def}, the stationarity equations that are relevant to problem~\eqref{psi:star} read
\begin{equation}
\bfsige - \Ce\dip\bfeps - \CmT\dip\bfal = \bfze, \qquad\quad -\bfA - \Cm\dip\bfeps - \Ca\dip\bfal = \bfze.
\label{psi:psi:star}
\end{equation}
Solving the system for $\bfeps$ and~$\bfal$ yields
\begin{equation}
\bfeps = (\CeS)^{-1}\dip\lpar \bfsige + \CmT\dip\Ca^{-1}\dip\bfA \rpar, \qquad~~ 
\bfal  = -(\CaS)^{-1}\dip\lpar \Cm\dip\Ce^{-1}\dip\bfsige + \bfA \rpar,
\end{equation}
where  
\begin{equation}
\CeS := \Ce - \CmT\dip\Ca^{-1}\dip\Cm, \qquad~~ \CaS := \Ca - \Cm\dip\Ce^{-1}\dip\CmT. \label{schur:def}
\end{equation}
denote the respective Schur complements of $\Ce$ and $\Ca$. \pagebreak[4]
As a result, from~\eqref{LF:gap:def:psi} with~$\epsilon_\psi=0$, the conjugated potential $\psi^{\star}$ is given by
\begin{equation}
\psi^{\star}(\bfsige,\bfA) = 
\demi \big( \bfsige\dip\Cs\dip\bfsige + 2\bfsige\dip\Cms\dip\bfA + \bfA\dip\CA\dip\bfA \big) \label{psi:star:expr}
\end{equation}
for any $(\bfsige,\bfA)$, with the fourth-order tensors $\Cs,\CA$ and~$\Cms$ defined by
\begin{equation}
\Cs = (\CeS)^{-1}, \qquad \CA = (\CaS)^{-1}, \qquad 
\Cms = (\CeS)^{-1}\dip\Cm\dip\Ca^{-1}. \label{C:coeff:conj}
\end{equation}
These tensors carry the  minor and (excluding~$\Cms$) major symmetries and obey the interrelations
\begin{equation}
\begin{aligned}
  \Cs\dip\Ce &= \Cms\dip\Cm + \IS, &\qquad \CA\dip\Ca &= \CmsT\dip\CmT + \IS, \\
  \Cs\dip\CmT &= \Cms\dip\Ca, &\qquad \CA\dip\Cm &= \CmsT\dip\Ce,
\end{aligned}  \label{schur:C}
\end{equation}
where~$\IS$ is the symmetric fourth-order identity tensor. For completeness, we note that~$\Cms$ attains major symmetry in the special case when the material tensors in~\eqref{phi:psi:def} are isotropic, see also Remark~\ref{symm1}. 

Similarly, making use of~(\ref{phi:star}), \eqref{LF:gap:def:phi} and~\eqref{LF:gap:prop:phi} we obtain
\begin{equation}
  \varphi^{\star}(\bfsigv,\bfA)
 = \demi \big( \bfsigv\dip\Ds\dip\bfsigv - 2\bfsigv\dip\Dms\dip\bfA + \bfA\dip\DA\dip\bfA \big), \label{phi:star:expr}
\end{equation}
where the fourth-order tensors $\Ds,\DA,\Dms$ are given by
\begin{equation}
 \Ds = (\DeS)^{-1}, \qquad \DA = (\DaS)^{-1}, \qquad 
 \Dms = (\DeS)^{-1}\dip\Dm\dip\Da^{-1} \label{D:coeff:conj}
\end{equation}
in terms of the Schur complements
\begin{equation}
\DeS := \De - \DmT\dip\Da^{-1}\dip\Dm, \qquad~~ \DaS := \Da - \Dm\dip\De^{-1}\dip\DmT, \label{schur:defD}
\end{equation}
and obey the counterparts of interrelations~\eqref{schur:C}. Recalling~\eqref{C:inst}, we note that $\DeS = \DS\Isub$ and $\Ds = \DS\Isub^{-1}$.\enlargethispage*{1ex}

\section{MECR-based minimization problem} \label{sec:ECR:min}

\subsection{Inverse problem} \label{sec:setting}

In this work, we seek to recover the SGM parameters featured by~\eqref{phi:psi:def} of a viscoelastic solid from the full-field measurements, $\bfum$, of the displacement field therein -- taken over an ``observation window" $\Omega\subset\mathbb{R}^d$, $d\in\{2,3\}$. Specifically, the potentials $\psi$ and~$\varphi$ are assumed to depend on a set of material parameters $\bsfp$, to be estimated from the data. For simplicity of discussion, we assume (i) no knowledge of the boundary data on $\G\!=\!\partial\OO$, (ii) absence of the body forces inside~$\OO$, and (iii) that the viscoelastic material occupying $\OO$ is initially at rest. These default settings are relaxed later in Section~\ref{sec:discuss}. We first tackle the general case of transient viscoelastic motions observed over a time interval $[0,T]$, and then specialize the formulation to  time-harmonic problems.

Since the boundary conditions on $\G$ are by premise left unspecified, we shall make use the displacement spaces\begin{equation}
  \Ucal:= \Vcal, \qquad
  \Wcal:=\lcb \bfv\shin\Vcal,\,\bfv\sheq\bfze\ \text{on}\ \G \rcb \subset \Ucal \label{u:spaces}
\end{equation}
where $\Vcal$ is a suitable energy space (typically $\Vcal=H^1(\OO;\Rbb^d)$ for the time-harmonic case and $\Vcal=H^1(\OO\shtimes[0,T];\Rbb^d)$ for the transient case). The displacement $\bfu$ and stress $\bfsig$ are required to verify the balance of linear momentum, whose weak form \emph{a priori} reads
\begin{equation}
\iO\int_0^T \big( \bfsig\dip\bfeps[\bfw] + \rho\ddot{\bfu}\sip\bfw \big) \dt\dV
= \iG\int_0^T \bfn\sip\bfsig\sip\bfw \dt\dS \qquad \forall\:\bfw\in\Ucal.
\end{equation}
In view of the assumed lack of information on boundary conditions on $\Gamma$, we henceforth restrict the above weak statement to $\bfw\shin\Wcal$, in effect exploiting only the interior balance equation:
\begin{equation}
\iO\int_0^T \big( \bfsig\dip\bfeps[\bfw] + \rho\ddot{\bfu}\sip\bfw \big) \dt\dV = 0 \qquad \forall\:\bfw\in\Wcal. \label{balance:weak}
\end{equation}
For completeness, we note that~\eqref{u:spaces} and~\eqref{balance:weak} may be altered to accommodate the specification of boundary conditions on (the whole or part of) $\Gamma$, in particular extending the present framework to settings where the forward problem is well-posed, see Section~\ref{sec:discuss}.

\subsection{Objective functional}

As examined in Section~\ref{sec:ECR}, the overall constitutive mismatch between $(\bfeps,\bfal)$ and $(\bfsig\sheq\bfsige\shp\bfsigv,\bfA)$ for a given material (i.e.~given $\bsfp$) can be quantified as
\begin{equation}
  \Ecal(\bfu,\bfal,\bfsige,\bfsigv,\bfA,\bsfp)
  \,=\, \Ecale(\bfeps[\bfu],\bfal,\bfsige,\bfA,\bsfp) \,+\, \Ecalv(\bfepsd[\bfu],\bfald,\bfsigv,\bfA,\bsfp) \label{ECR:def}
\end{equation}
in terms of the stored-energy and dissipation components of the global ECR, given respectively by
\begin{equation}
  \Ecale := \iO\int_0^T \epsilon_\psi(\bfeps,\bfal,\bfsige,\bfA,\bsfp
  ) \dt\dV, \qquad
  \Ecalv := \iO\int_0^T T\epsilon_\varphi(\bfepsd,\bfald,\bfsigv,\bfA,\bsfp) \dt\dV. \label{ECR:decomp}
\end{equation}

\begin{remark}
Equalities $\bfsige\!=\!\bfsige[\bfu]$ and $\bfsigv\!=\!\bfsigv[\bfu]$, whose right-hand sides are given by~\eqref{stress}, hold if and only if $\Ecal\!=\!0$, see~(\ref{LF:gap:prop:psi},b); this is a basic tenet of the ECR approach to the characterization of viscoelastic materials. Further, the appearance of the vector argument~$\bsfp$ in~\eqref{ECR:def} and thereon signifies an explicit dependence of germane functionals on the viscoelastic coefficients featured by the tensors~$\Ca,\Cm$ etc.~in~\eqref{phi:psi:def}. With reference to~\eqref{ECR:decomp}, this motivates rewriting~$\epsilon_\psi(\bfeps,\bfal,\bfsige,\bfA)$ given by~\eqref{LF:gap:def:psi} as $\epsilon_\psi(\bfeps,\bfal,\bfsige,\bfA,\bsfp)$, and similarly for~$\epsilon_\psi$.\enlargethispage*{1ex}
\end{remark}

In the sequel, we make use of the short-hand notation
\begin{equation} \label{X}
\bsfX=(\bfu,\bfal,\bfsige,\bfsigv,\bfA).   
\end{equation}
Since $\Ecal$ is to be minimized by taking advantage of the available sensory data, it is natural to introduce the weighted functional
\begin{equation}
  \Lambdak(\bsfX,\bsfp)
 := \Ecal(\bsfX,\bsfp) + \tdemi\kappa\exs\McalT (\bfu\shm\bfum,\bfu\shm\bfum), \label{MECR:def}
\end{equation}
where $\McalT$, given by
\begin{equation}
  \McalT(\bfr,{\bfs}) := \int_0^T \!\! \Mcal(\bfr(\bdot,t),{\bfs}(\bdot,t)) \dt \label{Mcal:def}    
\end{equation}
in terms of a positive, symmetric bilinear form $\Mcal$ acting on functions defined in $\OO$, quantifies the mismatch between the model displacements and their observations. A common choice for $\Mcal$ is the $L^2$ scalar product $\Mcal(\bfr,{\bfs})=\eps_0 \nes\times\nes (\bfr,{\bfs})_{\OO}$ over the measurement region, with the scalar multiplier $\eps_0$ inserted to ensure dimensional consistency. The functional $\Lambdak$, often called the modified-error-in-constitutive-relation (MECR) functional, is thus designed to seek values of the material parameters (and related mechanical variables) that achieve a compromise between verifying the constitutive equations and reproducing the data.

While the practical goal of minimizing~\eqref{MECR:def} is to identify $\bsfp$, the optimal value of~$\bsfX$ -- as determined by the optimal $\bsfp$ -- minimize the constitutive-data mismatch (measured in terms of $\Lambdak$, which is zero under ideal conditions). Inherently, the field variables must verify the fundamental requirements of continuum mechanics, namely the kinematic compatibility in terms of $\bfu$, and the balance of linear momentum in terms of $\bfu$ and $\bfsig=\bfsige\shp\bfsigv$. In this vein, the constitutive identification can be cast as a PDE-constrained minimization problem
\begin{equation}
  \min_{\bsfX,\bsfp} \Lambdak(\bsfX,\bsfp) \qquad
  \text{subject to $\bfu\shin\Ucal$ ~and~ \eqref{balance:weak}}. \label{full:min}
\end{equation}
By virtue of~\eqref{balance:weak}, the first-order necessary optimality conditions for problem~\eqref{full:min} can be conveniently formulated by introducing the Lagrangian
\begin{equation}
\Lcal(\bsfX,\bfw,\bsfp) :=\, \Lambdak(\bsfX,\bsfp) 
  - \iO\int_0^T \!\big( (\bfsige\shp\bfsigv)\dip\bfeps[\bfw] + \rho\ddot{\bfu}\sip\bfw \big) \dt\dV, \label{L:def}
\end{equation}
where the test function $\bfw\shin\Wcal$ in the balance equation~\eqref{balance:weak} plays the role of a Lagrange multiplier.

\subsection{Stationarity conditions} \label{sec:stat:cond}

The first-order necessary optimality conditions for the minimization of $\Lambdak$ require that the directional derivatives of $\Lcal$ with respect to all variables vanish. Those conditions can here be categorized into three groups. The first group gathers the directional derivatives w.r.t. the stress and internal variables, namely
\begin{equation}
\begin{aligned}
  &\text{(a) } & \lbra \del{\sige}\Lcal,\bfsigHe \rbra &= 0 \quad\forall\:\bfsigHe, &\qquad&
  &\text{(c) } & \lbra \del{A}\Lcal,\bfAH \rbra &=0 \quad\forall\:\bfAH, \\
  &\text{(b) } & \lbra \del{\sigv}\Lcal,\bfsigHv \rbra &= 0 \quad\forall\:\bfsigHv, &\qquad&
  &\text{(d) } & \lbra \del{\alpha}\Lcal,\bfalH \rbra &= 0 \quad\forall\:\bfalH.
\end{aligned} \label{group1}
\end{equation}
The second group collects the directional derivatives w.r.t. the kinematic variables, i.e. 
\begin{equation}
\begin{aligned}
  &\text{(a) } & \lbra \del{w}\Lcal,\bfwH \rbra &=0 \quad\forall\:\bfwH, \\
  &\text{(b) } & \lbra \del{u}\Lcal,\bfuH \rbra &= 0 \quad\forall\:\bfuH,
\end{aligned} \label{group2}
\end{equation}
while the directional derivatives w.r.t. the constitutive parameters 
\begin{equation}
  \lbra \del{\sfp}\Lcal,\bsfpH \rbra = 0 \quad\forall\:\bsfpH  \label{group3}
\end{equation}
 make up the last group. In~\eqref{group1}--\eqref{group3}, notation $\lbra \del{x}f,\hat{x} \rbra$ stands for the directional derivative of a multivariate function $f=f(x,y,\ldots)$ w.r.t. $x$ in direction $\hat{x}$, given by
\begin{equation}
  \lbra \del{x}f,\hat{x} \rbra = \lim_{h\to 0} \inv{h}\lsqb f(x+h\hat{x},y,\ldots) - f(x,y,\ldots) \rsqb
\end{equation}
and is treated as a linear functional on the perturbation direction $\hat{x}$.

Conditions~\eqref{group1} and~\eqref{group2} achieve the partial minimization of $\Lambdak$ for fixed $\bsfp$ (by solving, in the present context, linear equations arising from convex quadratic functionals). Accordingly, this step provides the constitutive-data mismatch for an assumed material, which is an essential component of the overall methodology. Following~\cite{B-2016-07}, we hereon refer to~\eqref{group1} and~\eqref{group2} as the stationarity system. As we will shortly see, the first group~\eqref{group1} yields local equations that can be solved explicitly, allowing for subsequent elimination of stresses and internal variables, while the second group~\eqref{group2} produces a system of global variational equations to be solved numerically.

\paragraph{Local stationarity equations} Recalling definition~\eqref{ECR:def} of the ECR component $\Ecal$ and introducing the respective shorthand notations $\bfeps$ and $\bfeta$ (with no argument) for the linearized strains $\bfeps[\bfu]$ and $\bfeps[\bfw]$, equations~\eqref{group1} can be rewritten more explicitly as
\begin{equation}
\begin{aligned}
  &\text{(a)} &   0
 &= \iO \int_0^T \Lcb \del{\sige}\psi^{\star} - \bfeps - \bfeta \Rcb \dip\bfsigHe \dt\dV
 &\ & \forall\bfsigHe, \\
  &\text{(b)} &   0
 &= \iO \int_0^T \Lcb T\lpar \del{\sigv}\varphi^{\star}-\bfepsd \rpar -\bfeta \Rcb \dip\bfsigHv \dt\dV
 &\ & \forall\bfsigHv, \\
  &\text{(c)} &   0
 &= \iO \int_0^T \Lcb \del{A}\psi^{\star} + \bfal + T\lpar \del{A}\varphi^{\star}-\bfald \rpar \Rcb \dip\bfAH \dt\dV
 &\ & \forall\bfAH, \\
  &\text{(d)} &   0
 &= \iO \int_0^T \Lcb \lpar \del{\alpha}\psi + \bfA \rpar\dip\bfalH + T\lpar \del{\dot{\alpha}}\varphi - \bfA \rpar\dip\dot{\bfalH} \Rcb \dt\dV
 &\ & \forall\bfalH,\: \bfalH(\dotp,0)\sheq\bfze.
\end{aligned} \label{group1:exp}
\end{equation}
Equations~(\ref{group1:exp}a-c) are the weak forms of the pointwise (in space and time) equations
\begin{equation}
\begin{aligned}
  &\text{(a) } &   \bfze
 &= \del{\sige}\psi^{\star} - \bfeps - \bfeta  \\
 &&&= \Cs\dip\bfsige+\Cms\dip\bfA - \bfeps - \bfeta, \\
  &\text{(b) } &   \bfze
 &=  T\lpar \del{\sigv}\varphi^{\star} - \bfepsd \rpar -\bfeta \\
 &&&= T\Ds\dip\bfsigv - T\Dms\dip\bfA - T\bfepsd - \bfeta, \\
  &\text{(c) } &   \bfze
 &= \del{A}\psi^{\star} + \bfal + T\lpar \del{A}\varphi^{\star}-\bfald \rpar \\
 &&&= \CmsT\dip\bfsige - T\DmsT\dip\bfsigv + (\CA\shp T\DA)\dip\bfA + \bfal-T\bfald,
\end{aligned} \label{group1:pointwise}
\end{equation}
which can be solved in closed form for $\bfsige,\bfsigv$ and~$\bfA$. As shown in Appendix~\ref{ssA:proof} (ESM), we find the optimal ``stress'' variables satisfying~(\ref{group1}a-c) to read
\begin{equation}
\begin{aligned}
  \bfsige
 &= \Ce\dip\bfeps + \CmT\dip\bfal + (\Ce-\CmT\dip\Da^{-1}\dip\Dm)\dip\bfeta -T\CmT\dip\bfbe, \\
  \bfsigv
 &= \DmT\dip\bfald + \De\dip\bfepsd + \Tinv \DS\Isub\dip\bfeta - \DmT\dip\bfbe, \\
  \bfA
 &= \Da\dip\bfald + \Dm\dip\bfepsd - \Da\dip\bfbe,
\end{aligned} \label{ssA}
\end{equation}
where the auxiliary tensor-valued function $\bfbe$ solves
\begin{equation}
  (\Da\shp T\Ca)\dip\bfbe = \Da\dip\bfald + \Dm\dip\bfepsd + \Ca\dip\bfal + \Cm\dip\bfeps + \CH\dip\bfeta \label{R:beta:def}
\end{equation}
and $\CH$ is given in~\eqref{alpha(t)}. 

The remaining local equation~(\ref{group1:exp}d) is used to compute $\bfal$ and, by way of~\eqref{R:beta:def}, $\bfbe$ in terms of the time histories of $\bfu$ and $\bfw$. On integrating by parts in time the term involving $\dot{\bfalH}$, (\ref{group1:exp}d) produces the pointwise equations
\begin{equation}
 0 = \del{\alpha}\psi + \bfA - T\partial_t\lpar \del{\dot{\alpha}}\varphi - \bfA \rpar, \qquad
 0 = \lpar \del{\dot{\alpha}}\varphi - \bfA \rpar(T), \label{ODE:beta}
\end{equation}
where the application of partial time derivative~$\partial_t$ assumes $\varphi=\varphi(\bfx,t)$ and~$\bfA=\bfA(\bfx,t)$, a spatiotemporal dependence that is implicit to~\eqref{group1:exp}. The terms $\del{\alpha}\psi + \bfA$ and $\del{\dot{\alpha}}\varphi - \bfA$ are then evaluated with the help of formulas~(\ref{ssA}c) and~\eqref{R:beta:def}, yielding
\begin{equation}
\begin{aligned}
\del{\alpha}\psi + \bfA
 &= \Ca\dip\bfal + \Cm\dip\bfeps + \Da\dip\bfald + \Dm\dip\bfepsd - \Da\dip\bfbe \hspace{-3mm}&&= T\Ca\dip\bfbe - \CH\dip\bfeta, \\
\del{\dot{\alpha}}\varphi - \bfA
 &= \Da\dip\bfald + \Dm\dip\bfepsd - \Da\dip\bfald - \Dm\dip\bfepsd + \Da\dip\bfbe \hspace{-3mm}&&= \Da\dip\bfbe.
\end{aligned} \label{ODE:beta:coefs}
\end{equation}
As a result, \eqref{ODE:beta} reduces to a pointwise ODE
\begin{equation}
  \Da\dip\dot{\bfbe} - \Ca\dip\bfbe + \Tinv \CH\dip\bfeta = \bfze, \qquad \bfbe(T)=0, \label{ODE:beta:exp}
\end{equation}
which can be solved explicitly for $\bfbe$. Specifically, for the convolution operator $\FS$ given by~\eqref{KS:def} and any tensor-valued density $\bfs$, we have 
\[\dot{\FSR}[\bfsR](t)=\partial_t\int_{0}^{T-t} \exp[(\theta-T+t)\QS]\Da^{-1}\dip\bfs(T-\theta) \dth \hh=\hh \QS\dip\FSR[\bfsR]-\bfs(t), 
\]
and consequently 
\begin{equation}
  \Da\dip\dot{\FSR}[\bfsR] - \Ca\dip\FSR[\bfsR] + \bfs = \bfze, \label{sol:prop:reversed}
\end{equation}
where $\bfsR(\dotp):=\bfs(T\shm\dotp)$ and $\FSR[\bfs](\dotp)=\FS[\bfs](T\shm\dotp)$. Noting in addition that $\FSR[\bfs](T)=\bfze$ for any $\bfs$, the solution of~\eqref{ODE:beta:exp} is computed as
\begin{equation}
  \bfbe(t) = \Tinv \FSR[\CH\dip\bfetaR](t). \label{beta:expr:reversed}
\end{equation}
With~\eqref{beta:expr:reversed} at hand, \eqref{R:beta:def} can be solved for the viscous strain history $\bfal$ that solves the stationarity conditions. To this end, we introduce the decomposition
\begin{equation}
  \bfal(t) = \bfal[\bfu](t) + \bfa(t), \label{alpha:a}
\end{equation}
where $\bfal[\bfu]$ is computed via~\eqref{alpha(t)} for a given set of constitutive parameters. Since $\bfal[\bfu]$ and $\bfal$ respectively satisfy~\eqref{evol} and~\eqref{R:beta:def} and $(\Da\shp T\Ca)\dip\bfbe-\CH\dip\bfeta=\Da\dip(\bfbe\shp T\dot{\bfbe})$ due to~\eqref{ODE:beta:exp}, the viscous strain misfit $\bfa\sheq\bfal[\bfu]\shm\bfal$ can be shown to satisfy the pointwise ODE
\begin{equation}
  \Da\dip\dot{\bfa} + \Ca\dip\bfa = \Da\dip(\bfbe\shp T\dot{\bfbe}), \qquad \bfa(0)=0 \label{a:ODE}
\end{equation}
 whose right-hand side depends linearly on $\bfw$. Invoking~\eqref{sol:prop} with $\bfs=\Da\dip(\bfbe\shp T\dot{\bfbe})$, we obtain 
\begin{equation}
  \bfa(t) = \FS[\Da\dip(\bfbe\shp T\dot{\bfbe})](t). \label{beta:expr:stat}
\end{equation}

\begin{remark}
A number of standard rheological models are characterized by $\De=\Dm=\bfze$, i.e. trivial irreversible stress $\bfsigv\!=\!\bfze$, see for instance Section~\ref{vispecial}. In those situations, the local stationarity condition targeting $\bfsigv$ (given alternatively by~(\ref{group1}b), (\ref{group1:exp}b) and~(\ref{group1:pointwise}b)) ceases to apply. The remaining stationarity equations, however, still form a closed system. Specifically, when $\De=\Dm=\bfze$ the foregoing results remain valid upon: (i) forsaking the cited equations and (ii) setting  $\bfsigv\!\!=\!\bfze$, $\,\CH\!=\!\Cm$, $\CS\Isub\!=\!\Ce$, $\DS\Isub\!=\!\De$, $\exs\Ds\!=\Dms\!=\bfze$, and~$\DA\!=\Da^{-1}$. 
\end{remark}

\paragraph{Constitutive gap} On using~\eqref{beta:expr:reversed}--\eqref{beta:expr:stat} in~\eqref{ssA}, all stress variables featured in the local stationarity equations~(\ref{group1}a-d) are obtained in terms of $\bfu$ and $\bfw$. In particular, summing the first two formulas in~\eqref{ssA} and making use of the expression~\eqref{sigma} for~$\bfsig[\bfu]$ and decomposition~\eqref{alpha:a}, the stress $\bfsig\sheq\bfsige+\bfsigv$ solving the local stationarity conditions is obtained as
\begin{equation}
  \bfsig
 = \bfsig[\bfu] + \CmT\dip\bfa + \DmT\dip\dot{\bfa} + (\CeS+\CmT\dip\Ca^{-1}\dip\CH + \Tinv \DS\Isub)\dip\bfeta
 - (T\CmT\shp\Dm)\dip\bfbe, \label{sigma:expr}
\end{equation}
where the instantaneous (elasticity and viscosity) tensors $\CS\Isub$ and $\DS\Isub$ are given by~\eqref{C:inst}. The above formula emphasizes the distinction between the constitutive model~\eqref{sig:expr} (which furnishes $\bfsig[\bfu]$ depending exclusively on~$\bfu$) and the stationarity solution $\bfsig$ (which depends on~$\bfu$ and~$\bfw$). On using~\eqref{beta:expr:reversed}--\eqref{beta:expr:stat} in~\eqref{sigma:expr}, after some algebra we obtain
\begin{equation}
  \bfsig(t) = \bfsig[\bfu](t) + \SS_t[\bfw](t), \label{sigma:expr:stat}
\end{equation}
where
\begin{equation}
 \SS_t[\bfw](t)
 = (\CS\Isub + \Tinv \DS\Isub) \dip\bfeta(t) + \CHT\dip\lpar \bfa(t)-T\bfbe(t) \rpar \label{SSorig}
\end{equation}
which, in light of~\eqref{beta:expr:reversed} and~\eqref{beta:expr:stat}, is linear in the history of $\bfeta\sheq\bfeps[\bfw]$ over $[t,T]$. The tensor-valued functional $\SS_t[\bfw]$ thus quantifies the constitutive gap $(\bfsig-\bfsig[\bfu])(t)$ between (i) the stress $\bfsig$ solving the stationarity system, and (ii) $\bfsig[\bfu]$ computed via the assumed constitutive model.

\paragraph{Global stationarity equations} We are now in a position to formulate the global stationarity equations arising from conditions~(\ref{group2}a,b) bearing in mind that the displacement fields $\bfu$ and $\bfw$, and hence their variations $\bfuH$ and $\bfwH$ in the directional differentiation process, are elements of $\Ucal$ and $\Wcal$, respectively. First, condition~(\ref{group2}a) reads
\begin{equation}
  \iO\int_0^T \big(\bfsig\dip\bfeps[\bfwH] + \rho\ddot{\bfu}\sip\bfwH \big) \dt\dV
 = 0 \qquad\forall\:\bfwH\shin\Wcal. \label{aux09}
\end{equation}
Since $\bfsig$ is given by~\eqref{sigma:expr:stat} while $\bfu$ has quiescent past, \eqref{aux09} yields
\begin{multline}
  \iO\int_0^T \big(\bfsig[\bfu]\dip\bfeps[\bfwH] + \rho\ddot{\bfu}\sip\bfwH \big) \dt\dV
 = -\iO\int_0^T \SS_t[\bfw]\dip\bfeps[\bfwH] \dt\dV \qquad\forall\:\bfwH\shin\Wcal, \\[-1ex]
   \bfu(\bdot,0) = \dot{\bfu}(\bdot,0) = \bfze ~~ \text{in $\OO$}, \label{stat:W}
\end{multline}
which can be interpreted as a viscoelastic \emph{forward} evolution problem for $\bfu\in\Ucal$ driven by the constitutive mismatch, $\SS_t[\bfw]$, between $\bfsig$ and $\bfsig[\bfu]$. We note that given $\bfw\in\Wcal$, \eqref{stat:W} constitutes an \emph{underdetermined problem} for $\bfu$ since $\Wcal$ (the space of test functions) is strictly contained in $\Ucal$.

By virtue of~\eqref{LF:gap:def:psi}, \eqref{LF:gap:def:phi}, \eqref{ECR:def}--\eqref{MECR:def} and~\eqref{L:def}, on the other hand, condition~(\ref{group2}b) becomes
\begin{multline}
  \iO \int_0^T \Lcb \lpar \del{\eps}\psi - \bfsige \rpar\dip\bfeps[\bfuH]
  + T\lpar \del{\dot{\eps}}\varphi - \bfsigv \rpar\dip\bfeps[\dot{\bfuH}] - \rho\bfw\sip\ddot{\bfuH} \Rcb \dt\dV \\
  + \kappa \int_0^T \!\! \Mcal(\bfu\shm\bfum,\bfuH) \dt = 0 \qquad \forall\:\bfuH\shin\Ucal. \label{aux10}
\end{multline}
To suppress the temporal differentiation of $\bfuH$, \eqref{aux10} is conveniently integrated by parts as
\begin{multline}
  \iO \int_0^T \Lcb \del{\eps}\psi - \bfsige - T\partial_t\lpar \del{\dot{\eps}}\varphi - \bfsigv \rpar \Rcb \dip\bfeps[\bfuH] \dt\dV
  - \iO \int_0^T \rho\ddot{\bfw}\sip\bfuH \dt\dV \\
  + \iO T\lpar \del{\dot{\eps}}\varphi - \bfsigv \rpar(T) \dip\bfeps[\bfuH](T) \dV
  - \iO \rho \Lcb \bfw(T)\sip\dot{\bfuH}(T) - \dot{\bfw}(T)\sip\bfuH(T) \Rcb \dV \\
  + \kappa \int_0^T \!\! \Mcal(\bfu\shm\bfum,\bfuH) \dt = 0 \qquad \forall\:\bfuH\shin\Ucal. \label{final:w}
\end{multline}
Classical arguments underpinning variational formulations then dictate that the resulting integrals over $\OO$ and over $\OO\shtimes[0,T]$ vanish separately. The first requirement results in $\bfw$ being at rest at the final time, i.e. $\bfw(\bdot,T) = \dot{\bfw}(\bdot,T) = \bfze$ in $\OO$. Then, the second requirement is reformulated using~\eqref{ssA} to express $\del{\eps}\psi - \bfsige$ and $\del{\dot{\eps}}\varphi - \bfsigv$ in terms of $\bfu$ and $\bfw$ (via their strains and strain rates). We find
\begin{equation}
\begin{aligned}
  \del{\eps}\psi - \bfsige
 &= T\CmT\dip\bfbe  + \lpar \CmT\dip\Da^{-1}\dip\Dm - \Ce \rpar\dip\bfeta,  \\
  \del{\dot{\eps}}\varphi - \bfsigv
 &= \DmT\dip\bfbe - \Tinv \DS\Isub\dip\bfeta,
\end{aligned} \label{aux07}
\end{equation}
which allows us to evaluate
\begin{align}
 \del{\eps}\psi - \bfsige - T\partial_t\lpar \del{\dot{\eps}}\varphi - \bfsigv \rpar
 &= T\big(\CmT\dip\bfbe-\DmT\dip\dot{\bfbe}\big) + \lpar \CmT\dip\Da^{-1}\dip\Dm - \Ce \rpar\dip\bfeta + \DS\Isub\dip\dot{\bfeta} \\[0.5ex]
 &\Eq{(a)} T\CHT \dip\bfbe + \DS\Isub\dip\dot{\bfeta} - \CS\Isub\dip\bfeta \\[0.5ex]
 &\Eq{(b)} \CHT \dip\dip\FSR[\CH\dip\bfetaR] + \DS\Isub\dip\dot{\bfeta} - \CS\Isub\dip\bfeta 
 \:\Eq{(c)} -\bfsigR[\bfwR], \label{aux03}
\end{align}
where (a) follows from the governing ODE~\eqref{ODE:beta:exp} for $\bfbe$ and the definition~\eqref{C:inst} of $\CS\Isub$; (b) uses the expression~\eqref{beta:expr:reversed} for $\bfbe$, and (c) stems from the constitutive relation~\eqref{sig:expr} in time-reversed form. In view of~\eqref{final:w} and~\eqref{aux03}, condition~(\ref{group2}b) yields the variational equation
\begin{multline}
  \iO \int_0^T \big( \bfsigR[\bfwR]\dip\bfeps[\bfuH] + \rho\ddot{\bfw}\sip\bfuH \big) \dt\dV
  = \kappa \int_0^T \!\! \Mcal(\bfu\shm\bfum,\bfuH) \dt \qquad \forall\:\bfuH\shin\Ucal, \\[-1ex]
  \bfw(\bdot,T) = \dot{\bfw}(\bdot,T) = \bfze \quad \text{in $\OO$.} \label{stat:u}
\end{multline}
which can be interpreted as a viscoelastic \emph{backward} evolution problem for $\bfw\in\Wcal$ driven by the measurement residuals. Equation~\eqref{stat:u} in isolation, however, defines an \emph{overdetermined problem}: given $\bfu\in\Ucal$, it has a unique solution $\bfw\shin\Ucal$ which in general fails to vanish on $\G$ as mandated by the definition of~$\Wcal$ in~\eqref{u:spaces}.

\paragraph{Coupled stationarity problem} Summarizing, the first-order conditions~\eqref{group1} and \eqref{group2} lead to the coupled variational stationarity system for $(\bfu,\bfw)\in\Ucal\shtimes\Wcal$ given by~\eqref{stat:W} and~\eqref{stat:u}. Based on the corresponding results~\cite{B-2016-07} for time-harmonic elastodynamics, coupled equations~\eqref{stat:W} and \eqref{stat:u} are expected to be uniquely solvable for $(\bfu,\bfw)\in\Ucal\shtimes\Wcal$ assuming: (i) the sufficiency of sensory data $\bfum$, and (ii) additional restrictions on the data functional $\McalT$. However, further investigation (that is beyond the scope of this study) is necessary to settle this issue. Given the time histories $(\bfu,\bfw)\in\Ucal\shtimes\Wcal$ solving~\eqref{stat:W} and~\eqref{stat:u}, the Cauchy stress tensor $\bfsig= \bfsige +\bfsigv$ and the tensor of thermodynamic tensions~$\bfA$ are then computed from~\eqref{ssA}, while the viscoelastic strain~$\bfal$ is obtained from~\eqref{alpha:a} with~\eqref{alpha(t)} and~\eqref{beta:expr:stat}.

\subsection{Reduced objective function and its derivative}

In a full-space version~\eqref{full:min} of the constitutive identification problem, stationarity equations~\eqref{group1}--\eqref{group3} are solved (iteratively) as a whole; see~\cite[e.g.]{epanomeritakis:08} in the context of geophysical full-waveform inversion. Alternatively, a \emph{reduced-space approach} can be formulated: let $\bsfXp\!=\!(\bfup,\bfalp,\bfsigep,\bfsigvp,\bfAp)$ solve~\eqref{group1}--\eqref{group2} for given $\bsfp$ (via partial constrained minimization of $\Lambdak$ with fixed $\bsfp$). Then, reduced versions of the ECR and data-misfit components of $\Lambdak$ that depend only on $\bsfp$ can be defined as
\begin{equation} \label{ECRed}
  \EcalR(\bsfp) := \Ecal(\bsfXp,\bsfp), \qquad\qquad
  \McalRT(\bsfp) := \McalT(\bfup\shm\bfum,\bfup\shm\bfum).
\end{equation}
As a result, letting $\LambdaRk(\bsfp)=\Lambdak(\bsfXp,\bsfp)$, problem~\eqref{full:min} can be recast in reduced minimization form as
\begin{equation}
  \min_{\bsfp} \LambdaRk(\bsfp), \qquad\qquad
  \LambdaRk(\bsfp)
 \,:=\, \EcalR(\bsfp) + \tdemi\kappa\exs\McalRT(\bsfp)
 \,=\, \min_{\bsfX} \Lambdak(\bsfX,\bsfp). \label{min:reduced}
\end{equation}
The next result, whose proof is given in Appendix~\ref{MECR:eval:proof} (ESM), provides the key expressions of practical importance in the application of gradient-based minimization method to solving~\eqref{min:reduced}.

\begin{prop}[reduced MECR functional and its gradient]\label{MECR:eval}
Let $\bfup,\bfwp$ and the dependent quantities $\bfalp,\bfsigep,\bfsigvp$ and $\bfAp$ solve the stationarity system~\eqref{group1}--\eqref{group2} for given $\bsfp$. Then the reduced MECR functional in~\eqref{min:reduced} is given by
\begin{equation} \label{Lammin}
\LambdaRk(\bsfp) = -\tdemi\kappa\exs\McalT(\bfup\shm\bfum,\bfum),
\end{equation}
and its derivative by
\begin{equation}
\LambdaRk'(\bsfp) \,=\, \del{\sfp}\Lcal(\bsfXp,\bfwp,\bsfp) \,=\, \del{\sfp}\Ecal(\bsfXp,\bsfp), \label{RECR-der}
\end{equation}
where $()'$ denotes the derivative of a univariate function (that accounts for implicit dependencies); $\Ecal(\bsfXp,\bsfp)$ is given by~\eqref{ECR:def}--\eqref{ECR:decomp}, and $\del{\sfp}$ is the partial derivative w.r.t.~$\bsfp$.
\end{prop}

Expression~\eqref{RECR-der} in particular demonstrates that evaluation of the MECR gradient requires no additional computation of the global stationarity solutions. For completeness, we remark that the constitutive gap can be alternatively expressed as $\Ecal(\bsfXp,\bsfp)=\Ecalc(\bfwp,\bsfp)$ as elucidated in the following claim (see Appendix~\ref{MECR:eval:proof2}, ESM for proof).

\begin{prop}\label{MECR:eval2}
For any $\bfw\shin\Wcal$, we have
\begin{equation}
  \int_0^T\! \SS_t[\bfw]\dip\bfeta \dt
 \,=\! \int_0^T \Lcb \bfeta\dip\big( \CeS + \Tinv \DS\Isub \big) \dip\bfeta
  + T \bfbe\dip\Da\dip\bfbe + T^2 \dot{\bfbe}\dip\Da\dip\Ca^{-1}\dip\Da\dip\dot{\bfbe} \Rcb \dt \:\geqslant\, 0, \label{SS:integr}
\end{equation}
where $\CeS$ is the Schur complement of $\Ce$ given by~\eqref{schur:def}. The ECR functional $\Ecal=\Ecale+\Ecalv$ given by~\eqref{ECR:def}--\eqref{ECR:decomp} permits an alternative representation
\begin{equation}
\Ecalc(\bfwp,\bsfp) = \demi\iO\int_0^T \SS_t[\bfwp]\dip\bfeps[\bfwp] \dV\dt, \label{aux12main}
\end{equation}
with its stored-energy and dissipation components given respectively by
\begin{equation}
\begin{aligned}
\Ecalec(\bfwp,\bsfp)
 &= \tdemi\iO\int_0^T \Lcb \bfetap\dip\CeS\dip\bfetap + T^2\dot{\bfbe}_{\mathsf{p}}\dip\Da\dip\Ca^{-1}\dip\Da\dip\dot{\bfbe}_{\mathsf{p}} \Rcb \dt\dV, \\
\Ecalvc(\bfwp,\bsfp)
 &= \tdemi \iO\int_0^T T^{-1} \Lcb \bfetap\dip\DS\Isub\dip\bfetap + T^2\bfbep\dip\Da\dip\bfbep \Rcb \dt\dV.
\end{aligned} \label{Eevfinal}
\end{equation}
\end{prop}
Expressions~\eqref{aux12main}--\eqref{Eevfinal} highlight the fact that $\Ecal(\bsfXp,\bsfp)\!=\!\check\Ecal(\bfwp,\bsfp)$ carries a quadratic dependence on~$\bfwp$. As in previous MECR studies pertaining to linear elasticity, the Lagrange multiplier field $\bfw$ thus plays a key role in quantifying the constitutive gap. However, we note that representation \eqref{aux12main}--\eqref{Eevfinal} is useful for the evaluation of individual ECR components, but (recalling Proposition~\ref{MECR:eval}) not particularly so for evaluating the gradient $\del{\sfp}\Ecal(\bsfXp,\bsfp)=\del{\sfp}\Ecalc(\bfw(\bsfXp,\bsfp),\bsfp)$ for the computation of~$\del{\sfp}\bfw(\bsfXp,\bsfp)$ inherently requires solving an additional \emph{global} stationarity problem.

\subsection{L-curve of the MECR functional} \label{L-curve}

Let $(\bsfXp(\kappa),\bsfp(\kappa))$ be the solution (provisionally assuming uniqueness) of the minimization problem~\eqref{full:min} for given $\kappa\!>\!0$, which is in this section treated as a function of $\kappa$ to highlight its role. Let then
\begin{equation}
E(\kappa):= \EcalR(\bsfp(\kappa)), \qquad M(\kappa) := \tdemi\McalRT(\bsfp(\kappa))
\end{equation}
denote the MECR components at the minimizer, so that $\LambdaRk(\bsfp(\kappa))=E(\kappa)+\kappa M(\kappa)$. For the case of time-harmonic  elasticity~\cite{B-2016-07}, the limiting behavior of $E(\kappa)$ and $M(\kappa)$ is found as
\begin{equation}
\begin{aligned}
  E(\kappa) &= O(\kappa), &\quad M(\kappa) &= O(1) &\quad& \text{as }~\kappa\to 0, \\
  E(\kappa) &= O(1), &\quad M(\kappa) &= O(\kappa^{-2}) &\quad& \text{as }~\kappa\to\infty,
\end{aligned} \label{EM:limit}
\end{equation}
and we provisionally assume that~\eqref{EM:limit} also applies to the present problem. In~\cite{B-2016-07}, the limiting case as $\kappa\to 0$ was shown (assuming time-harmonic elasticity) to recover the conventional PDE-constrained $L^2$ minimization framework (without regularization).

We now study the L-curve defined by
\begin{equation}
  \kappa\mapsto \lpar E(\kappa),\, M(\kappa) \rpar \qquad \kappa\in\Rbb^+, \label{Lcurve:def}
\end{equation}
which is (explicitly or implicitly) involved in many methods for identifying an optimal~$\kappa$. 

\begin{prop}\label{Lcurve:prop}
Function $\kappa\mapsto E(\kappa)$ (resp. $\kappa\mapsto M(\kappa)$) is increasing (resp. decreasing), and the L-curve~\eqref{Lcurve:def} is a convex arc in the $(E,M)$-plane. See Appendix~\ref{Lcurve:prop:proof} (ESM) for proof. 
\end{prop}

An important consequence of Proposition~\ref{Lcurve:prop}, combined with the knowledge of the limiting cases~\eqref{EM:limit}, is  the following: for any $\mathfrak{n}\shg0$ such that $M(0)\!>\!\mathfrak{n}^2$, there exists $\kappa$ such that $M(\kappa)\!=\!\mathfrak{n}^2$ (and that value is usually unique). In practice, $\mathfrak{n}$ stands for the (estimated) measurement noise level (such that $\tdemi\McalT(\bfu\shm\bfum,\bfu\shm\bfum)=\mathfrak{n}^2$ for the displacement $\bfu$ arising in the true material. The selection rule for selecting the optimal value $\kappa=\kappa^\star$ based on achieving $M(\kappa^\star)=\gamma\hh\mathfrak{n}^2$ at the minimizer for some $\gamma\shgeq1$ is known as the Morozov discrepancy criterion~\cite{Morozov1984}. In general, selection rules that do not require prior information on $\mathfrak{n}$, e.g.~those based on the L-curve criterion, also exist. They usually consist in seeking $\kappa=\kappa^\star$ such that $(E,M)(\kappa^\star)$ is closest to the origin; the precise meaning of ``closest'' being definable in many ways, e.g. the point of maximum curvature~\cite{hansen2010discrete} in the $\log M$ vs.~$\log E$ diagram. We note for completeness that the latter curve is not guaranteed to be convex everywhere; as will be seen shortly, however, there may exist an inner convex region that caters for the selection of $\kappa^\star$ via the L-curve criterion. 

\begin{figure}[b]
\centering\includegraphics[width=0.49\textwidth]{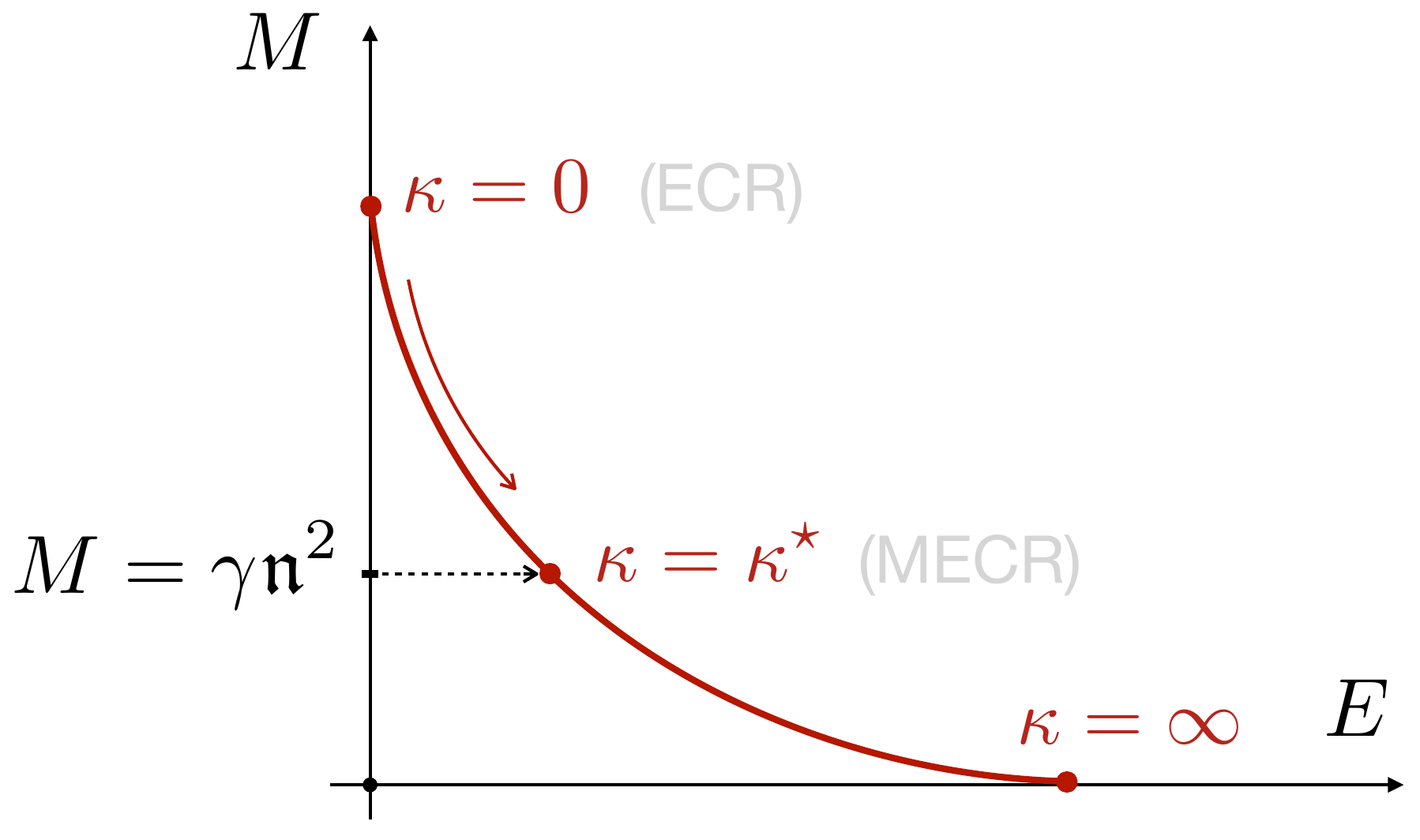}
\caption{\small\textup{L-curve in the $(E,M)$ plane and Morozov criterion for selecting~$\kappa$.}}\label{morozov}
\end{figure}

\subsection{Discussion} \label{sec:discuss}

In this section, we collect a few remarks regarding comparison of methods, or direct extensions or modifications of the foregoing MECR-based methodology.

\paragraph{Boundary conditions and excitations}
The MECR-based formulation presented in Section~\ref{sec:ECR:min} can easily be modified to accommodate other boundary condition and excitation settings. For example, we may assume the displacement $\bfu$ to be constrained on $\GD\shsubs\G$, and a surface traction $\bfg$ to be prescribed on another part $\GN$ of $\G$, with the boundary subsets $\GN$ and $\GD$ only required to not overlap (i.e., $\GN \cap \GD \sheq \emptyset$), so that the remaining part $\Gc\shdeq\G\shsetm(\GN \cup \GD)$ where no information is available may still be non-empty. In such instances,  the displacement spaces $\Ucal,\Wcal$ are redefined as
\begin{equation}
  \Ucal:=\lcb \bfv\shin\Vcal,\,\bfv\sheq\bfze\ \text{on}\ \GD \rcb, \qquad
  \Wcal:=\lcb \bfv\shin\Vcal,\,\bfv\sheq\bfze\ \text{on}\ \GD\shcup\Gc \rcb \subset \Ucal. \label{u:spaces:gen}
\end{equation}
In particular, for well-posed boundary conditions, $\Gc \sheq \emptyset$ and~\eqref{u:spaces:gen} simplifies to $\Wcal=\Ucal$.

In situations where the prescribed excitation is applied directly to~$\overline{\Omega}$, the balance of linear momentum~\eqref{balance:weak} becomes
\begin{equation}
  \iO\int_0^T \big( \bfsig\dip\bfeps[\bfw] + \rho\ddot{\bfu}\sip\bfw \big) \dt\dV
  =  \Fcal(\bfw) \qquad \forall\:\bfw\in\Wcal, \label{balance:weak:gen}
\end{equation}
where the linear functional $\Fcal$ gathers all local excitations (such as surface tractions on $\GD$ or body forces).

The MECR-based identification method of Section~\ref{sec:ECR:min} then still applies under the above conditions, provided (i) the spaces $\Ucal,\Wcal$ are defined by~\eqref{u:spaces:gen} and (ii) the linear functional $\Fcal(\bfw)$ is added to the right-hand sides of~\eqref{L:def} and~\eqref{stat:W}.

\paragraph{Initial and final conditions} 
For simplicity, this study assumes quiescent past for the field quantities, whereby $\bfu$ and $\bfw$ are subject to the homogeneous initial and final conditions, respectively. This feature is shared with the usual PDE-constrained minimization of $L^2$ data residuals. In situations where the sensory data $\bfum$ are available only after some later time, the ability to restrict computations over the actual time duration of measurements would be beneficial. In such situations, the initial values of $\bfu$ and $\bfal$ are usually not known. It might be possible to overcome this lack of initial data by a treatment akin to that of the boundary conditions on $\G$; specifically, one would expect (by analogy) to have to impose the homogeneous initial conditions, in addition to the homogeneous final conditions, on $\bfw$ via re-specification of the test function space~$\Wcal$ in~\eqref{u:spaces}.

\paragraph{Weighted ECR functional}
A modified definition~\eqref{ECR:def} of the ECR component of the form $\Ecal=\Ecale+\gamma\Ecalv$ (where $\Ecale,\Ecalv$ are defined by~\eqref{ECR:decomp} and $\gamma\shg0$ is a dimensionless weight) may easily be implemented, by simply changing $T$ to $\gamma T$ in the formulation of the coupled stationarity problem. This allows one to reweigh the contribution the stored-energy and dissipation components of the constitutive model.


\section{Stationarity system for the time-harmonic case} \label{tharmonic}

\noindent We next derive the stationarity system for the MECR functional applied to time-harmonic motions. The latter are assumed to have the form $\bsfu(\bfx,t)=\Re(\bfu(\bfx)e^{-\rmi\oo t})$, where $\oo$ is a prescribed frequency and $\bfu$ is a complex-valued function. The reference signal duration $T$ is taken as one period, i.e. $T=2\pi/\oo$, and the featured spatiotemporal integrals of bilinear (or quadratic) expressions reduce as
\begin{equation} \label{tharm1}
  \iO\int_0^T  \bsfu\sip\bsfv \dV\dt \;=\; \frac{\pi}{\oo} \iO \Re( \bfu\sip\bfvbb ) \dV, \label{time2freq}
\end{equation}
where the overbar denotes complex conjugation. In addition, for real-valued functions $f$ of a generic complex-valued variable $\bfx=\bfx\Rsub\shp\rmi\bfx\Isub$ (e.g. the potentials) we adopt the following convention for their directional derivatives:
\begin{equation}
\lbra \del{\bfx}f,\bfxH \rbra = \Re\lcb (\del{\bfx\Rsub}f+\rmi\del{\bfx\Isub}f)\sip\bfxbH \rcb, \label{dirder:conv}
\end{equation}
which has the property  
\begin{equation}
\begin{aligned}
\lbra \del{\rmi\bfx}f,\bfxH \rbra 
&= \Re\lcb (-\del{\bfx\Isub}f+\rmi\del{\bfx\Rsub}f)\sip\bfxbH \rcb \\
&= \Re\lcb \rmi(\del{\bfx\Rsub}f+\rmi\del{\bfx\Isub}f)\sip\bfxbH \rcb   \quad \rightarrow \quad \del{\rmi\bfx}f = \rmi \del{\bfx}f. 
\end{aligned}\label{dirder:conv:scaling}
\end{equation}
In this setting, all quadratic potentials become Hermitian forms, e.g.
\begin{equation}
\psi(\bfeps,\bfal) = 
\demi\bigl( \bfeps\dip\Ce\dip\bfepsb + \bfal\dip\Cm\dip\bfepsb + \bfeps\dip\CmT\dip\bfalb + \bfal\dip\Ca\dip\bfalb \bigr),
\end{equation}
with the same tensor-valued coefficients as in the transient case. The constitutive relations~\eqref{stress} also retain the same form (with e.g. $\bfepsd=-\rmi\oo\bfeps$ for strain rates), so that
\begin{equation}
  \bfsige[\bfu] = \Ce\dip\bfeps+\CmT\dip\bfal, \qquad \bfsigv[\bfu] = -\rmi\oo(\De\dip\bfeps+\DmT\dip\bfal). \label{sigev:freq}
\end{equation}
By virtue of the above results, we further have
\begin{equation}
 \varphi(\bfepsd,\bfald) = \oo^2\varphi(\bfeps,\bfal), \qquad \del{\dot\alpha}\varphi(\bfepsd,\bfald) = -\rmi\oo \del{\alpha}\varphi(\bfeps,\bfal), \label{o2:phi}
\end{equation}
and we shall in this section use the short-hand notation $\varphi :=\varphi(\bfeps,\bfal)$. Similarly, the conjugate potentials $\psi^{\star}$ and~$\varphi^{\star}$ are given by the Hermitian-form versions of~\eqref{psi:star:expr} and~\eqref{phi:star:expr}, with their tensor coefficients still given by~\eqref{C:coeff:conj} and~\eqref{D:coeff:conj}. 

\begin{remark}
As all objective functionals and PDE constraints in variational form involve integration over time, the multiplicative factor $\pi/\oo$ featured in~\eqref{tharm1} is carried over by each relevant frequency-domain counterpart of the general formulation and is hereon omitted for brevity.
\end{remark}

The linear viscoelastic constitutive relation results from noting that the differential equation~\eqref{evol} on $\bfal$ becomes algebraic and yields
\begin{equation}
  \bfal[\bfu] = - (\Ca\shm\rmi\oo\Da)^{-1}\dip(\Cm\shm\rmi\oo\Dm)\dip\bfeps, \label{alpha:u:freq}
\end{equation}
allowing for easy elimination of $\bfal=\bfal[\bfu]$ in~\eqref{sigev:freq}. 
As a result, the stress is related linearly to the strain as
\begin{equation} \label{complexC}
\bfsig[\bfu] = (\bfsige+\bfsigv)[\bfu] = \CS(\oo)\dip\bfeps,
\end{equation}
with the tensor of complex viscoelastic moduli $\CS(\oo)$ given by
\begin{equation} \label{C:freq}
  \CS(\oo)
 = (\Ce\shm\rmi\oo\De) - (\CmT\shm\rmi\oo\DmT)\dip(\Ca\shm\rmi\oo\Da)^{-1}\dip(\Cm\shm\rmi\oo\Dm) 
\end{equation}
which carries major and minor symmetries. Straightforward calculations allow us to show that $\CS(\oo)$ permits an alternative expression
\begin{equation}  
  \CS(\oo) = \CS\Isub - \rmi\oo\DS\Isub - \CHT\!\dip(\Ca\shm\rmi\oo\Da)^{-1}\dip\CH. \label{C:freq:alt}
\end{equation}

\paragraph{Objective functional}
Recalling~\eqref{X}, the frequency-domain counterpart of the ECR functional reads
\begin{equation}
  \Ecal(\bsfX,\bsfp;\oo)
 := \iO \Lcb \epsilon_\psi(\bfeps,\bfal,\bfsige,\bfA,\bsfp) + T\epsilon_\varphi(\bfeps,\bfal,\bfsigv,\bfA,\bsfp;\oo) \Rcb \dV, \label{ECR:freq}
\end{equation}
where the Legendre-Fenchel gaps are given (using the first of~\eqref{o2:phi} for $\epsilon_\varphi$) by
\begin{equation}
\begin{aligned}
  \epsilon_\psi(\bfeps,\bfal,\bfsige,\bfA,\bsfp)
 &:= \psi(\bfeps,\bfal,\bsfp) + \psi^{\star}(\bfsige,\bfA,\bsfp) - \Re\lsqb\bfsige\dip\bfepsb - \bfA\dip\bfalb\rsqb, \\
\epsilon_\varphi(\bfeps,\bfal,\bfsigv,\bfA,\bsfp;\oo)
 &:= \oo^2\varphi(\bfeps,\bfal,\bsfp) + \varphi^{\star}(\bfsigv,\bfA,\bsfp)
- \Re\lsqb \rmi\oo(\bfsigv\dip\bfepsb + \bfA\dip\bfalb) \rsqb,
\end{aligned} \label{LF:gap:freq}
\end{equation}
with the potentials and their conjugates understood as Hermitian forms. The MECR functional $\Lambdak$ is then defined by
\begin{equation}
  \Lambdak(\bsfX,\bsfp;\oo)
 := \Ecal(\bsfX,\bsfp;\oo) + \tdemi\kappa\exs \Mcal(\bfu\shm\bfum,\overline{\bfu\shm\bfum}), \label{MECR:def:freq}
\end{equation}
where~$\Mcal(\bfr,\bfs)$ is the positive symmetric bilinear form introduced in~\eqref{Mcal:def}. The minimization of~\eqref{MECR:def:freq} subject to the time-harmonic version of the balance constraint~\eqref{balance:weak} leads to introducing the Lagrangian
\begin{equation}
\Lcal(\bsfX,\bfw,\bsfp;\oo) :=\, \Lambdak(\bsfX,\bsfp;\oo) 
-\hh \Re\,\Lcb \iO \!\big( (\bfsige\shp\bfsigv)\dip\bfeps[\bfwb] - \rho\oo^2\bfu\sip\bfwb \big) \dV \Rcb
\label{L:def:freq}
\end{equation}
where, consistently with the convention~\eqref{dirder:conv}, the test function $\bfw\shin\Wcal$ is conjugated. The resulting first-order optimality conditions are the time-harmonic versions of conditions~\eqref{group1}-\eqref{group3}.

\paragraph{Local stationarity equations} Recalling the short-hand notation $\bfeps=\bfeps[\bfu]$ and $\bfeta=\bfeps[\bfw]$, equations~(\ref{group1}a-c) lead to the time-harmonic versions of the pointwise equations~\eqref{group1:pointwise}, i.e.
\begin{equation}
\begin{aligned}
&\text{(a)}  &  \bfze  &= \del{\sige}\psi^{\star} - \bfeps - \bfeta, \\
&\text{(b)}  &  \bfze  &=  T\lpar \del{\sigv}\varphi^{\star} + \rmi\oo\bfeps \rpar -\bfeta, \\
&\text{(c)}  &  \bfze  &= \del{A}\psi^{\star} + \bfal + T\lpar \del{A}\varphi^{\star} + \rmi\oo\bfal \rpar.
\end{aligned} \label{group1:pointwise:freq}
\end{equation}
Solving them for $\bfsige,\bfsigv,\bfA$ yields
\begin{equation}
\begin{aligned}
\bfsige &= \Ce\dip\bfeps + \CmT\dip\bfal + \lpar \Ce - \CmT\dip\Da^{-1}\dip\Dm \rpar\dip\bfeta -T\CmT\dip\bfbe, \\
\bfsigv &= - \rmi\oo\DmT\dip\bfal -\rmi\oo\De\dip\bfeps + \Tinv \DS\Isub\dip\bfeta - \DmT\dip\bfbe, \\
\bfA &= - \rmi\oo\Da\dip\bfal  - \rmi\oo\Dm\dip\bfeps - \Da\dip\bfbe,
\end{aligned} \label{ssA:freq}
\end{equation}
with $\bfbe$ given by
\begin{equation}
(\Da\shp T\Ca)\dip\bfbe = (\Ca\shm\rmi\oo\Da)\dip\bfal + (\Cm\shm\rmi\oo\Dm)\dip\bfeps + \CH\dip\bfeta.
\label{beta:freq}
\end{equation}

Then, the time-harmonic counterpart of the stationarity condition~(\ref{group1:exp}d) is
\begin{equation}
  0 = \iO \Lcb \del{\alpha}\psi + \bfA - \rmi\oo T(\rmi\oo\del{\alpha}\varphi + \bfA) \Rcb \dip\bfalbH \dV \qquad
  \forall\:\bfalH
\end{equation}
with $\bfA$ given by~\eqref{ssA:freq}, implying (after some algebra) the pointwise algebraic equation
\begin{equation}
  (\Ca\shp\rmi\oo\Da)\dip\bfbe - \Tinv \CH\dip\bfeta = \bfze, \label{ODE:beta:freq}
\end{equation}
which provides the value of $\bfbe$ in terms of $\bfw$. We next substitute the last result into~\eqref{beta:freq} and set $\bfal=\bfal[\bfu]\shp\bfa$ (where $\bfa$ evaluates, as in Sec.~\ref{sec:stat:cond}, the constitutive gap between $\bfal[\bfu]$ given by~\eqref{alpha:u:freq} and $\bfal$ solving the local stationarity equations); $\bfa$ is hence given in terms of $\bfw$ by
\begin{equation}
  (\Ca\shm\rmi\oo\Da)\dip\bfa = (\Da\shp T\Ca)\dip\bfbe - \CH\dip\bfeta
 = (\Tinv \shm\rmi\oo)\Da\dip(\Ca\shp \rmi\oo \Da)^{-1}\dip\CH\dip\bfeta. \label{aux04}
\end{equation}

\paragraph{Constitutive gap} From~\eqref{ssA:freq}, the stress $\bfsig=\bfsige\shp\bfsigv$ solving the stationarity system is computed as
\begin{equation}
  \bfsig
 = \CS(\oo)\dip\bfeps + (\CmT\shm\rmi\oo\DmT)\dip\bfa
  + \lpar \CeS + \Tinv \DS\Isub + \Cm\dip\Ca^{-1}\dip\CH \rpar\dip\bfeta
  -(T\CmT+\DmT)\dip\bfbe,
\end{equation}
where~$\CS(\oo)$ is given by~\eqref{C:freq}. On substituting the values of $\bfbe$ and $\bfa$ respectively provided by~\eqref{ODE:beta:freq} and~\eqref{aux04}, $\bfsig$ can be written in terms of $\bfu$ and $\bfw$ by
\begin{equation}
  \bfsig = \CS(\oo)\dip\bfeps + \SS(\oo)\dip\bfeta, \label{sigma:expr:freq}
\end{equation}
where~$\CS(\oo)$ is given by~\eqref{C:freq} and
\begin{equation}
  \SS(\oo)
 = \CS\Isub + \Tinv \DS\Isub + \CHT\!\dip(\Ca-\rmi\oo\Da)^{-1}\dip(\Tinv \Da-\Ca)\dip(\Ca+\rmi\oo\Da)^{-1}\dip\CH, 
\end{equation}
which is seen to define a Hermitian sesquilinear form over the field of second-order tensors. In fact, using definition~\eqref{schur:def} of the Schur complement $\CeS$ in~\eqref{C:inst} which yields $\CS\Isub=\CeS+\CHT\!\dip\Ca^{-1}\dip\CH$, we obtain an  alternative expression
\begin{equation}
  \SS(\oo)
 = \CeS + \Tinv \DS\Isub + \CHT\!\dip(\Ca-\rmi\oo\Da)^{-1}\dip(\Tinv \Da+\oo^2\Da\dip\Ca^{-1}\dip\Da)\dip(\Ca+\rmi\oo\Da)^{-1}\dip\CH,
\end{equation}
which demonstrates that $\SS(\oo)$ is Hermitian positive definite (which is consistent with its time-integrated counterpart~\eqref{SS:integr}).

\paragraph{Global stationarity equations} Condition~(\ref{group2}a) now reads
\begin{equation}
  \iO \big( \bfsig\dip\bfeps[\bfwbH] - \rho\oo^2\bfu\sip\bfwbH \big) \dV = 0 \qquad\forall\:\bfwH\shin\Wcal. 
\end{equation}
The stress being given by~\eqref{sigma:expr:freq}, we hence obtain
\begin{equation}
  \iO \Lcb \big(\CS(\oo)\dip\bfeps[\bfu]\big)\dip\bfeps[\bfwbH] - \rho\oo^2\bfu\sip\bfwbH \Rcb \dV
 = -\iO \big(\SS(\oo)\dip\bfeps[\bfw]\big)\dip\bfeps[\bfwbH] \dV \qquad\forall\:\bfwH\shin\Wcal \label{stat:W:freq}
\end{equation}
which, as in the transient case, takes the form of a viscoelasticity problem for $\bfu$ driven by an initial strain field that depends linearly on $\bfeps[\bfw]$.

The time-harmonic version of condition~(\ref{group2}b) \emph{a priori} takes the form 
\begin{multline}
  \iO \Lcb \big( \del{\eps}\psi - \bfsige - \rmi\oo T(\rmi\oo \del{\eps}\varphi + \bfsigv) \big) \dip\bfeps[\bfubH]
  + \rho\oo^2\bfw\sip\bfubH \Rcb \dV =
  - \kappa\exs \Mcal(\bfu\shm\bfum,\bfubH) \quad \forall\:\bfuH\shin\Ucal. \label{aux05}
\end{multline}
Moreover, $\bfsige$ and $\bfsigv$ being given by~\eqref{ssA:freq}, we have
\begin{align}
  \del{\eps}\psi - \bfsige - \rmi\oo T(\rmi\oo \del{\eps}\varphi + \bfsigv)
 &= \big\{(\CmT\shp \rmi\oo\DmT)\dip(\Ca \shp \rmi\oo\Da)^{-1}\dip\CH
   + ( \CmT\dip\Da^{-1}\dip\Dm - \Ce - \rmi\oo \DS\Isub )\big\} \dip\bfeta \\
 &= \big\{(\CmT\shp \rmi\oo\DmT)\dip \big((\Ca\shp \rmi\oo\Da)^{-1}\dip\CH + \Da^{-1}\dip\Dm \big) - (\Ce\shp \rmi\oo\De)\big\} \dip\bfeta \\
 &= \big\{(\CmT\shp \rmi\oo\DmT)\dip(\Ca\shp \rmi\oo\Da)^{-1}\dip( \Cm + \rmi\oo\Dm )\dip\bfeta - (\Ce + \rmi\oo\De)\big\}\dip\bfeta  \\
 &= -\overline{\CS(\oo)}\dip\bfeta 
\end{align}
Accordingly, the variational equation~\eqref{aux05} reduces to
\begin{equation}
  \iO \Big( \big(\overline{\CS(\oo)}\dip\bfeps[\bfw]\big)\dip\bfeps[\bfubH] - \rho\oo^2\bfw\sip\bfubH \Big) \dV =
  \kappa\exs\Mcal(\bfu\shm\bfum,\bfubH) \qquad \forall\:\bfuH\shin\Ucal. \label{aux06}
\end{equation}
Indeed, since the time reversal amounts to complex conjugation in the frequency domain, \eqref{aux06} has the form (as in the transient case~\eqref{stat:u}) of a time-reversed time-harmonic viscoelasticity problem driven by the measurement residuals.

\paragraph{Coupled stationarity problem} Summarizing, the stationarity conditions~\eqref{group1}--\eqref{group2} lead to the coupled variational problem on $(\bfu,\bfw)\in\Ucal\shtimes\Wcal$ given by~\eqref{stat:W:freq} and~\eqref{aux06}. Once the solution pair $(\bfu,\bfw)$ is computed, components~$\bfsige$ and~$\bfsigv$ of the Cauchy stress tensor, tensor of thermodynamic tensions~$\bfA$,  and viscoelastic strain tensor~$\bfal$ are  given explicitly by~\eqref{ssA:freq} and~\eqref{aux04}, with $\bfbe$ solving~\eqref{ODE:beta:freq}.

\begin{remark}\label{multifreq}
The foregoing formulation allows for a straightforward extension to multi-frequency sensing where the constitutive parameters are sought by minimizing a cumulative MECR functional 
\begin{equation} \label{cumulx1}
  \LambdakL \,=\, \sum_{\ell=1}^L {c_\ell} \hh\hh\Lambdak(\bsfX_{\ell},\bsfp;\oo_{\ell}),    
\end{equation}
where $\oo_{\ell}$, $\ell\!=\!\overline{1,L}$ are the acquisition frequencies; $\Lambdak(\dots;\oo_{\ell})$ is given by~\eqref{MECR:def:freq}, and each set $\bsfX_{\ell}$ of the field variables is determined, for given~$\bsfp$, by solving the stationarity conditions for the Lagrangian~\eqref{L:def:freq} with $\oo=\oo_{\ell}$. The weights ${c_\ell}\!\in\!\mathbb{R}^+$ can be chosen arbitrarily, and could be taken for example as multiples of the characteristic times $2\pi/\oo_{\ell}$. 
\end{remark}

\section{Example implementations of the SGM framework} \label{vispecial}

\begin{figure}[ht]
\centering\includegraphics[width=0.75\linewidth]{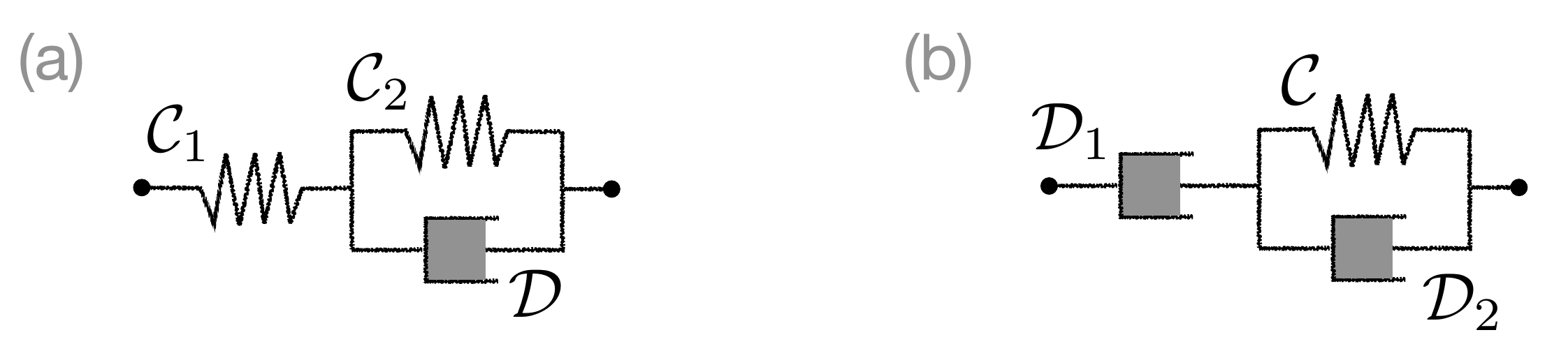} \vspace*{-2mm}
\caption{\small\textup{1D mechanical analogues of two rudimentary viscoelastic models: (a) standard linear solid, and (b) Jeffreys material.}}\label{viscomodels}
\end{figure}

\noindent For lucency of presentation, we illustrate the SGM framework by extrapolating upon two basic viscoelastic models depicted in Fig.~\ref{viscomodels}. 

\subsection{Standard linear solid} \label{SecSLS}

We first consider the spring-and-dasphot arrangement in Fig.~\ref{viscomodels}(a), whose 1D stress-strain relationship reads $\sigma + \tau_1\hh\dot\sigma = \gamma(\eps + \tau_2\hh\dot{\eps})$ featuring the coefficients $\gamma=\mathcal{C}_1 \mathcal{C}_2/(\mathcal{C}_1\!+\mathcal{C}_2)$, $\tau_1=\mathcal{D}/(\mathcal{C}_1\!+\mathcal{C}_2)$, and $\tau_2=\mathcal{D}/\mathcal{C}_2$.
In this vein, we focus on the class of SGM models verifying
\begin{equation}
\Ce=-\Cm=\CS_1, \qquad \Ca=\CS_1\shp\CS_2, \qquad
\Da=\DS, \qquad \De=\Dm=\bfze,
\label{standard:gen}
\end{equation}
where the fourth-order tensors~$\CS_1,\,\CS_2$ (synthesizing the elastic response) and $\DS$ (representing the viscous response) are positive and carry the usual major and minor symmetries. We note that the positivity of $\DS$ ensures positive dissipation as per the second law of thermodynamics. The free energy and dissipation potentials corresponding to~\eqref{standard:gen} are given by
\begin{equation}
  \psi(\bfeps,\bfal)
 = \demi(\bfeps\shm\bfal)\dip\CS_1\dip(\bfeps\shm\bfal) + \demi\bfal\dip\CS_2\dip\bfal, \qquad\qquad
  \varphi(\bfepsd,\bfald) = \demi\bfald\dip\DS\dip\bfald, \label{auxfeee1}
\end{equation}
where $\bfeps-\bfal$ and $\bfal$ generalize respectively upon the axial strains undergone by the springs~$\mathcal{C}_1$ and~$\mathcal{C}_2$ in Fig.~\ref{viscomodels}(a). In this setting, we obtain
\begin{subequations}
\begin{align}
\bfsige[\bfu] &= \del{\eps}\psi = \CS_1\dip(\bfeps\shm\bfal), \label{standard:zener} \\
\bfsigv[\bfu] &= \del{\epsd}\varphi = \bfze, \\
\bfA[\bfu] &= -\del{\alpha}\psi = \CS_1\dip\bfeps - (\CS_1\shp\CS_2)\dip\bfal \label{A:standard:1} \,=\, \del{\alphad}\varphi = \DS\dip\bfald. \label{A:standard:2}
\end{align}
\end{subequations}
The evolution equation~\eqref{evol} becomes
\begin{equation}
  \bfald + \QS\dip\bfal = (\DS^{-1}\dip\CS_1)\dip\bfeps, \label{evol:standard}
\end{equation}
with $\QS:=\Da^{-1}\dip\Ca =\DS^{-1}\dip(\CS_1\shp\CS_2)$, and expression~(\ref{alpha(t)}) for the viscous strain reduces to
\begin{equation}
\bfal(t) = \FS[\CS_1\dip\bfeps](t),  \label{alpha:zener}
\end{equation}
where the convolution operator $\bfs\mapsto\FS[\bfs]$ is given by~\eqref{KS:def} with $\Da=\DS$.
By virtue of~\eqref{sig:expr}, the stress response of a standard linear solid due to displacement history $\tau\mapsto\bfu(\tau),\,\tau\shleq t$ reduces to 
\begin{equation}
\bfsig[\bfu](t) = \CS_1\dip\bfeps(t) - \CS_1\dip\FS[\CS_1\dip\bfeps](t). \label{sig:standard}
\end{equation}

\paragraph{Isotropic case}

In the general anisotropic case, tensor $\QS$ has up to six distinct eigenvalues, which define as many characteristic relaxation times. For an isotropic solid, on the other hand, tensors $\CS_1,\CS_2$ and~$\DS$ are of the form
\begin{equation}
\CS_1 = 3\upkappa\JS + 2\upmu\hh\KS, \qquad
\CS_2 = 3\upkappa'\JS + 2\upmu'\KS, \qquad
\DS = 3\upeta\JS + 2\upchi\KS \label{CD:isotropic}
\end{equation}
featuring the four elastic moduli $\{\upkappa,\upkappa',\upmu,\upmu'\}$ and two viscous moduli $\{\upeta,\upchi\}$. The fourth-order tensors $\JS$ and~$\KS$, representing respectively the volumetric and deviatoric components of an isotropic fourth-order tensor, are given by
\[
  \JS = \tiers\bfI\tens\bfI, \qquad \KS=\IS-\JS
\]
where~$\bfI=\delta_{ij}\,\bfe_i\!\otimes\!\bfe_j$ and $\IS=\tdemi(\delta_{ik}\delta_{jl}+\delta_{il}\delta _{jk})\,\bfe_i\!\otimes\!\bfe_j\!\otimes\!\bfe_k\!\otimes\!\bfe_l$ denote respectively the second-order and symmetric fourth-order identity tensors. It is easily shown that $\JS$ and~$\KS$ verify
\begin{equation}
  \JS\dip\JS=\JS, \qquad \KS\dip\KS=\KS, \qquad \JS\dip\KS=\bfze. \label{JK:orth}
\end{equation}
On introducing the bulk and shear relaxation times $\tau\Bsub\shdeq\upeta/(\upkappa\shp\upkappa')$ and $\tau\Ssub\shdeq\upchi/(\upmu\shp\upmu')$, one has
\[
  \QS \,=\, \tau\Bsub^{-1}\JS + \tau\Ssub^{-1}\KS.
\]
As a result, the tensor exponential $t\mapsto\exp(-\QS t)$ can be written explicitly as
\begin{equation}
\exp(-\QS t) = e^{-t/\tau\Bsub}\JS + e^{-t/\tau\Ssub}\KS,
\end{equation}  
whereby 
\begin{equation}
\CS_1\dip\exp(-\QS t)\dip(\DS^{-1}\dip\CS_1) = (3\upkappa^2/\upeta)e^{-t/\tau\Bsub} \JS + (2\upmu^2/\upchi)e^{-t/\tau\Ssub} \KS.
\end{equation}
The constitutive relation~(\ref{sig:standard}) then becomes
\begin{equation}
  \bfsig[\bfu](t)
 = 3\upkappa \Lcb \Tr\bfeps(t)
 - \frac{\upkappa}{\upeta} \int_{0}^{t} e^{(\tau-t)/\tau\Bsub}\Tr\bfeps(\tau) \dtau \Rcb \bfI
   + 2\upmu \Lcb \bfe(t) - \frac{\upmu}{\upchi} \int_{0}^{t} e^{(\tau-t)/\tau\Ssub}\bfe(\tau) \dtau \Rcb,
\label{sig:standard:iso}
\end{equation}
where~$\Tr\bfeps$ and $\bfe\sheq\KS\dip\bfeps$ denote respectively the volumetric and deviatoric strain. Note that the volumetric and deviatoric components of~(\ref{sig:standard:iso}) each carry the structure of the parent 1D model.

\subsection{Jeffreys material} \label{SecJF}

We next examine the spring-and-dashpot assembly shown in Fig.~\ref{viscomodels}(b), a three-parameter specialization of the Burgers material~\cite{findley2013creep} endowed with the 1D stress-strain relationship $\sigma + \tau_1\hh\dot\sigma = \upeta(\dot{\eps} + \tau_2\hh\ddot{\eps})$ with $\upeta = \mathcal{D}_1$, $\tau_1 = (\mathcal{D}_1\!+\mathcal{D}_2)/\mathcal{C}$, and $\tau_2 =\mathcal{D}_2/\mathcal{C}$. In this spirit, we consider the class of SGM models verifying
\begin{equation}
\Ce=\Ca=-\Cm=\CS, \qquad\qquad 
\De = -\Dm = \DS_{2}, \qquad \Da=\DS_{1}\!+\DS_{2},
\label{jeff:gen}
\end{equation}
where the fourth-order tensors $\CS_1,\CS_2$ and $\DS$ satisfy the same assumptions as before. Using assumptions~\eqref{jeff:gen} in~\eqref{phi:psi:def}, the free energy and the dissipation potential become
\begin{equation}
\psi(\bfeps,\bfal) = \demi(\bfeps\shm\bfal)\dip\CS\dip(\bfeps\shm\bfal), \qquad\qquad
\varphi(\bfepsd,\bfald) = \demi\bfald\dip\DS_1\dip\bfald + \demi(\bfepsd\shm\bfald)\dip\DS_2\dip(\bfepsd\shm\bfald), \label{phi:psi:def:zener}
\end{equation}
where $\bfeps-\bfal$ and $\bfal$ mirror respectively the axial strains undergone by the dashpots~$\mathcal{D}_2$ and~$\mathcal{D}_1$ in Fig.~\ref{viscomodels}(b). As a result, we obtain
\begin{equation} 
\begin{aligned}
\bfsige[\bfu] &= \del{\eps}\psi = \CS\dip(\bfeps\shm\bfal),  \\
\bfsigv[\bfu] &= \del{\epsd}\varphi = \DS_2\dip(\bfepsd\shm\bfald), \\
\bfA[\bfu] &= -\del{\alpha}\psi = \CS\dip(\bfeps\shm\bfal) 
 \,=\, \del{\alphad}\varphi = \DS_1\dip\bfald -\DS_2\dip(\bfepsd\shm\bfald). 
\end{aligned}
\label{A:jeff}
\end{equation}
The evolution equation~\eqref{evol} becomes
\begin{equation}
\bfald + \QS\dip\bfal = (\IS-\PS)\dip\bfepsd + \QS\dip\bfeps, \label{evol:jeff}
\end{equation}
where $\PS=(\DS_1\shp\DS_2)^{-1}\dip\DS_1$ and $\QS=(\DS_1\shp\DS_2)^{-1}\dip\CS$. Formula~(\ref{alpha(t)}) for the viscous strain reduces to
\begin{equation}
  \bfal(t) = (\IS\shm\PS)\dip\bfepsd(t) + \FS[\CS\dip\PS\dip\bfeps](t),  \label{alpha:jeff}
\end{equation}
and the viscoelastic stress tensor~\eqref{sig:expr} reads
\begin{equation}
\bfsig[\bfu](t) = \PS\Tsup\!\dip\CS\dip\PS\dip\bfeps(t) + \DS_2\dip\PS\dip\bfepsd(t)  - \PS\Tsup\!\dip\CS\dip\FS[\CS\dip\PS\dip\bfeps](t). \label{sig:jeff}
\end{equation}

\begin{remark}    
From the first and the last of~\eqref{A:jeff}, we obtain the constraint $\bfsige\!=\!\bfA$, which should be compared against its counterpart $\bfsigv\!=\!\bfze$ featured by the standard linear solid. As a result, compliance tensors specifying the conjugate potential $\psi^\star(\bfsige,\bfA)$ in~\eqref{psi:star:expr} describing Jeffreys material are indeterminate (and can be chosen arbitrarily) up to the constraint $\Cs+2\Cms+\CA=\CS^{-1}$. For more general SGM descriptions that feature linearly-independent $\bfsige, \bfsigv$ and~$\bfA$, one may consider spring-and-dashpot models of higher complexity, e.g. Burgers material.
\end{remark}

\paragraph{Isotropic case} For an isotropic Jeffreys material, we let
\begin{equation}
\CS = 3\upkappa\JS + 2\upmu\hh\KS, \qquad
\DS_1 = 3\upeta\JS + 2\upchi\KS, \qquad
\DS_2 = 3\upeta'\JS + 2\upchi'\KS, \label{CD:isotropic:jeff}
\end{equation}
featuring the two elastic moduli $\{\upkappa,\upmu\}$ and four viscous moduli $\{\upeta,\upeta',\upchi,\upchi'\}$. Accordingly, we have
\[
\QS = (1/\tau\Bsub)\JS+(1/\tau\Ssub)\KS
\]
in terms of the relaxation times $\tau\Bsub\shdeq(\upeta+\upeta')/\upkappa$ and $\tau\Ssub\shdeq(\upchi+\upchi')/\upmu$. As before, the tensor exponential $t\mapsto\exp(\QS t)$ can  be written in explicit form, and one obtains
\begin{align}
  \exp(-\QS t)
 &= e^{-t/\tau\Bsub} \JS + e^{-t/\tau\Ssub} \KS, \\[1ex]
  \PS\Tsup\dip\CS\dip\exp(-\QS t)\dip(\DS_1\shp\DS_2)^{-1}\dip\CS\dip\PS
 &= \frac{3\upkappa^2\upeta^2}{(\upeta+\upeta')^3}\hh e^{-t/\tau\Bsub} \JS + \frac{2\upmu^2\upchi^2}{(\upchi+\upchi')^3}\hh e^{-t/\tau\Ssub} \KS.
\end{align}
The constitutive relationship~(\ref{sig:jeff}) then becomes
\begin{multline}
  \bfsig[\bfu](t)
  = \Bigl\{ \frac{3\upkappa\upeta^2}{(\upeta+\upeta')^2}\Tr\bfeps(t) + \frac{3\upeta\upeta'}{(\upeta+\upeta')}\Tr\dot{\bfeps}(t)
 - \frac{3\upkappa^2\upeta^2}{(\upeta+\upeta')^3} \int_{0}^{t} e^{(\tau-t)/\tau\Bsub}\Tr\bfeps(\tau) \dtau \Bigr\}\bfI \\[-1ex]
   + \Bigl\{\frac{2\upmu\upchi^2}{(\upchi+\upchi')^2} \bfe(t) + \frac{2\upmu\upchi\upchi'}{(\upchi+\upchi')} \dot{\bfe}(t)
   - \frac{2\upmu^2\upchi^2}{(\upchi+\upchi')^3} \int_{0}^{t} e^{(\tau-t)/\tau\Ssub}\bfe(\tau) \dtau \Bigr\},
\label{sig:zener:iso}
\end{multline}
whose volumetric and deviatoric parts each carry the structure of the 1D Jeffreys material.

\section{Numerical results} \label{numres}

\noindent For computational purposes, we assume that the full-field sensory data are available at each finite element (FE) node of the numerical model used for material characterization, and we proceed with the latter in a piecemeal fashion by introducing a set of (possibly overlapping) \emph{subzones}~\cite{mcgarry2022}, $\{S_n\!\subset\!\Omega\}$, which tile the domain of interest $\Omega\!\subset\!\mathbb{R}^d$. In practical terms, the ``full-field" assumption can be met by interpolating the physical motion measurements (e.g. via cubic splines) and projecting the interpolated field onto an FE grid used for MECR computations. We further let each subzone $S_n\!\subset\mathbb{R}^d$ be an equilateral $d$-parallelepiped composed of $\mathcal{N}^d$ pixels. In this setting, MECR material characterization over a subzone becomes a discrete inverse problem featuring $\mathfrak{p}\hh\mathcal{N}^d$ parameters, $\mathfrak{p}\!=\!\dim\bsfp$ being the number of independent viscoelastic moduli in~\eqref{phi:psi:def}. The key advantage of the subzone approach is that, for a given resolution requirement (i.e. pixel size), it reduces the dimension of the parametric space and so moderates the non-convexity of the cost functional. 

For clarity, all physical parameters are hereon presented in a dimensionless form -- as normalized by (i) the characteristic size of a viscoelastic body; (ii) its mass density that is taken as being constant throughout, and (iii) the reference shear modulus.

\subsection{Subzone and pixel size considerations}\label{subzone}

In general, the absence of boundary conditions in the global stationarity equations~\eqref{stat:W} and~\eqref{stat:u} (or equivalently \eqref{aux05}--\eqref{aux06} for time-harmonic problems), rewritten for $\Omega\mapsto S_n$, is compensated for by sufficient full-field measurements taken over a subzone. When the characteristic linear size $|S_n|^{1/d}$ of a subzone is reduced, we encounter a diminishing ratio between the number of subzone \emph{interior} DOFs (featured by an FE model) and that of subzone \emph{boundary} DOFs. In other words, a lesser fraction of interior data remains available for material identification. Qualitatively speaking, this behavior indicates that a subzone should be sufficiently large as to provide sufficient ``net'' amount of data for the reconstruction of sought material parameters. 

To illustrate the concept, consider a square subzone $S_n\!\subset\!\mathbb{R}^2$ whose underpinning FE mesh has $N\!\times\! N$ nodes and so $2N^2$ degrees of freedom, since $\bfu=(u_1,u_2)$ in this case. Its boundary then carries $8(N\!-\!1)$ degrees of freedom. Out of the $2N^2$ measurements available, $8(N\!-\!1)$ data are implicitly used up to compensate for the missing boundary conditions, which leaves $2(N\!-\!2)^2$ data for the  identification of $\mathfrak{p}\hh\hh\mathcal{N}^2$ parameters. Accordingly, we require that  
\begin{equation} \label{densities}
 2(N\!-\!2)^2 > \mathfrak{p}\hh\hh\mathcal{N}^2 \quad \Rightarrow \quad 
\frac{(N\!-\!2)^2}{\mathcal{N}^2} > \frac{\mathfrak{p}}{2}. 
\end{equation}
For a given finite element size ($h$) that is driven by relevant considerations -- for instance the observed wavelength in a physical experiment or computer memory limitations, relationship~\eqref{densities} imposes an interdependent: (i) lower bound on the characteristic subzone size $N$ (measured in the units of~$h$), and (ii) upper bound on the number of pixels per subzone ($\mathcal{N}^2$) that must hold for the MECR inversion to make sense. In situations where $N\!\gg\!1$ (i.e. where the boundary conditions ``tax" can be ignored), we obtain a simplified requirement $(N/\mathcal{N})^2\gtrsim \mathfrak{p}/2$ for the total number of sensory data over a subzone to be larger than the number of material unknowns therein.

\begin{remark} \label{general1}
Clearly, interpolating the available sensory data over a subzone may not introduce significant ``new'' (i.e. linearly independent) information, and satisfying~\eqref{densities} may be insufficient to guarantee over-determinacy of the problem. Such is the case with the long-wavelength data  ($\lambda\!>\!|S_n|^{1/d}$), whose variation over significant swaths of a subzone may be amenable to a linear approximation. To overcome the impediment, it is often necessary to deploy multiple ``illuminating'' fields, $\bfum^\ell$ ($\ell\!=\!\overline{1,L}$), generated by having the excitation sources placed at distinct locations or carrying distinct temporal signatures. In either situation, the resulting cost functional has the form and properties as in Remark~\ref{multifreq}, with $\omega_\ell$ being replaced by a suitable vector synthesizing the spatiotemporal source characteristics.   
\end{remark}

\subsection{Computational treatment} \label{comptre}

{\color{black}
In what follows, time-harmonic synthetic sensory data $\bfum$ are generated via \mbox{NGSolve} -- an open-source, Python-based finite element computational platform~\cite{NGSolve}. To illustrate the developments, viscoelastic simulations are performed using triangular elements (order $p\!=\!3$) within the framework of \emph{plane strain} kinematics by letting $\bfx\!\in\mathbb{R}^2$ and $\bfu(\bfx)\!\in\!\mathbb{C}^2$. The characteristic element size is taken as $h\!=\!0.0025$, which is sufficient to accurately simulate the wave motion at all frequencies featured in the sequel. For a given viscoelastic body~$D$, the region of interest~$\Omega\subset D$ is illuminated by point sources $\bff(\bfx) = \hat{\bff}\hh \delta(\bfx\!-\!\bfx^\circ)$ with $\|\hat{\bff}\|\!=\!1$ and $\bfx^\circ\!\!\in\!D$ assuming a variety of source locations, force directions, and excitation frequencies. For computational purposes, the Dirac delta function is approximated as a 2D Gaussian distribution endowed with a small variance. The FE mesh used for solving the global stationarity equations~\eqref{aux05}--\eqref{aux06} over each subzone is generated with $p\!=\!3$ and $h\!=\!0.05$. In this setting, synthetic full-field data $\bfum$ is projected onto the latter FE mesh using a linear VoxelCoefficient method available through NGSolve.

\subsubsection{Minimization of the MECR functional} \label{Minimi}

The non-quadratic minimization of~\eqref{cumulx1} with weights ${c_\ell}\!=\!2\pi/\oo_{\ell}$, generalized to account for all sources of illumination at each frequency (see Remark~\ref{general1}), is performed over each subzone~$S_n$ via the sequential least squares programming (SLSQP) algorithm available through the open-source Python library Scipy \cite{Scipy2020}. For $\mathcal{N}\!>\!1$, we tile the interrogation window~$\Omega$ with equisized square subzones that overlap by a single pixel in each direction. As a result, for a fixed number of subzones, $|S_n|$ will decrease with increasing~$\mathcal{N}$. The inversion starts with a uniform initial guess for each subzone, $S_n^{\text{\tiny{(1)}}}$, by letting a single pixel ($\mathcal{N}^{\text{\tiny{(1)}}}\!=\!1$) cover the entire subzone. The result of such inversion is then used to generate the initial profile of viscoelastic properties for a refined inversion ($\mathcal{N}^{\text{\tiny{(2)}}}\!>\!1$) over the successor subzone $S_n^{\text{\tiny{(2)}}}$. In this way, the subsequent refinements from $\mathcal{N}^{\text{\tiny{(j)}}}$ to $\mathcal{N}^{\text{\tiny{(j+1)}}}$ are applied recursively until the sought resolution is achieved. In the examples to follow, we deploy a two-step resolution refinement over each~$S_n$ with $\mathcal{N}^{\text{\tiny{(2)}}}\!=\!5$ and $\mathcal{N}^{\text{\tiny{(3)}}}\!=\!7$.

\subsection{Piecewise-homogeneous standard linear solid} \label{num-sls}

With reference to Fig.~\ref{test-config}(a), we seek to characterize the interior $\Omega\!\subset\!D$ of a square viscoelastic body $D$ housing an elliptic inclusion. The viscoelastic model used to generate $\bfum$ is that of an isotropic standard linear solid described in Sec.~\ref{SecSLS}, whose ``true'' values of $\bsfp\!=\!(\upmu,\upmu',\upchi,\upkappa,\upkappa',\upeta)$ are listed in Table~\ref{tab1}. For this model, the fourth-order tensor of complex moduli~\eqref{C:freq} reads  
\begin{equation} \label{rational2}
\CS(\bsfp,\oo) = 3 K(\bsfp,\oo) \JS + 2 G(\bsfp,\oo) \KS, 
\end{equation}
where 
\begin{equation} \label{KG-SLS}
K(\bsfp,\oo) = \frac{\upkappa\hh(\upkappa'\!-\rmi\oo\hh\upeta)}{\upkappa+\upkappa'\!-\rmi\oo\hh\upeta}, \qquad 
G(\bsfp,\oo) = \frac{\upmu\hh(\upmu'\!-\rmi\oo\hh\upchi)}{\upmu+\upmu'\!-\rmi\oo\hh \upchi}.
\end{equation}
The interrogation window $\Omega$ has dimensions $0.8 \times 0.8$ and is centered inside~$D$. As shown in the figure, $D$ is subjected to the homogeneous Dirichlet boundary conditions along a portion of its bottom edge and is traction-free on the remainder of~$\partial D$. For the present problem, we select four (either mid-height or mid-width) point source locations near $\partial D$; at each location, we excite the body by a unit force $\hat{\bff}=\tfrac{1}{\sqrt{2}}(1,1)$ at four separate frequencies: $\{\omega_\ell\} = \{16,24,36,54\}$. The latter choice, generated by the geometric progression $\omega_{\ell+1}=1.5\hh\omega_\ell$, helps $\{\omega_\ell\}$ to be sparse while having a broadband coverage, which caters for linear independence of the respective data sets. On taking the regularization parameter as $\kappa^\star\sheq 10$, the inversion is performed by tiling~$\Omega$ with a grid of $4\times 4$ equisized subzones $S_n$, $n\!=\!\overline{1,16}.$

\begin{figure}[ht]
\centering\includegraphics[width=0.8\linewidth]{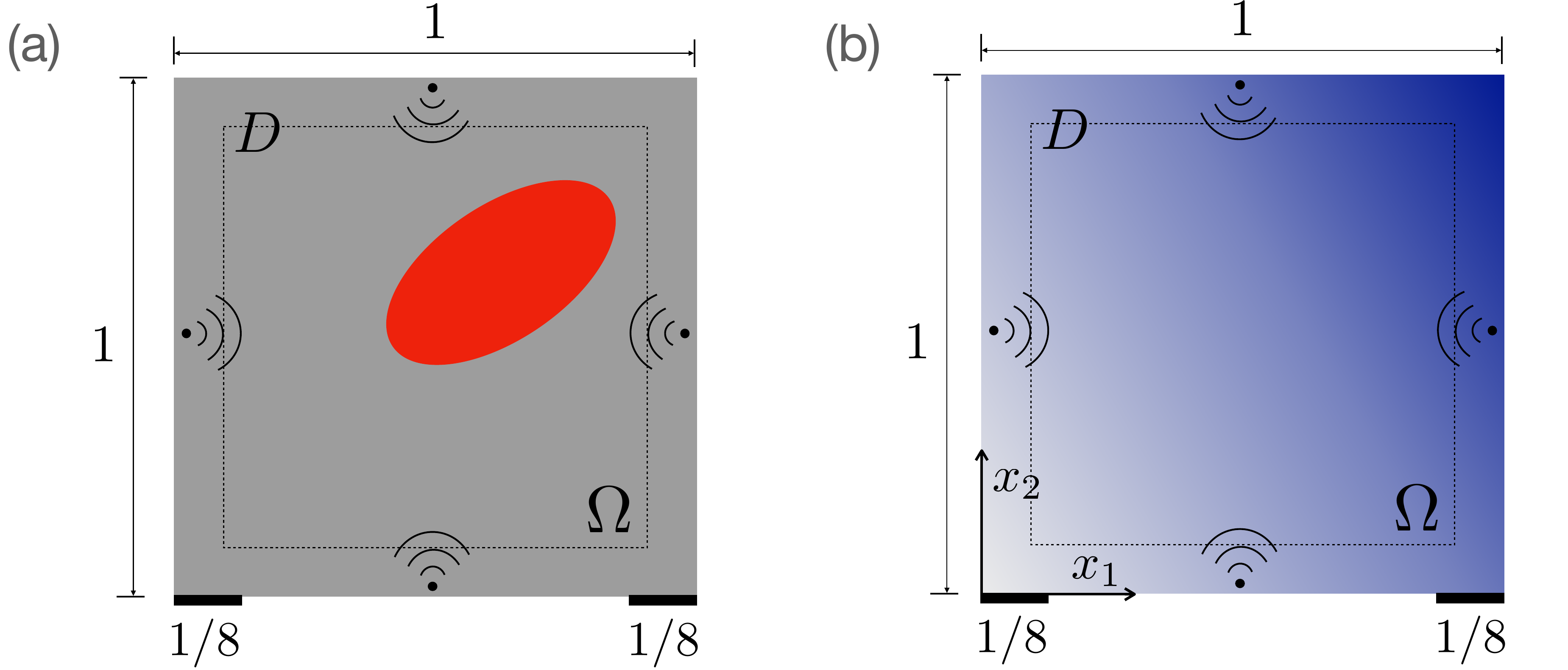} \vspace*{-2mm}
\caption{\small\textup{Test configurations: (a) piecewise-homogeneous standard linear solid, and (b) smoothly graded Jeffreys material.}}\label{test-config}
\end{figure}

\begin{table}
\caption{\small\textup{Standard linear solid example, Fig.~\ref{test-config}(a): piecewise-constant viscoelastic parameters and initial guess applied to each subzone $S_n^{\text{\tiny{(1)}}}$ $(n\!=\!\overline{1,16}$).}} 
\centering
\begin{tabular}{|c|c|c|c|c|c|c|} \hline
{Parameter}     & $\upmu$ & $\upmu'$ & $\upchi$ & $\upkappa$ & $\upkappa'$ & $\upeta$ \\ \hline\hline
{Background}    &   3   &   1    &   0.3  &     8    &    3      &  0.01  \\ \hline
{Inclusion}     &   5   &  2.5   &   0.7  &    10    &    5      &  0.05  \\ \hline
{Uniform initial guess} &   2   &   2    &  0.1 &     2    &    2      &  0.1 \\ \hline 
\end{tabular}
\label{tab1}
\end{table}

The result of viscoelastic reconstruction, obtained via a two-step resolution refinement over each~$S_n$ (with $\mathcal{N}^{\text{\tiny{(2)}}}\!=\!5$ and $\mathcal{N}^{\text{\tiny{(3)}}}\!=\!7$) is shown in the top two rows of Fig.~\ref{recon-sls1}. Using the local colorbar scale, the top left insert for each parameter depicts the true configuration, while the bottom left insert describes the initial guess according to Table~\ref{tab1}. As can be seen from the display, MECR reconstruction of the shear parameters is very good, while that of the bulk parameters is poor. Upon inspection, the primary reason for such drawback was found to reside in the smallness of the ``true'' background bulk viscosity, $\upeta\!=\!0.01$. Specifically, when $\upeta\hh\oo=o(1)$, Taylor series expansion of~$K$ in~\eqref{KG-SLS} de facto yields a \emph{two-parameter} Kelvin-Voigt model  
\begin{equation} \label{KG-SLSa}
K(\bsfp,\oo) = \upkappa_a - \rmi\hh\upeta_a \hh \oo + O((\upeta\oo)^2), \qquad 
\upkappa_a = \frac{\upkappa\hh \upkappa'}{\upkappa+\upkappa'}, \quad \upeta_a = \frac{\upkappa^2 \hh \upeta}{(\upkappa+\upkappa')^2}. 
\end{equation} 
Indeed, for the background material we have $\{\upeta\hh\omega_\ell\} = \{0.16,0.24,0.36,0.54\}$, where two out of four frequencies provide insufficient information to independently resolve~$\upkappa,\upkappa'$ and $\upeta$. To verify this hypothesis, we repeat minimization~\eqref{ECRed} with the MECR functional~$\LambdaRk(\bsfp)$ \emph{re-parameterized} in terms of~$\bsfp_a\!=\!(\upmu,\upmu',\upchi,\upkappa_a,\upeta_a)$ due to~\eqref{KG-SLSa}. The result in terms of~$\upkappa_a$ and~$\upeta_a$ is shown in the bottom row of Fig.~\ref{recon-sls1}, featuring the quality of reconstruction that is commensurate with that of the shear parameters. For all parameters, maximum deviation from the respective ``true'' distributions is localized near the edge of the inclusion. This is to be expected, for the MECR reconstruction inherently assumes locally-constant viscoelastic moduli. 

\begin{figure}[ht]
\centering\includegraphics[width=0.85\linewidth]{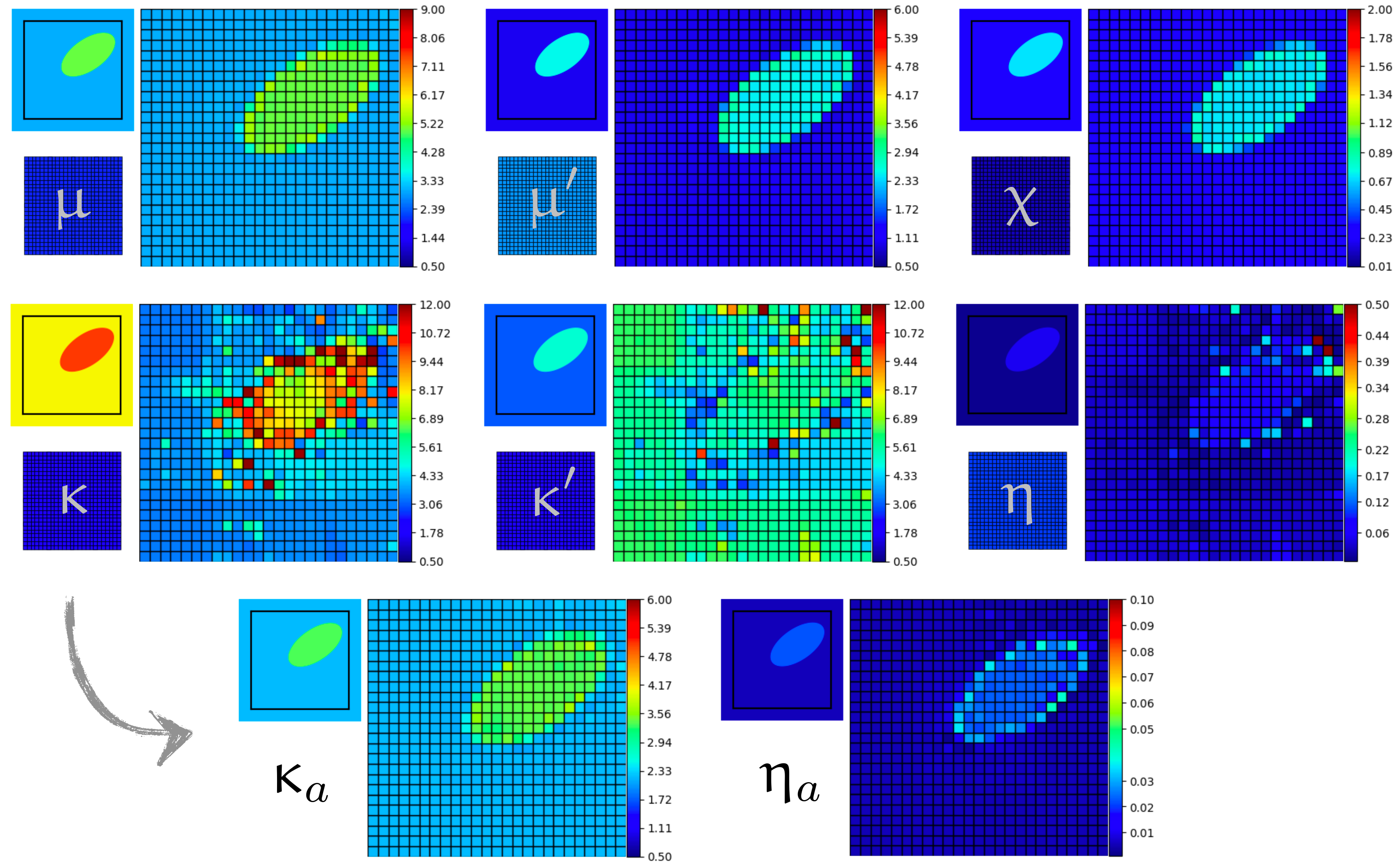} \vspace*{-2mm}
\caption{\small\textup{Reconstruction of the standard linear solid, ``noise-free'' data: shear parameters $\upmu, \upmu'$ and~$\upchi$ (top row), bulk parameters $\upkappa, \upkappa'$ and $\upeta$ (middle row), and resolvable bulk parameters $\upkappa_a$ and $\upeta_a$ (bottom row). The bottom left insert in each panel describes the initial guess (see Table~\ref{tab1}) using the local color scale.}}
\label{recon-sls1}
\end{figure}

\subsubsection{Application to noisy measurements} \label{noise}

In the case of ``noise-free'' observations, the choice of the regularization parameter is practically immaterial (as verified numerically). To deal with noisy data, on the other hand, one must deploy a suitable selection rule for~$\kappa^\star$; see the discussion in Section~\ref{L-curve}. In this work, we adopt the L-curve criterion~\cite{hansen2010discrete}, which carries an advantage of not requiring \emph{a priori} knowledge of the noise level. Recalling~\eqref{min:reduced}, we thus minimize the reduced MECR functional $\LambdaRk\!=\!E(\kappa) +\kappa\hh M(\kappa)$ over a broad range of $\kappa$ values, and examine the resulting $M(\kappa)$ vs.~$E(\kappa)$ relationship (see Fig.~\ref{morozov}) on a log-log scale. In the examples to follow, the interior data $\bfum$ are perturbed by uncorrelated errors, taking the relative noise levels as $\delta\nes\simeq\nes 2\%$ and $\delta\nes\simeq\nes 10\%$ where $\delta=\|\bfum^\delta\!-\bfum\|_{\Omega}/\|\bfum\|_{\Omega}$ and $\|\bdot\|_{\Omega}$ refers to the L2 norm. Specifically, for a simulated displacement component $\uiobs$  ($i\!=\!\overline{1,2}$) at a given observation point, we let $\uiobs^\delta\!=\nes\uiobs(1\shp \delta r_i)$, where $r_i$ is a standard normal random variable and $\delta\in\{0.02,0.1\}$.

One particular feature of the present examples, driven in part by (i) limited subzone size $|S_n|^{\frac{1}{2}}$ and (ii) restriction that the adjoint field $\bfw\!\in\!\mathcal{W}$ must vanish on $\partial S_n$ (see~\eqref{u:spaces}), is that $E(\kappa)\!\ll\!M(\kappa)$ at the minimizer. In principle, such disparity does not pose systemic problems; however, care must be taken to discard any portion of the $L$-curve where $E(\kappa)$ drops below an effective precision of the minimization scheme (the SLSQP algorithm stops~\cite{Scipy2020} when the absolute change in the cost functional between two successive iterations drops below $10^{-10}$). In this ``small~$E$'' range, the $M(\kappa)$ vs.~$E(\kappa)$ relationship is expected to be random noise realization-dependent and thus not helpful in selecting~$\kappa$. To cater for such behavior, we select the regularization parameter as 
\begin{equation} \label{kappacrit}
 \kappa^\star = \max\{\kappa_c,\kappa_s\},   
\end{equation}
where $\kappa_c$ corresponds to a (suitably defined) ``corner" of the $L$-curve, and $\kappa_s$ is the value of~$\kappa$ below which the $L$-curve begins to fluctuate with realizations of the random noise. 

In this vein, panels (a) and (b) in Fig.~\ref{lcurves} plot the $\log M-\log E$ relationship for $\delta\sheq2\%$ and~$\delta\sheq 10\%$, respectively. For computational expediency, the diagrams are evaluated using one-step resolution refinement over each~$S_n$ with $\mathcal{N}^{\text{\tiny{(2)}}}\!=\!5$. In the panels, also included are the reconstruction maps of~$\upeta_a$ (the most sensitive parameter) for $\kappa\in\{0.001,0.01,0.1,1,10,100\}$. As can be seen from the displays, the ``corner" of the $L$-curve $\kappa_c$ appears to be located between $0.01$ and~$0.1$. On the other hand, the shape of the $L$-curve is found to visibly vary with realizations of the random noise for $\kappa\!\in\!\{0.001,0.01,0.1\}$, which then via~\eqref{kappacrit} identifies~$\kappa^\star\!=\!\kappa_s\!=\!1$  from the available grid of trial~$\kappa$ values. This is illustrated in Fig.~\ref{lcurves}(b), which plots the $L$-curve for two realizations ($R_{1/2}$) of the 5\% random noise. The selection rule~\eqref{kappacrit} is verified \emph{a posteriori} via visible deterioration of the reconstructed $\upeta_a$ maps for $\kappa\!\in\!\{0.001,0.01,0.1\}$ at both noise levels. With such result in place, Fig.~\ref{recon-sls2} compares the reconstruction maps of~$\bsfp_a$ for the ``noise-free'' data, $\delta\!\simeq\!0.01$, and $\delta\!\simeq\!0.05$ (obtained via two-step resolution refinement with $\mathcal{N}^{\text{\tiny{(2)}}}\!=\!5$ and $\mathcal{N}^{\text{\tiny{(3)}}}\!=\!7$). From the displays, it is seen that the five-parameter viscoelastic reconstruction is reasonably robust in the presence of measurement errors, with the exception of~$\upeta_a$ at 5\% noise. For a better insight into the quality of viscoelastic reconstruction, Table~\ref{sls-moderr} lists the relative (L2 norm) error values 
\begin{equation} \label{reconerr}
\Delta \sfp_j = \frac{\|\sfp_j^{\text{recon}} - \sfp_j^{\text{exact}}\|_\Omega}{\|\sfp_j^{\text{exact}}\|_\Omega}, \qquad j=\overline{1,\dim\bsfp} 
\end{equation}
for $\bsfp=\bsfp_a$ obtained at all three noise levels. 

In contrast to the over-parameterization issue seen in Fig.~\ref{recon-sls1}, the lack of resolution in the bottom-right panel of Fig.~\ref{recon-sls2} stems from the relative smallness of~$\upeta_a$ and does not significantly affect reconstruction of the remaining parameters. For the present problem, $\upeta_a/(\tfrac{1}{5}\|\bsfp_a\|)$ equals approximately 0.3\% and 0.9\% in the background and the inclusion, respectively. We also note that the map of $\upeta_a$ for $\kappa\sheq 1$ and $\delta\sheq 0.05$ in Fig.~\ref{lcurves}(b) fares better for it is obtained in a reduced parametric space ($5\!\times\!5$ vs. $7\!\times\!7$ pixels per subzone). 

\begin{figure}
\centering\includegraphics[width=1.0\textwidth]{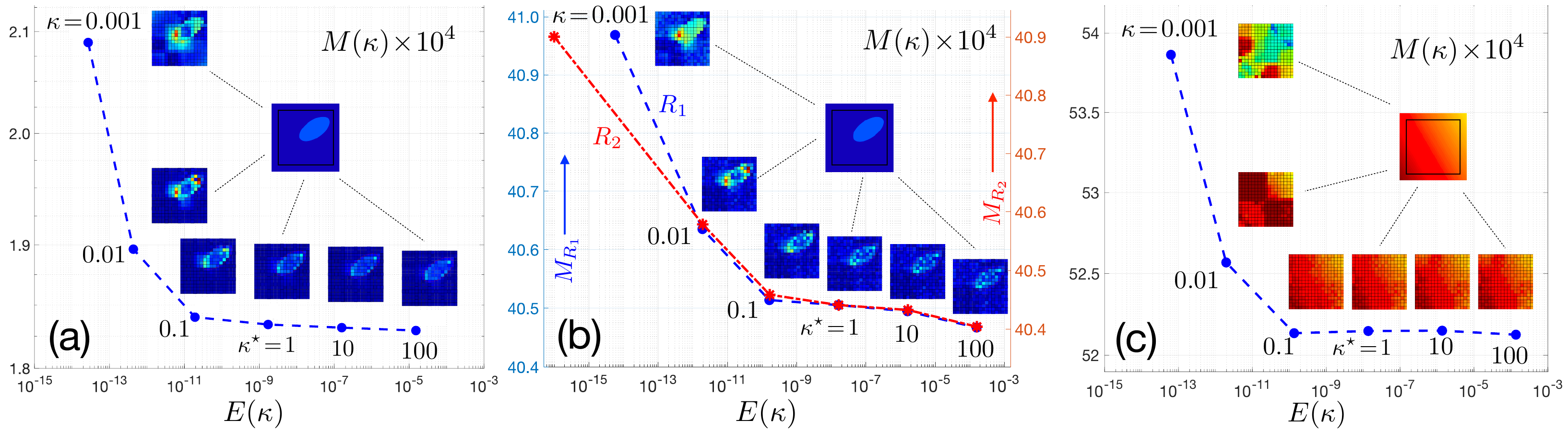} 
\caption{\small\textup{$\log M-\log E$ relationship computed for (a) standard linear solid, 2\% noise, (b) standard linear solid, 10\% noise, and (c) Jeffreys material, 2\% noise. The reconstruction inserts in panels (a)  and (b) refer to parameter $\upeta_a$; those in panel (c) refers to $\upmu$.}} \label{lcurves}
\end{figure}

\begin{figure}
\centering\includegraphics[width=0.84\textwidth]{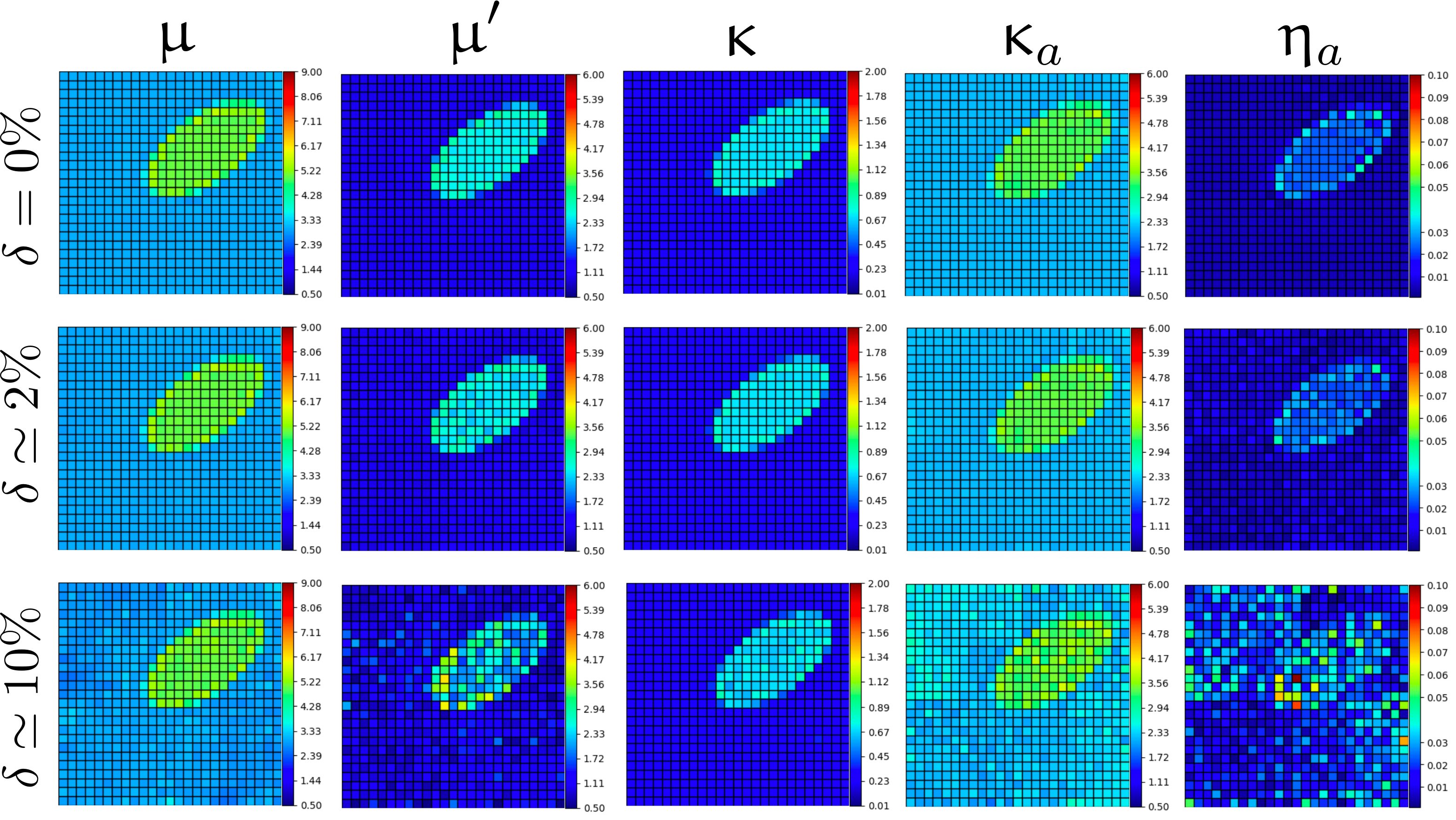}
\caption{\small\textup{Reconstructions of the standard linear solid for three levels of random noise.}} \label{recon-sls2}
\end{figure}

\begin{table}[h] 
\caption{\small\textup{Relative error~\eqref{reconerr} of viscoelastic moduli reconstruction for the standard linear solid.}} 
\centering
\begin{tabular}{|c|c|c|c|c|c|c|} \hline
$\Delta\sfp_j$ & $\upmu$ & $\upmu'$ & $\upchi$ & $\upkappa_a$ & $\upeta_a$   \\ \hline\hline
{noise-free}    & 0.039   & 0.081   & 0.076  & 0.034    & 0.278       \\ \hline
{2\% noise}     & 0.040   & 0.095   & 0.077  & 0.035    & 0.365       \\ \hline
{10\% noise}     & 0.055   & 0.272   & 0.097  & 0.084    & 1.756       \\ \hline
\end{tabular}
\label{sls-moderr}
\end{table}

\subsection{Smoothly graded Jeffreys material} \label{num-jm}

We next pursue MECR reconstruction of the Jeffreys material shown in Fig.~\ref{test-config}(b) whose complex bulk and shear modulus are linearly graded according to 
\begin{equation} \label{KG-JM}
K(\bfx;\bsfp,\oo) = \frac{\omega\hh\upeta(\bfx)(\upkappa(\bfx)\!-\rmi\omega\upeta'(\bfx))}{\omega(\upeta(\bfx)\!+\upeta'(\bfx)) + \rmi\upkappa(\bfx)}, \qquad 
G(\bfx;\bsfp,\oo) = \frac{\omega\hh\upchi(\bfx)(\upmu(\bfx)\!-\rmi\omega\upchi'(\bfx))}{\omega(\upchi(\bfx)\!+\upchi'(\bfx)) + \rmi \upmu(\bfx)},
\end{equation}
where
\[
\sfp_j(\bfx) = \sfp_j\blsup + (\sqrt{3}\!-\!1)(\sfp_j\trsup\!-\!\sfp_j\blsup)\hh \bfd\sip\bfx , \quad \bfd = (\cos\tfrac{\pi}{6},\sin\tfrac{\pi}{6}), \qquad \bsfp=(\upmu, \upchi, \upchi',\upkappa, \upeta, \upeta') 
\]
with $\sfp_j\blsup\!=\!\sfp_j((0,0))$ and $\sfp_j\trsup\!=\!\sfp_j((1,1))$. The values of~$\sfp_j\blsup$ and~$\sfp_j\trsup$ for each of the six viscoelastic parameters are listed in Table~\ref{tab2}. 

\begin{table}[h]
\caption{\small\textup{Jeffreys material example, Fig.~\ref{test-config}(b): linearly-graded viscoelastic parameters and initial guesses applied to each subzone $S_n^{\text{\tiny{(1)}}}$ $(n\!=\!\overline{1,16}$).}} 
\centering
\begin{tabular}{|c|c|c|c|c|c|c|} \hline
{Parameter}     & $\upmu$ & $\upchi$ & $\upchi'$ & $\upkappa$ & $\upeta$ & $\upeta'$ \\ \hline\hline
{Bottom left}: $\sfp_j\blsup$    &   5   &   0.5    &   0.08  &     7    &    0.7      &  0.05  \\ \hline
{Top right}: $\sfp_j\trsup$     &   4   &  1   &   0.25  &    5    &    1.2      &  0.2  \\ \hline
{Uniform initial guess} &   2   &   1    &  0.1 &     2    &    1      &  0.1 \\ \hline 
\end{tabular}
\label{tab2}
\end{table}

The testing and inversion parameters, including the choice of excitation sources, reconstruction subzones, and pixel density are borrowed from the previous example (Section~\ref{num-sls}) except for the excitation frequencies which are taken as $\{ \omega_\ell \} = \{ 4, 8, 16, 32 \}$. Fig.~\ref{recon-jeff1} shows the reconstruction  with $\kappa^\star\sheq 10$ of the Jeffreys viscoelastic parameters from ``noise-free'' data. As can be seen from the display, all six parameters of the smoothly-graded material are reasonably well reconstructed.    

To examine the effect of measurement noise, the interior data $\bfum$ are perturbed by uncorrelated errors, taking the relative noise level as $\delta\nes\simeq\nes 2\%$. With reference to the L-curve shown in Fig.~\ref{lcurves}(c), the regularization parameter is taken as $\kappa^\star\sheq\kappa_s\sheq1$, which corresponds to a threshold range $E(\kappa_c)\sheq O(10^{-9})$ below which the L-curve fluctuates with realizations of random noise due to finite precision of the SLSQP algorithm. This ``stability'' threshold is reflected by the fact that $E(0.1)\!<\!E(1)$ in Fig.~\ref{lcurves}(c), which is inadmissible according to Proposition~\ref{Lcurve:prop}. The viscoelastic reconstructions for~$\delta\nes\simeq\nes 0\%$ and~$\delta\nes\simeq\nes 2\%$ are compared in the top two rows of Fig.~\ref{recon-jeff2}, from which one observes a marked sensitivity (in terms of~$\upeta$ and~$\upeta'$) of the MECR inversion to measurement errors. Again, this sensitivity is driven in part by the relative smallness of~$\upeta'$, see Table~\ref{tab2}. To combat the problems caused by trying to resolve ``weak'' viscoelastic parameters (if any), one may forgo the re-parameterization approach exercised in Section~\ref{num-sls} and resort instead to a coarsened resolution of MECR reconstruction, which carries an advantage of reducing the dimension of the parameter space and so constraining the optimal solution. This is illustrated in the bottom row of Fig.~\ref{recon-jeff2}, which shows the reconstruction at 2\% noise with $\mathcal{N}\sheq5$ ($5\!\times\!5$ pixels per subzone) instead of~$\mathcal{N}\sheq7$ deployed for generating the top two rows in the same figure. From the reconstructed maps, it is seen that the coarsened inversion is more robust to noisy data, and may in fact be preferred when characterizing specimens where a smooth gradation of constitutive properties is expected. For completeness, Table~\ref{jeff-moderr} lists the relative error values $(\Delta \sfp_j)$ at both noise and resolution levels, which further confirms this observation. From Table~\ref{sls-moderr} and Table~\ref{jeff-moderr}, we also note that the values of $\Delta \sfp_j$ featured by the two examples are \emph{comparable} despite a qualitative difference in the respective constitutive behaviors and types of heterogeneity.\enlargethispage*{5ex}

\begin{figure}[ht]
\centering\includegraphics[width=0.84\linewidth]{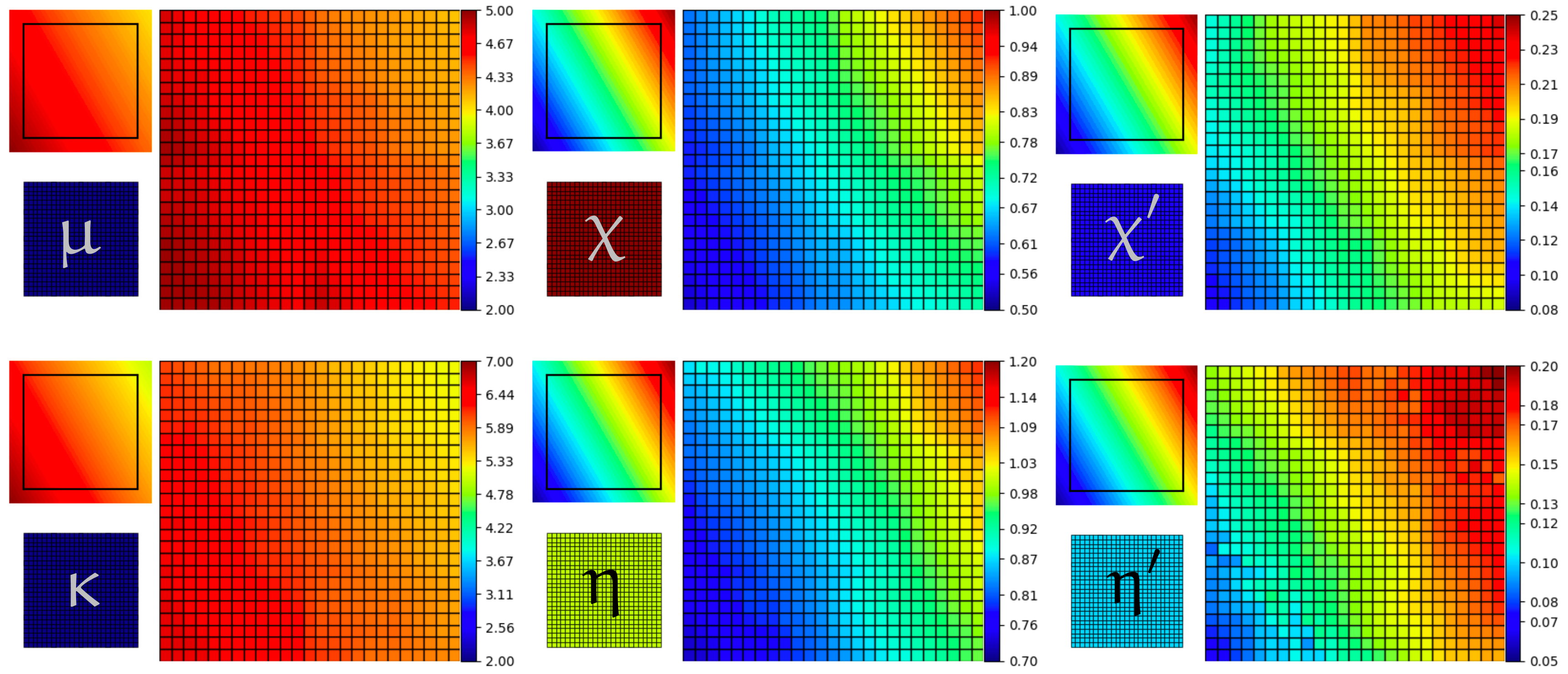} \vspace*{-2mm}
\caption{\small\textup{Reconstruction of the Jeffreys material, ``noise-free'' data: shear parameters $\upmu, \upchi$ and~$\upchi'$ (top row) and bulk parameters $\upkappa, \upeta$ and $\upeta'$ (bottom row). The bottom left insert in each panel describes the initial guess (see Table~\ref{tab2}) using the local color scale.}}
\label{recon-jeff1}
\end{figure}

\begin{figure}
\centering\includegraphics[width=1.0\textwidth]{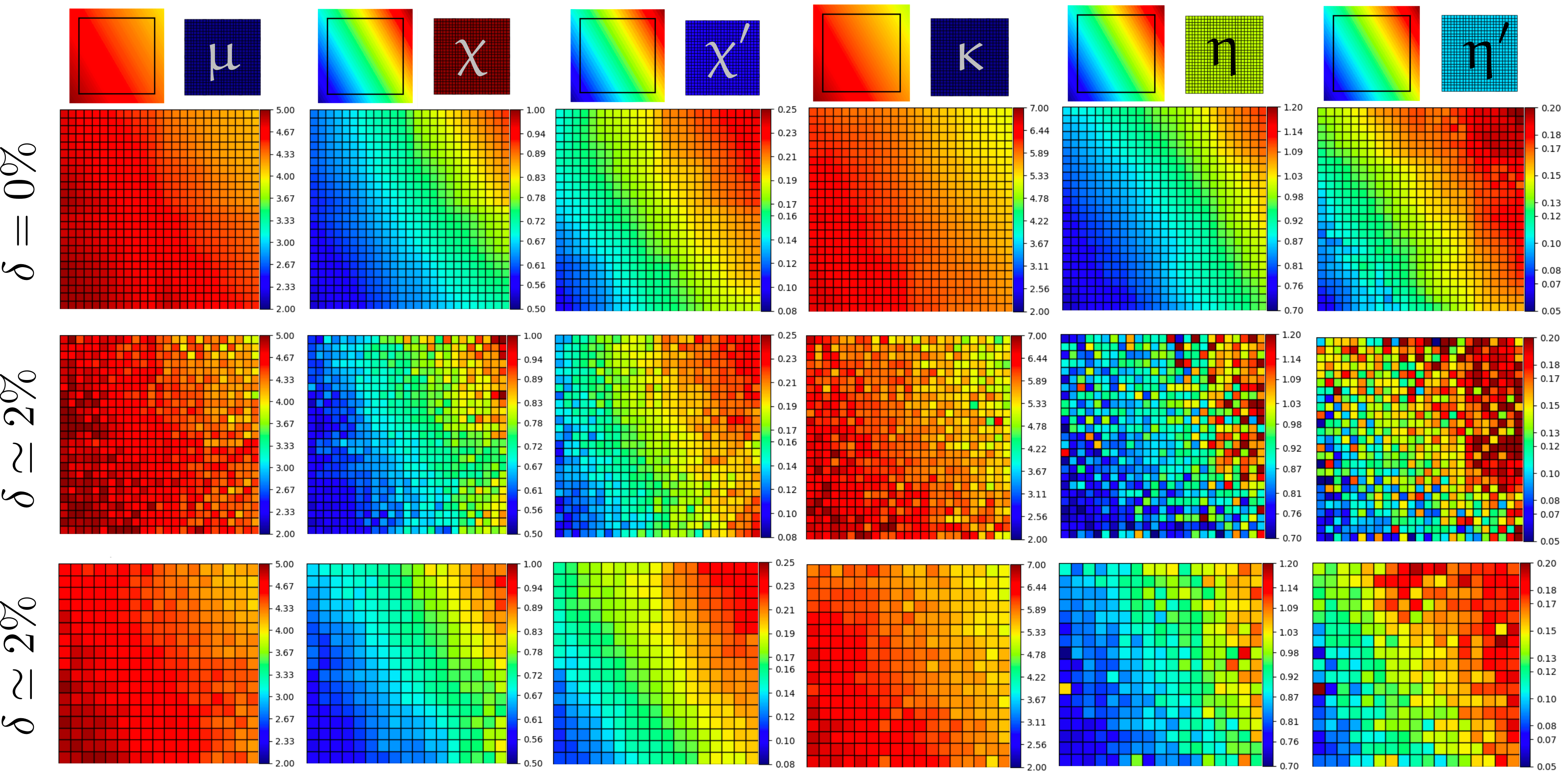}
\caption{\small\textup{Reconstructions of the Jeffreys material: ``noise-free'' data, $\mathcal{N}\sheq7$ (top row); 2\% noise,  $\mathcal{N}\sheq7$ (middle row), and 2\% noise,  $\mathcal{N}\sheq5$ (bottom row). The initial guesses of Table~\ref{tab2} are shown in inserts using the local color scales.}} \label{recon-jeff2}
\end{figure}

\begin{table}[h] 
\caption{\small\textup{Relative error~\eqref{reconerr} of viscoelastic moduli reconstruction for the Jeffreys material.}} 
\centering
\begin{tabular}{|c|c|c|c|c|c|c|c|} \hline
$\Delta\sfp_j$\ & $\upmu$ & $\upchi$ & $\upchi'$ & $\upkappa$ & $\upeta$ & $\upeta'$  \\ \hline\hline
{noise-free}, $\mathcal{N}\sheq7$  &   0.033   &  0.101   &   0.099  &    0.031    &    0.079      &   0.16  \\ \hline
{2\% noise},  $\mathcal{N}\sheq7$  &   0.048   &  0.112   &   0.122  &    0.053    &    0.111      &   0.244  \\ \hline
{2\% noise},  $\mathcal{N}\sheq5$  &   0.031   &  0.099   &   0.102  &    0.038    &    0.091      &   0.18  \\ \hline
\end{tabular}
\label{jeff-moderr}
\end{table}

\section{Summary and outlook}

\noindent In this study, we develop a modified error-in-constitutive-relation (MECR) approach to the full-field characterization of linear viscoelastic solids, formulated by way of thermodynamic (free energy and dissipation) potentials. Assuming the availability of full-field interior kinematic data, the constitutive mismatch between the kinematic quantities (strains and internal variables) and their “flux” companions (Cauchy stress tensor and that of thermodynamic tensions), is established with the aid of the Legendre-Fenchel transform linking the thermodynamic potentials to their energetic conjugates. In this setting, the MECR functional is formulated as a linear combination between the constitutive gap and the kinematic data misfit, evaluated for a trial set of constitutive parameters. The affiliated stationarity conditions are shown to yield two coupled evolution problems, namely (i) the forward evolution problem for the (trial) displacement field driven by the constitutive mismatch, and (ii) the backward evolution problem for the adjoint field driven by the data mismatch. The formulation is established in a general setting, catering for both time- and frequency-domain sensory data. We illustrate the developments by pursuing time-harmonic MECR reconstruction of (a) piecewise-homogeneous standard linear solid, and (b) smoothly-varying Jeffreys material. 

One particular application motivating our developments is that of mineral CO$_2$ storage~\cite{snaebjornsdottir2020carbon}, where the evolution of mechanical rock properties due to reactive flow is of major interest. In particular, the algorithm developed herein targets high-fidelity viscoelastic characterization of mafic and ultramafic rock specimens undergoing carbonation, excited ultrasonically under the plane stress condition --  which caters for optical monitoring of the interior kinematic data via laser Doppler vibrometry~\cite{pourahmadian2018elastic}. Beyond their immediate application, the developments in this study may facilitate small-strain mechanical characterization of additively manufactured materials, whose as-made (homogeneous or heterogeneous) properties are largely an open question. 

\paragraph{Acknowledgment} This work was supported as part of the \emph{Center on Geo-processes in Mineral Carbon Storage}, an Energy Frontier Research Center funded by the U.S. Department of Energy, Office of Science, Basic Energy Sciences at the University of Minnesota under award \# DE-SC0023429. Thanks are extended to MTS Systems Corporation for providing an opportunity for M. Bonnet to visit the University of Minnesota as an MTS Visiting Professor of Geomechanics. The support of the Minnesota Supercomputing Institute during the course of this investigation is kindly acknowledged.\enlargethispage*{5ex}

\bibliography{marcbibs}%

\begin{thebibliography}{10}

\bibitem{allix:05}
Allix O., Feissel P., Nguyen H.
\newblock Identification strategy in the presence of corrupted measurements.
\newblock \emph{Eng. Computations}, \textbf{22}:487--504 (2005).

\bibitem{B-2016-07}
Aquino W., Bonnet M.
\newblock Analysis of the error in constitutive equation method for elasticity
  imaging.
\newblock \emph{SIAM J. Appl. Math.}, \textbf{79}:822--849 (2019).

\bibitem{B-2012-2}
Banerjee B., Walsh T., Aquino W., Bonnet M.
\newblock Large scale parameter estimation problems in frequency-domain
  elastodynamics using an error in constitutive equation functional.
\newblock \emph{Comp. Meth. Appl. Mech. Eng.}, \textbf{253}:60--72 (2013).

\bibitem{charara2005full}
Charara M., Barnes C., Tarantola A.
\newblock Full waveform inversion of seismic data for a viscoelastic medium.
\newblock In P.~Hansen, B.~Jacobsen, K.~Mosegaard (editors), \emph{Methods and
  Applications of Inversion}, vol.~92 of \emph{Lecture Notes in Earth
  Sciences}, pages 68--81. Springer (2005).

\bibitem{chen2018estimating}
Chen H., Innanen K., Chen T.
\newblock Estimating {P- and S-wave} inverse quality factors from observed
  seismic data using an attenuative elastic impedance.
\newblock \emph{Geophysics}, \textbf{83}:R173--R187 (2018).

\bibitem{diaz2015modified}
Diaz M., Aquino W., Bonnet M.
\newblock A modified error in constitutive equation approach for
  frequency-domain viscoelasticity imaging using interior data.
\newblock \emph{Comp. Meth. Appl. Mech. Eng.}, \textbf{296}:129--149 (2015).

\bibitem{epanomeritakis:08}
Epanomeritakis I., Akcelik V., Ghattas O., Bielak J.
\newblock A {Newton-CG} method for large-scale three-dimensional elastic
  full-waveform seismic inversion.
\newblock \emph{Inverse Problems}, \textbf{24}:034015 (2008).

\bibitem{feissel:allix:06}
Feissel P., Allix O.
\newblock Modified constitutive relation error identification strategy for
  transient dynamics with corrupted data: the elastic case.
\newblock \emph{Comp. Meth. Appl. Mech. Eng.}, \textbf{196}:1968--1983 (2007).

\bibitem{findley2013creep}
Findley W., Lai J., Davis F.
\newblock \emph{Creep and Relaxation of Nonlinear Viscoelastic Materials}.
\newblock Dover (2013).

\bibitem{ger:nqs:suq:83}
Germain P., Nguyen Q., Suquet P.
\newblock Continuum thermodynamics.
\newblock \emph{ASME J. Appl. Mech.}, \textbf{50}:1010--1020 (1983).

\bibitem{greenleaf2003selected}
Greenleaf J., Fatemi M., Insana M.
\newblock Selected methods for imaging elastic properties of biological
  tissues.
\newblock \emph{Ann. Rev. Biomed. Eng.}, \textbf{5}:57--78 (2003).

\bibitem{guzina2005spectral}
Guzina B., Madyarov A.
\newblock On the spectral analysis of {L}ove waves.
\newblock \emph{Bull. Seism. Soc. Amer.}, \textbf{95}:1150--1169 (2005).

\bibitem{Halp75}
Halphen B., Nguyen Q.
\newblock Sur les mat\'eriaux standards g\'en\'eralis\'es.
\newblock \emph{J. M\'ecanique}, \textbf{14}:39--63 (1975).

\bibitem{hansen2010discrete}
Hansen P.
\newblock \emph{Discrete Inverse Problems: Insight and Algorithms}.
\newblock SIAM (2010).

\bibitem{ladeveze:83}
Ladev\`eze P., Leguillon D.
\newblock Error estimate procedure in the finite element method and
  applications.
\newblock \emph{SIAM J. Numer. Anal.}, \textbf{20}:485--509 (1983).

\bibitem{lad:moe:97}
Ladevèze P., Moës N.
\newblock A new a posteriori error estimation for nonlinear time-dependent
  finite element analysis.
\newblock \emph{Comp. Meth. Appl. Mech. Eng.}, \textbf{157}:45--68 (1997).

\bibitem{lad:ned:rey:94}
Ladevèze P., Nedjar D., Reynier M.
\newblock Updating of finite element models using vibration tests.
\newblock \emph{AIAA J.}, \textbf{32}:1485--1491 (1994).

\bibitem{marchand:18}
Marchand B., Chamoin L., Rey C.
\newblock Parameter identification and model updating in the context of
  nonlinear mechanical behaviors using a unified formulation of the modified
  Constitutive Relation Error concept.
\newblock \emph{Comp. Meth. Appl. Mech. Eng.}, \textbf{345}:1094--1113 (2019).

\bibitem{mariappan2010magnetic}
Mariappan Y., Glaser K., Ehman R.
\newblock Magnetic resonance elastography: a review.
\newblock \emph{Clinical Anatomy}, \textbf{23}:497--511 (2010).

\bibitem{maugin:92}
Maugin G.
\newblock \emph{Thermomechanics of Plasticity and Fracture}.
\newblock Cambridge University Press (1992).

\bibitem{mcgarry2022}
McGarry M., Van~Houten E., Sowinski D., et~al.
\newblock Mapping heterogenous anisotropic tissue mechanical properties with
  transverse isotropic nonlinear inversion {MR} elastography.
\newblock \emph{Med. Image Analy.}, \textbf{78}:102432 (2022).

\bibitem{Morozov1984}
Morozov V.
\newblock \emph{Methods for Solving Incorrectly Posed Problems}.
\newblock Springer Verlag (1984).

\bibitem{son:book}
Nguyen Q.
\newblock \emph{Stability and Nonlinear Solid Mechanics}.
\newblock Wiley (2000).

\bibitem{oliphant2001complex}
Oliphant T., Manduca A., Ehman R., Greenleaf J.
\newblock Complex-valued stiffness reconstruction for magnetic resonance
  elastography by algebraic inversion of the differential equation.
\newblock \emph{Magn. Reson. Med.}, \textbf{45}:299--310 (2001).

\bibitem{parker2005unified}
Parker K., Taylor L., Gracewski S., Rubens D.
\newblock A unified view of imaging the elastic properties of tissue.
\newblock \emph{J. Acoust. Soc. Amer.}, \textbf{117}:2705--2712 (2005).

\bibitem{pourahmadian2018elastic}
Pourahmadian F., Guzina B.
\newblock On the elastic anatomy of heterogeneous fractures in rock.
\newblock \emph{Int. J. Rock Mech. Min. Sci.}, \textbf{106}:259--268 (2018).

\bibitem{NGSolve}
Sch{\"o}berl J.
\newblock C++11 {I}mplementation of {F}inite {E}lements in {NGSolve}.
\newblock \emph{{Institute for analysis and scientific computing, Vienna
  University of Technology}}, \textbf{30} (2014).

\bibitem{sigrist2017ultrasound}
Sigrist R., Liau J., El~Kaffas A., Chammas M., Willmann J.
\newblock Ultrasound elastography: review of techniques and clinical
  applications.
\newblock \emph{Theranostics}, \textbf{7}:1303 (2017).

\bibitem{simo}
Simo J., Hughes T.
\newblock \emph{Computational Inelasticity}.
\newblock Springer-Verlag (1998).

\bibitem{sinkus2005imaging}
Sinkus R., Tanter M., Catheline S., Lorenzen J., Kuhl C., Sondermann E., Fink
  M.
\newblock Imaging anisotropic and viscous properties of breast tissue by
  magnetic resonance-elastography.
\newblock \emph{Magn. Reson. Med.}, \textbf{53}:372--387 (2005).

\bibitem{snaebjornsdottir2020carbon}
Sn{\ae}bj{\"o}rnsd{\'o}ttir S., Sigf{\'u}sson B., Marieni C., Goldberg D.,
  Gislason S., Oelkers E.
\newblock Carbon dioxide storage through mineral carbonation.
\newblock \emph{Nature Rev. Earth \& Envir.}, \textbf{1}:90--102 (2020).

\bibitem{tan2016gradient}
Tan L., McGarry M., Van~Houten E., Ji M., Solamen L., Weaver J., Paulsen K.
\newblock Gradient-based optimization for poroelastic and viscoelastic {MR}
  elastography.
\newblock \emph{IEEE Trans. Med. Imag.}, \textbf{36}:236--250 (2016).

\bibitem{tokmashev2013experimental}
Tokmashev R., Tixier A., Guzina B.
\newblock Experimental validation of the topological sensitivity approach to
  elastic-wave imaging.
\newblock \emph{Inverse Problems}, \textbf{29}:125005 (2013).

\bibitem{Scipy2020}
Virtanen P., Gommers R., Oliphant T., et~al.
\newblock {SciPy} 10: fundamental algorithms for scientific computing in
  {P}ython.
\newblock \emph{Nature Methods}, \textbf{17}:261--272 (2020).

\bibitem{xia2014estimation}
Xia J.
\newblock Estimation of near-surface shear-wave velocities and quality factors
  using multichannel analysis of surface-wave methods.
\newblock \emph{J. Appl. Geophys.}, \textbf{103}:140--151 (2014).

\bibitem{yuan2010reconstruction}
Yuan H., Guzina B.
\newblock Reconstruction of viscoelastic tissue properties from {MR}
  elastography-type measurements.
\newblock \emph{Compt. Rend. Mecanique}, \textbf{338}:480--488 (2010).

\bibitem{zhang2012solution}
Zhang Y., Oberai A., Barbone P., Harari I.
\newblock Solution of the time-harmonic viscoelastic inverse problem with
  interior data in two dimensions.
\newblock \emph{Int. J. Num. Meth. Eng.}, \textbf{92}:1100--1116 (2012).

\bibitem{zong2015complex}
Zong Z., Yin X., Wu G.
\newblock Complex seismic amplitude inversion for {P-wave} and {S-wave} quality
  factors.
\newblock \emph{Geophys. J. Int.}, \textbf{202}:564--577 (2015).

\end{thebibliography}

\newpage 

\appendix

\setcounter{equation}{0}
\def\theequation{A.\arabic{equation}}

\section{Proofs}

\subsection{Proof of Lemma~\ref{reciprocity}}
\label{reciprocity:proof}

Expressing $\bfsig[\bfu]$ via~\eqref{sig:expr} and integrating by parts in time the term featuring $\dot{\bfu}$ and using the shorthand notations $\bfeps=\bfeps[\bfu]$, $\bfeta=\bfeps[\bfw]$, we have
\begin{equation}
  \int_0^T \bfeta\dip\bfsig[\bfu] \dt
  = \int_0^T \Lcb \lpar \bfeta\dip\CS\Isub - \dot{\bfeta}\dip\DS\Isub \rpar\dip\bfeps
    - \bfeta\dip\lpar \CHT\!\dip\FS[\CH\dip\bfeps(\bfu)] \rpar \Rcb \dt
    + \lpar\bfeta\dip\DS\Isub\dip\bfeps\rpar \rabs^T_0, \label{aux01}
\end{equation}
where $\CH$ is defined in~\eqref{alpha(t)}. Then, recalling the definition~\eqref{KS:def} of the convolution operator $\FS$, we obtain
\begin{align}
  \int_0^T \bfeta(t)\dip\CHT\!\dip\FS[\CH\dip\bfeps(\bfu)](t) \dt
 &= \int_0^T \bfeta(t)\dip\CHT\!\dip \Lcb \int_{0}^{t} \exp[-\QS(t\shm\tau)]\dip\Da^{-1}\dip
    \CH\dip\bfeps(\tau) \dtau \Rcb \dt \\
 &\Eq{(a)} \int_0^T \bfeps(\tau)\dip\CHT\!\dip \Lcb \int_{\tau}^{T} \exp[-\QS(t\shm\tau)]\dip\Da^{-1}\dip
    \CH\dip\bfeta(t) \dt \Rcb \dtau \\
 &\Eq{(b)} \int_0^T \bfeps(\tau)\dip\CHT\!\dip \Lcb \int_{0}^{T-\tau} \exp[-\QS(T\shm\tau-\theta)]\dip\Da^{-1}\dip
    \CH\dip\bfeps[\bfwR](\theta) \dth \Rcb \dtau \\
 &= \int_0^T \bfeps(t)\dip\CHT\!\dip\FS[\CH\dip\bfeps(\bfwR)](T\shm t) \dt,
\end{align}
where (a) results from transposition (using the fact that $\exp[-\QS t]\dip\Da^{-1}$ is symmetric
and reversing the order of the successive time integrations, and (b) stems from the change of variable $t\mapsto T\shm\theta$. Used in~\eqref{aux01}, the above identity then produces
\begin{align}
  \int_0^T \bfeta\dip\bfsig[\bfu] \dt
  &= \int_0^T \bfeps(t)\dip\Lcb \CS\Isub\dip\bfeps[\bfwR] - \DS\Isub\dip\bfeps[\bfwRd]
  - \CHT\!\dip\FS[\CH\dip\bfeps(\bfwR)] \Rcb(T\shm t) \dt +  \lpar \bfeta\dip\DS\Isub\dip\bfeps \rpar \rabs^T_0 \\
  &= \int_0^T \bfeps(t)\dip\bfsig[\bfwR](T\shm t) \dt +  \lpar \bfeta\dip\DS\Isub\dip\bfeps \rpar \rabs^T_0,
\end{align}
since the tensors $\CS\Isub$ and $\DS\Isub$ are symmetric. The claim of the lemma follows.

\subsection{Proof of formulas~\eqref{ssA}} \label{ssA:proof}

We start by solving~(\ref{group1:pointwise}a) for $\bfsige$ and~(\ref{group1:pointwise}b) for $\bfsigv$, which yields
\begin{equation}
\begin{aligned}
  \bfsige
 &= \Cs\nes^{-1}\dip\lpar \bfeps+\bfeta-\Cms\dip\bfA \rpar &\qquad
  \bfsigv
 &= \Ds\nes^{-1}\dip\lpar \bfepsd + \Tinv \bfeta + \Dms\dip\bfA \rpar \\
 &= \CeS\dip(\bfeps\shp\bfeta) - \CmT\dip\Ca^{-1}\dip\bfA,
 &&= \DeS\dip(\bfepsd \shp \Tinv \bfeta) + \DmT\dip\Da^{-1}\dip\bfA,
\end{aligned}\label{aux16}
\end{equation}
with the latter expressions resulting from the identities $\Cs\nes^{-1}\dip\Cms=\CmT\dip\Ca^{-1}$ (due to~\eqref{C:coeff:conj}) and $\Ds\nes^{-1}\dip\Dms=\DmT\dip\Da^{-1}$ (thanks to~\eqref{D:coeff:conj}). The above expressions are then substituted into~(\ref{group1:pointwise}c). Taking advantage of the identities $\CA\shm\CmsT\dip\CmT\dip\Ca^{-1}=\Ca^{-1}$ (due to~\eqref{schur:C}) and $\DA\shm\DmsT\dip\DmT\dip\Da^{-1}=\Da^{-1}$ (obtained similarly), we find
\begin{equation}
  \lpar \Ca^{-1}\shp T\Da^{-1} \rpar\dip\bfA
 = \DmsT\dip\DeS\dip(\bfeta\shp T\bfepsd) - \CmsT\dip\CeS\dip(\bfeps\shp\bfeta) + T\bfald - \bfal. \label{aux15}
\end{equation}
To solve~\eqref{aux15} for $\bfA$, we note that $\Ca^{-1}\shp T\Da^{-1} = \Ca^{-1}\dip(\Da\shp T\Ca)\dip\Da^{-1}$; moreover we also have $\CmsT\dip\CeS=\Ca^{-1}\dip\Cm$ and $\DmsT\dip\DeS=\Da^{-1}\dip\Dm$ due to~\eqref{C:coeff:conj}, \eqref{D:coeff:conj} and the fact hat~$\CeS$ and~$\DeS$ are symmetric. Exploiting the foregoing identities in~\eqref{aux15}, we obtain
\begin{align}
  \bfA
 &= \Da\dip(\Da\shp T\Ca)^{-1}\dip\lcb T\Ca\dip\Da^{-1}\dip\Dm\dip\bfepsd - \CH\dip\bfeta -\Cm\dip\bfeps
    - \Ca\dip\bfal + T\Ca\dip\bfald \rcb \\
 &= \Da\dip(\Da\shp T\Ca)^{-1}\dip\lcb T\Ca\dip\Da^{-1}\dip\Dm\dip\bfepsd + T\Ca\dip\bfald - (\Da\shp T\Ca)\dip\bfbe + \Da\dip\bfald + \Dm\dip\bfepsd \rcb \\
 &= \Da\dip\bfald + \Dm\dip\bfepsd - \Da\dip\bfbe,
\end{align}
with $\bfbe$ given by~\eqref{R:beta:def}, which completes the proof of~(\ref{ssA}c). The above result is then used in~\eqref{aux16}, which produces~(\ref{ssA}a,b) after some rearrangement.

\subsection{Proof of Proposition~\ref{MECR:eval}}\label{MECR:eval:proof}

As all potentials considered in this study are quadratic (i.e. homogeneous of degree 2) functions, application of the Euler's theorem for homogeneous functions yields
\begin{equation}
  2\psi(\bfeps,\bfal) = \del{\eps}\psi\dip\bfeps + \del{\alpha}\psi\dip\bfal \label{aux11}
\end{equation}
and similarly for $\psi^{\star}$, $\varphi$ and $\varphi^{\star}$. Adding equations~(\ref{group1:exp}a) with $\bfsigHe\sheq\bfsige$, (\ref{group1:exp}b) with $\bfsigHv\sheq\bfsigv$, (\ref{group1:exp}c) with $\bfAH\sheq\bfA$, (\ref{group1:exp}d) with $\bfalH\sheq\bfal$, \eqref{aux09} with $\bfwH\sheq\bfw$ and \eqref{aux10} with $\bfuH\sheq\bfu$ and using property~\eqref{aux11} for all potentials, we find
\begin{eqnarray}
  0 &=& 2\EcalR(\bsfp) + \kappa\exs \McalT(\bfup\shm\bfum,\bfup)  \notag \\
  &=& 2\big\{\EcalR(\bsfp) + \tdemi\kappa\exs\McalRT(\bsfp) + \tdemi\kappa\exs\McalT(\bfup\shm\bfum,\bfum)\big\},
\end{eqnarray}
which yields the claimed expression for $\LambdaRk(\bsfp)$.

Then, by the definitions of $\LambdaRk'$ and~$\LambdaRk$, we have
\begin{equation}
\LambdaRk'(\bsfp) := D_{\mathsf{p}}\Lambdak(\bsfXp,\bsfp)  \label{aux71}
\end{equation}
where the short-hand notation $\bsfX$ is defined by~\eqref{X}, $\bsfXp$ is the solution of the stationarity system for given~$\bsfp$ and $D_{\sfp}$ signifies the \emph{total derivative} w.r.t.~$\bsfp$. Setting the Lagrangian~\eqref{L:def} in concise form as
\begin{equation}
  \Lcal(\bsfX,\bfw,\bsfp) = \Lambdak(\bsfX,\bsfp) - \Ccal(\bsfX,\bfw), \label{L:concise}
\end{equation}
where the functional $\Ccal$ is bilinear and symbolizes the PDE constraint~\eqref{balance:weak}, the total derivative of $\Lcal$ w.r.t. $\bsfp$ is given by
\begin{equation}
  D_{\mathsf{p}}\Lcal(\bsfXp,\bfwp,\bsfp)
 = D_{\mathsf{p}}\Lambdak\lpar \bsfXp,\bsfp \rpar - D_{\mathsf{p}}\Ccal\lpar \bsfXp,\bfwp \rpar
 = D_{\mathsf{p}}\Lambdak\lpar \bsfXp,\bsfp \rpar, \label{aux13}
\end{equation}
since by the definition of the dependence of $\bsfXp$ and $\bfwp$ on $\bsfp$, the PDE constraint is satisfied for each $\bsfp$ (and hence has a zero total derivative).
Letting $(\bsfXp^\prime,\bfwp^\prime)$ denote the derivative w.r.t.~$\bsfp$ of the stationarity solution $(\bsfXp,\bfwp)$ and making use of~\eqref{MECR:def} and~\eqref{L:concise}, the total derivative of $\Lcal$ can alternatively be expressed via chain rule to obtain
\begin{equation}
\begin{aligned}
   \lbra D_{\mathsf{p}}\Lcal(\bsfXp,\bfwp,\bsfp), \bsfpH\rbra
 &=\lbra \del{\bsfX}\Lcal,\bsfXp^\prime\sip\bsfpH \rbra + \lbra \del{\bfw}\Lcal,\bfwp^\prime\sip\bsfpH \rbra + 
 \lbra \del{\mathsf{p}}\Lcal,\bsfpH \rbra \\
 &= \lbra \del{\mathsf{p}}\Lcal(\bsfXp,\bfwp,\bsfp),\bsfpH \rbra
  =  \lbra \del{\mathsf{p}}\Lambdak(\bsfXp,\bsfp),\bsfpH \rbra
  = \lbra \del{\mathsf{p}}\Ecal(\bsfXp,\bsfp),\bsfpH \rbra, \quad 
 \end{aligned} \label{aux72}
\end{equation}
since all equations of the stationarity system, \eqref{group1} and~\eqref{group2}, are by premise verified and (for the last two equalities) the explicit dependence of $\Lcal$ on $\bsfp$ is confined to $\Ecal$, see~\eqref{MECR:def} and~\eqref{L:def}. Combining~\eqref{aux71}, \eqref{aux13} and~\eqref{aux72} gives the claimed expression for $\LambdaRk'(\bsfp)$.

\subsection{Proof of Proposition~\ref{MECR:eval2}}\label{MECR:eval:proof2} 

\paragraph{Proof of formula~\eqref{SS:integr}}
Recalling~\eqref{SSorig}, we have
\begin{equation}
  \SS_t[\bfw]\dip\bfeta(t)
 = \bfeta\dip\lpar \CeS + \Tinv \DS\Isub \rpar \dip\bfeta + \bfeta\dip\CHT\dip\Ca^{-1}\dip\CH\dip\bfeta + \lpar \bfa-T\bfbe \rpar\dip\CH\dip\bfeta. \label{aux18}
\end{equation}
We begin by using~\eqref{ODE:beta:exp} to express $\CH\dip\bfeta$ in terms of $\bfbe$ in the above sum, whose last two terms become
\begin{equation}
  \bfeta\dip\CHT\dip\Ca^{-1}\dip\CH\dip\bfeta + \lpar \bfa-T\bfbe \rpar\dip\CH\dip\bfeta
 = T \lpar \bfa-T\Ca^{-1}\dip\Da\dip\dot{\bfbe}\rpar \dip\lpar \Ca\dip\bfbe\shm\Da\dip\dot{\bfbe} \rpar \label{aux17}
\end{equation}
Moreover, the ODE~\eqref{a:ODE} verified by $\bfa$ provides
\begin{equation}
  T\bfa\dip\Ca\dip\bfbe = T(\bfbe+T\dot{\bfbe})\dip\Da\dip\bfbe - T\dot{\bfa}\dip\Da\dip\bfbe.
\end{equation}
Using this in~\eqref{aux17}, expanding and rearanging, we obtain
\begin{equation}
  \bfeta\dip\CHT\dip\Ca^{-1}\dip\CH\dip\bfeta + \lpar \bfa-T\bfbe \rpar\dip\CH\dip\bfeta
 = T\bfbe\dip\Da\dip\bfbe - T\del{t}(\bfa\dip\Da\dip\bfbe) + T^2\dot{\bfbe}\dip\Da\dip\Ca^{-1}\dip\Da\dip\dot{\bfbe}
\end{equation}
We now use the above in~\eqref{aux18} and integrate the resulting equality over $t\shin[0,T]$ and use $\bfbe(T)=\bfa(0)=\bfze$, to obtain the claimed formula\enlargethispage*{1ex}
\begin{equation}
  \int_0^T \SS_t[\bfw]\dip\bfeta(t) \dt
 = \int_0^T \bfeta\dip\lpar \CeS + \Tinv \DS\Isub \rpar \dip\bfeta
  + T\bfbe\dip\Da\dip\bfbe + T^2\dot{\bfbe}\dip\Da\dip\Ca^{-1}\dip\Da\dip\dot{\bfbe}
  \dt.
\end{equation}

\paragraph{Evaluation of $\Ecalc$, $\Ecalec$ and $\Ecalvc$} 
We proceed by obtaining first $\Ecalc$, then $\Ecalec$. Using the Euler's theorem~\eqref{aux11} for all relevant potentials in~\eqref{LF:gap:def:psi}, \eqref{LF:gap:def:phi} and~\eqref{ECR:decomp} and rearranging terms, the ECR functional is obtained as 
\begin{align}
\Ecal
&= \demi\iO\int_0^T \Lcb (\del{\eps}\psi \shm \bfsige)\dip\bfeps + T\lpar \del{\dot{\eps}}\varphi \shm \bfsigv)\dip\bfepsd
 + (\del{\alpha}\psi \shp \bfA)\dip\bfal + T(\del{\dot{\alpha}}\varphi \shm \bfA)\dip\dot{\bfal} \suite\qquad
 + (\del{\sige}\psi^{\star} \shm \bfeps)\dip\bfsige + T\lpar \del{\sigv}\varphi^{\star} \shm \bfepsd)\dip\bfsigv
 + (\del{A}\psi^{\star} \shp \bfal)\dip\bfA + T(\del{A}\varphi^{\star} \shm \bfald)\dip\bfA \Rcb \dt\dV
\end{align}
wherein the field variables solve the stationarity system~\eqref{group1}--\eqref{group2}. Invoking equations~(\ref{group1:exp}c,d) eliminates all terms involving $\bfA$ in the above formula, while the remaining terms are evaluated by means of~(\ref{group1:exp}a,b), temporal integration of $T\lpar \del{\dot{\eps}}\varphi - \bfsigv)$ by parts, the second of~\eqref{aux07}, and~\eqref{aux03}. Using further the fact that~$\bfbe(T)=\bfze$ and the assumption of quiescent past (in that $\bfeps[\bfu](0)=\bfze$), we obtain
\begin{equation}
\Ecal = \demi\iO\int_0^T \Lcb \! -\bfsigR[\bfwR]\dip\bfeps[\bfu] + (\bfsige\shp\bfsigv)\dip\bfeps[\bfw] \Rcb \dV\dt - 
\demi\iO \lpar \bfeps[\bfw]\dip\DS\Isub\dip\bfeps[\bfu] \rpar \rabs^T_0 \dV.
\end{equation}
Finally, invoking Lemma~\ref{reciprocity} and the expression for $\bfsig=\bfsige\shp\bfsigv$ from~\eqref{sigma:expr:stat}, we find  
\begin{equation}
\Ecal \,=\, \check\Ecal(\bfwp,\bsfp) := \demi\iO\int_0^T \SS_t[\bfw]\dip\bfeps[\bfw] \dV\dt. \label{aux12}
\end{equation}

We proceed in a similar way to compute $\Ecalec$. Using again~\eqref{aux11}, we have
\begin{equation}
\Ecale = \demi\iO\int_0^T \Lcb (\del{\eps}\psi \shm \bfsige)\dip\bfeps + (\del{\alpha}\psi \shp \bfA)\dip\bfal
+ (\del{\sige}\psi^{\star} \shm \bfeps)\dip\bfsige + (\del{A}\psi^{\star} \shp \bfal)\dip\bfA \Rcb \dt\dV.
\end{equation}
Making use of~\eqref{schur:C}, \eqref{ssA} and eliminating $\bfald$ via~\eqref{R:beta:def} we obtain 
\begin{equation}
\del{A}\psi^{\star} + \bfal = \CmsT\dip\bfsige + \CA\dip\bfA + \bfal = \Da^{-1}\dip\Dm\dip\bfeta + T\bfbe,
\end{equation}
and similarly
\begin{align}
\del{\sige}\psi^{\star} - \bfeps &= \bfeta, \\
\del{\alpha}\psi + \bfA &= T\Ca\dip\bfbe - \CH\dip\bfeta, \\
\del{\eps}\psi - \bfsige &= T\CmT\dip\bfbe  + \lpar \CmT\dip\Da^{-1}\dip\Dm - \Ce \rpar\dip\bfeta, &\qquad
\end{align}
by way of~(\ref{group1:pointwise}a), \eqref{ODE:beta:coefs} and~\eqref{aux07}. As a result, we find
\begin{align}
\Ecale &= \demi\iO\int_0^T \Lcb T\lsqb \Cm\dip\bfeps + \Ca\dip\alpha + \bfA \rsqb\dip\bfbe
+ \lsqb \bfeps\dip\lpar \CmT\dip\Da^{-1}\dip\Dm - \Ce \rpar - \bfal\dip\CH + \bfA\dip\Da^{-1}\dip\Dm + \bfsige \rsqb\dip\bfeta \Rcb \dt\dV \\
& \Eq{(a)}  \demi\iO\int_0^T \Lcb \bfeta\dip\CS\Isub\dip\bfeta - 2T\bfeta\dip\CHT\!\dip\bfbe + T^2\bfbe\dip\Ca\dip\bfbe \Rcb \dt\dV \\
&=\, \Ecalec(\bfwp,\bsfp) := \demi\iO\int_0^T \Lcb \bfeta\dip\CeS\dip\bfeta + (T\Ca\dip\bfbe-\CH\dip\bfeta)\dip\Ca^{-1}\dip(T\Ca\dip\bfbe-\CH\dip\bfeta) \Rcb \dt\dV, \label{ERe:expr}
\end{align}
where (a) follows from using the expressions for $\bfsige$ and~$\bfA$ in~\eqref{ssA}, eliminating $\bfald$ by means of~\eqref{R:beta:def}, and subsequent rearrangements. We finally use the ODE~\eqref{ODE:beta:exp} governing $\bfbe$ in~(a) to obtain the sought expression for $\Ecalec(\bfwp,\bsfp)$. Then, the formula for $\Ecalvc(\bfwp,\bsfp)$ follows immediately by subtracting the last result from from~\eqref{aux12} and making use of~\eqref{SS:integr}.
}

\subsection{Proof of Proposition~\ref{Lcurve:prop}}
\label{Lcurve:prop:proof}

For future reference, we conveniently write 
\begin{equation} \label{auxxx}
L(\kappa) \:=\: \LcalB(\bsfY(\kappa),\bfw(\kappa),\kappa) \,:=\: \Lcal(\bsfX,\bfw,\bsfp), \qquad \bsfY:=(\bsfX,\bsfp),
\end{equation}
where $\bsfY$ gathers all (field and constitutive) variables sought by the full-space minimization problem~\eqref{full:min} and $\bsfY,\bfw$ are treated as functions of $\kappa$, noting that the ensuing proof assumes sufficient differentiability of $\kappa\mapsto\bsfY(\kappa)$. In this vein, we also define $E(\kappa)=\Ecal(\bsfY(\kappa))$ and $M(\kappa)=\tdemi\McalT( \bsfX(\kappa))$.

For item (i), taking the total derivative of $L(\kappa)$ \emph{at the $\kappa$-dependent (constrained) minimizer of $\Lambdak$} and writing $D_{\kappa}(\dotp) = (\dotp)'$ for univariate functions, we obtain
\begin{equation}
 L'(\kappa)
 \,=\, \lbra\del{\bsfY}\LcalB,\bsfY'\rbra + \lbra\del{\bfw}\LcalB,\bfw'\rbra + \del{\kappa}\LcalB
 \,=\, \del{\kappa}\LcalB = M(\kappa), \label{aux001}
\end{equation}
since (a) $\bsfY$ solves  the first-order optimality conditions by premise, and (b) entries $\bsfY'$ and  $\bfw'$ belong to the requisite function spaces. On the other hand, as the minimizer of $\Lambdak=\Ecal+\tdemi\kappa\exs\McalT$ satisfies the PDE constraint for any $\kappa$, we have
\begin{equation}
  L'(\kappa) = \lpar E(\kappa) + \kappa M(\kappa) \rpar' = E'(\kappa) + \kappa M'(\kappa) + M(\kappa).\label{aux002}
\end{equation}
From~\eqref{aux001}--\eqref{aux002}, we deduce the property
\begin{equation}
  E'(\kappa) + \kappa M'(\kappa) = 0. \label{DM:der}
\end{equation}
Next, we use the fact that $\lbra\del{\bsfY}\LcalB,\bsfY'\rbra=0$ remains satisfied for any $\kappa$ as a result of the stationarity condition $\del{\bsfY}\LcalB=\bfze$.
Recalling~\eqref{auxxx} and taking the total derivative of this identity, we find
\begin{multline}
  0 = \lbra\del{\bsfY}\LcalB,\bsfY'\rbra'
  = \lbra\del{\bsfY\bsfY}\LcalB,\bsfY'\otimes\bsfY'\rbra + 
  \lbra\del{\bsfY\bfw}\LcalB,\bsfY'\otimes\bfw' \rbra +
  \lbra\del{\bsfY\kappa}\LcalB,\bsfY' \rbra + 
  \lbra\del{\bsfY}\LcalB,\bsfY''\rbra \\
  = \lbra\del{\bsfY\bsfY}\LcalB,\bsfY'\otimes\bsfY'\rbra + \tdemi\lbra\del{\bsfY}\McalT,\bsfY' \rbra
  = \lbra\del{\bsfY\bsfY}\LcalB,\bsfY'\otimes\bsfY'\rbra + M'(\kappa),
\end{multline}
where the second equality exploits: (i) $\bsfY''$ as a suitable set differentiation directions for the stationarity condition $\del{\bsfY}\LcalB=\bfze$, (ii) the last equality in~\eqref{aux001}, and (iii) the relationship
\begin{equation}
\begin{aligned}
\lbra\del{\bsfY\bfw}\LcalB,\bsfY'\otimes\bfw' \rbra &=
\lbra\del{\bsfY\bfw}\lcb\Lambdak\lpar \bsfX,\bsfp \rpar - \Ccal\lpar \bsfX,\bfw \rpar\rcb,\bsfY'\otimes\bfw' \rbra \\
& = -\lbra\del{\bsfX\bfw}\Ccal\lpar \bsfX,\bfw \rpar,\bsfX'\otimes\bfw' \rbra = -\Ccal(\bsfX',\bfw')=0
\end{aligned}
\end{equation}
due to~\eqref{L:concise}, since $\Ccal$ is a bilinear functional, $\bsfX(\kappa)$ always satisfies the PDE constraint, and $\bfw'$ is a valid test function. On recalling the convexity of~$\Lambdak$, we similarly have
\begin{equation}
\begin{aligned} 
\lbra\del{\bsfY\bsfY}\LcalB,\bsfY'\otimes\bsfY' \rbra 
&= \lbra\del{\bsfY\bsfY}\Lambdak(\bsfY),\bsfY'\otimes\bsfY'\rbra -\lbra\del{\bsfX}\Ccal(\bsfX',\bfw),\bsfX'\rbra \\
&= \lbra\del{\bsfY\bsfY}\Lambdak(\bsfY),\bsfY'\otimes\bsfY'\rbra \,\geqslant\, 0
\end{aligned}
\end{equation}
by the second-order Karush-Kuhn-Tucker optimality conditions. Hence $M'(\kappa)\leqslant0$ and, by property~\eqref{DM:der}, $E'(\kappa)\geqslant0$. This completes the proof of part (i).

We establish part (ii) of the claim by verifying that the curvature $\varrho=(M''E'-M'E'')/(M'{}^2+E'{}^2)^{3/2}$ of the L-curve is everywhere positive (i.e. the curvature center lies on the positive side of the normal under the present conditions, see Fig.~\ref{morozov}), which results from
\begin{equation}
\begin{aligned}  
M''E'-M'E'' &= M''E'-M'(E''\shp\kappa M''\shp M') + \kappa M'M'' + M'{}^2 \\
&= M''(E'\shp \kappa M') + M'{}^2 = M'{}^2 > 0
\end{aligned}
\end{equation}
since $E'\shp\kappa M'=0$ and $(E'\shp\kappa M')'=E''\shp\kappa M''\shp M'=0$.\enlargethispage*{1ex}

\end{document}